\documentclass[a4paper,fleqn,usenatbib]{mnras}

\usepackage{newtxtext,newtxmath}

\usepackage[T1]{fontenc}
\usepackage{ae,aecompl}

\usepackage{graphicx}	

\usepackage{epsfig}

\newcommand{\spitzer}{{\sl Spitzer}}
\newcommand{\servs}{{SERVS}}
\newcommand{\deepdrill}{{DeepDrill}}

\newcommand{\changed}[1]{{#1}}
\newcommand{\changedtwo}[1]{{#1}}

\title[near-infrared galaxy clustering]{The {\sl Spitzer} Extragalactic Representative Volume Survey
and DeepDrill extension: clustering of near-infrared galaxies}

\author[van Kampen et al.]{Eelco van Kampen$^{1}$\thanks{e-mail: evkampen@eso.org}, Mark Lacy$^{2}$, 
Duncan Farrah$^{3,4}$,  
Claudia del P. Lagos$^{5,6,7}$,  \newauthor
Matt Jarvis$^{8}$, Claudia Maraston$^{9}$,
Kristina Nyland$^{10}$, Seb Oliver$^{11}$, Jason Surace$^{12}$, \newauthor
Jessica Thorne$^{5}$  \newauthor
\\
$^1$ European Southern Observatory, Karl-Schwarzschild-Str. 2, 85748 Garching bei M\"unchen, Germany\\
$^2$ NRAO, 520 Edgemont Road, Charlottesville, VA 22903, USA\\
$^3$ Department of Physics and Astronomy, Watanabe 416, 2505 Correa Road, Honolulu, HI 96822, USA\\
$^4$ Institute for Astronomy, 2680 Woodlawn Drive, Honolulu, HI 96822-1897, USA\\
$^5$ International Centre for Radio Astronomy Research (ICRAR), M468, University of Western Australia,
     35 Stirling Hwy, Crawley, WA 6009, Australia. \\
$^6$ ARC Centre of Excellence for All Sky Astrophysics in 3 Dimensions (ASTRO 3D), {\tt https://astro3d.org.au/ }\\
$^7$ Cosmic Dawn Center (DAWN), Niels Bohr Institute, University of Copenhagen, Jagtvej 128, 2200 Copenhagen N, Denmark\\
$^8$ Oxford Astrophysics, Denys Wilkinson Building, Keble Road, Oxford OX1 3RH, UK\\
$^9$ Institute of Cosmology and Gravitation, Dennis Sciama Building, Burnaby Road, Portsmouth, PO1 3FX, UK\\
$^{10}$ U.S. Naval Research Laboratory, 4555 Overlook Ave SW, Washington, DC 20375, USA \\
$^{11}$ Astronomy Centre, Department of Physics and Astronomy, University of Sussex, Falmer, Brighton, BN1 9QH, UK\\
$^{12}$ IPAC, California Institute of Technology, Pasadena, CA 91125, USA\\
}

\date{Accepted 2023 May 11. Received 2023 April 28; in original form 2022 April 29}

\pubyear{2023}

\begin{document}
\label{firstpage}
\pagerange{\pageref{firstpage}--\pageref{lastpage}}
\maketitle

\begin{abstract}
We have measured the angular auto-correlation function of near-infrared galaxies in
\servs+\deepdrill, the \spitzer\ Extragalactic Representative Volume Survey and its
follow-up survey of the Deep Drilling Fields, in three large fields totalling
over 20 deg$^2$ on the sky, observed in two bands centred on 3.6 and 4.5 $\mu$m.
We performed this analysis on the full sample
as well as on sources selected by [3.6]-[4.5] colour in order to probe clustering
for different redshift regimes. We estimated the spatial correlation strength as well,
using the redshift distribution from S-COSMOS with the same source selection.
The strongest clustering was found for our bluest subsample, with <$z$>$\sim$0.7,
which has the narrowest redshift distribution of all our subsamples.
We compare these estimates to previous results from the literature, but also to
estimates derived from mock samples, selected in the same way as the
observational data, using deep light-cones generated from the SHARK semi-analytical model
of galaxy formation. For all simulated (sub)samples we find a slightly steeper slope than
for the corresponding observed ones, but the spatial clustering length is comparable in
most cases.
\end{abstract}



\section{Introduction}\label{intro}
Studying galaxy environments and their effects on galaxy evolution is one of the main motivations
for large, wide-area galaxy surveys. Galaxies are not spread evenly throughout the Universe, but 
cluster in pairs, groups and clusters: their distribution is clustered.
One consequence of that is that they interact, where the amount of clustering, which is usually
quantified using the 2-point correlation function for galaxies \citep{Peebles1980}, 
and its characteristic length-scale, the clustering strength $r_0$,
will determine the frequency and severeness of galaxy-galaxy or galaxy-group interaction,
and thus the evolutionary history of the galaxy population. 

We employ two recent "warm \spitzer" surveys to estimate the 2-point galaxy correlation function
(usually just called the 'galaxy correlation function' for short) for catalogues extracted from
the \spitzer\ Extragalactic Representative Volume Survey (\servs), which covers
18 deg$^2$ to 5$\sigma$ depths of 1$\mu$Jy at 3.6$\mu$m and 2$\mu$Jy at 4.5$\mu$m.
\servs, a \spitzer\ "Exploration Science" program (for full details this survey,
see \citealt{Mauduit2012}), and its extension, \deepdrill, performed in a similar fashion
(see \cite{Lacy2021} for details).
Both are deep enough and wide enough to contain a truly representative volume of the
Universe, sampling 0.8 Gpc$^3$ over $1 < z < 6$ in the case of \servs, and a similar volume
for \deepdrill, which was aimed at doubling the original survey size of \servs.
The three \servs+\deepdrill\ fields studied here are large enough to reliably estimate
the (rest-frame) NIR galaxy correlation function for each field individually, focusing on
the \spitzer\ IRAC bands centred on 3.6 and 4.5 $\mu$m.
In addition, the three fields taken together provide us a large total area
for a combined clustering analysis, in order to suppress any remaining sample variance
(often called 'cosmic variance' in a cosmological context).

\servs\ and \deepdrill\ cover a fairly wide redshift range, peaking at $z\approx 1$. 
Photometric redshifts are now available (Pforr 2019) for \servs, employing 12-15 bands
(depending on the field) of supplementary data, but their coverage and depth are unfortunately
still inhomogeneous and we lack redshifts for the \deepdrill\ extension. We also have few
(less than 2 percent) spectroscopic redshifts. So instead of using these, we use colour cuts
to select redshift distributions, allowing us to estimate spatial clustering lengths through
the Limber equation inversion technique.

Galaxies in the visual wavebands have clustering lengths of order 2--10 Mpc,
depending on colour, luminosity, and redshift interval, e.g. \cite{Coil2008} for the DEEP2 sample,
\cite{Zehavi2011} for the SDSS DR7 sample, and \cite{Christodoulou2012} for the GAMA sample. 

In the far-IR wavebands, for sources with $S_{250 \mu m}> 33\ \mu Jy$,
$r_0\approx 3-4\ h^{-1}$Mpc below $z<0.3$ \citep{vanKampen2012,Amvrosiadis2019},
while at high redshifts, sub-mm galaxies are more strongly clustered with
correlation lengths $r_0$ of $8.1 \pm 0.5\ h^{-1}$Mpc, $8.8 \pm 0.8\ h^{-1}$Mpc, and
$13.9 \pm 3.9\ h^{-1}$ Mpc at $z=$ 1-2, 2-3, and 3-5, respectively \cite{Amvrosiadis2019}. 
In the mid-IR, a clustering estimate at 24 micron is given by \cite{Gilli2007}:
$r_0 = 4.0 \pm 0.4\ h^{-1}$ Mpc for $f_{24} >20 \mu$Jy,
with no redshift selection. Later, \cite{Starikova2012} found $r_0 = 4.98 \pm 0.28\ h^{-1}$Mpc and
$r_0 = 8.04 \pm 0.69\ h^{-1}$Mpc for $<z> = 0.7$ and $<z> = 1.7$ populations,
respectively, for sources with $S_{24 \mu m}> 310\ \mu Jy$.

The first clustering estimate in the near-infrared was obtained in the K-band (i.e.\ around 2.2 micron)
for a galaxy population at relatively low redshifts by \cite{Baugh1996}.
Estimates from deeper and wider surveys followed, e.g.\ \cite{Waddington2007},
\cite{delaTorre2007}, \cite{Furusawa2011}, \cite{Hatfield2016}, \cite{Cochrane2018}, and others.
These were performed for a variety of NIR bands, survey sizes, and depths. Our current \servs+\deepdrill\
sample is larger than these previous efforts, with good survey homogeneity and completeness down to AB
magnitude 22 in both the 3.6 micron and 4.5 micron bands.

The purpose of this paper is to measure the spatial clustering length of flux-selected NIR galaxies
in \servs\ + \deepdrill\ to the best accuracy possible, for the full survey as well as for specific
subsamples, but also to compare to recent modelling efforts for this population. 
A very suitable set of simulations for this purpose is provided by the SHARK semi-analytical model
(\citealt{Lagos2018},2019), which make use of the SURFS N-body simulations \citep{Elahi2018},
the SED modelling software ProSpect \citep{Robotham2020} and the radiative transfer results of
the EAGLE hydrodynamical simulations to model attenuation \citep{Trayford2020}.

These multi-wavelength mock data sets include \spitzer\ bands to a depth exceeding our observational
data. In fact, the full simulated volume comprises a (wide) light-cone with an area of 107.9 deg$^2$,
a flux selection at the [3.6] band $> 0.575 \mu$Jy (equivalent to an $AB$ magnitude of 24.5),
and a redshift range $0 \leq z \leq 6$.

In a first comparison of these simulations to the \deepdrill\ data, \cite{Lacy2021} found that
for number counts the agreement between the model and observations is good overall,
particularly at faint magnitudes ($AB > 22$).
At intermediate magnitudes, $\sim$ 18 -- 21, the model light-cone predicts a lower number of galaxies
compared to both our counts and those in the literature of up to a factor of 1.8 in the [3.6] band
and 2.2 in the [4.5] band. We find the models somewhat too faint ($\sim$0.6 mag), with a small
colour difference of 0.05 mag. The main reason for this is probably that 
the SHARK simulations contain only the emission of galaxies, without the inclusion of active
galactic nuclei (AGN), which are expected to make a larger contribution at intermediate magnitudes.
Other aspects of the model, like the star formation history, dust treatment, or the adopted population
synthesis model, could also contribute to the number count mismatch. 

This paper is organized as follows: in Section 2 we describe in more detail the observational
data and simulations used, in Section 3 we elaborate on the methods used, and present in Section 4
the results of applying these methods to both the observed and mock data sets, including a range
of subsamples constructed by selecting on colour and/or luminosity. In Section 5 we
summarize and discuss our findings. We adopt the same cosmological parameters as used for
the SHARK simulations (which we compare to),
who used the \cite{PlanckColl2016} $\Lambda$CDM cosmology, so we set
$H_0 = 67.51$ km s$^{-1}$ Mpc$^{-1}$, $\Omega_\Lambda$ = 0.6751,
and $\Omega_{m}$ = 0.3121. Clustering length are quoted in $h^{-1}$Mpc, to facilitate
comparison of our estimates to various previous ones found in the literature.

\section{Observational and model data used}

\subsection{\spitzer\ imaging and catalogues}

The \spitzer\ data, taken in the 3.6 and 4.5 $\mu$m bands, were collected in two major efforts,
both in the post-cryogenic ('warm') part of the \spitzer\ mission. In the early months of that,
\servs\ \citep{Mauduit2012} imaged five highly observed astronomical fields (ELAIS-N1, ELAIS-S1,
Lockman Hole, Chandra Deep Field South, and XMM-LSS), requiring 1400 hours of observing time.
Three of these fields \changed{(listed in \autoref{tab:flux-ratios})} form the central
areas of the three \deepdrill\ fields used for this paper:
the DeepDrill Survey (Program ID 11086) was observed between 2015-05-04 and 2016-12-26
and specifically designed to extend three \servs\ fields, again requiring 1400 hours,
and doubling the original \servs\ survey area. Due to scheduling constraints, the coverage
is non-uniform, with areas around the edges of the \servs\ fields in particular receiving
additional coverage, and some outlying regions not receiving the full coverage.
We take this into account in our clustering analysis by excluding these outlying regions.

\deepdrill\ data processing was similar to that what was done for the \servs\ dataset
\citep{Mauduit2012}, using an existing data cleaning pipeline derived from processing
SWIRE and COSMOS data \citep{Lonsdale2003, Sanders2007}. In order to achieve homogeneous
coverage across each full \deepdrill\ field, all \servs\ image data contained therein
were recalibrated to the final warm mission calibration.
Full details on the final imaging, source extraction, and matching of the two catalogs
(the [3.6] and [4.5] bands) are found in \cite{Lacy2021}, and references therein
(notably \citealt{Vaccari2016}).

The total area of the three fields covered by \servs+\deepdrill\ is 28 sq. degrees,
of which just over 20 sq. degrees is used for the clustering estimates in this paper.
The area not included consists of border regions where the survey coverage is inhomogeneous
or incomplete. Circular regions around bright stars,
in which the source extraction was very incomplete, have also been excluded.
This was done by placing circular masks around stars brighter than $K=8$ in the
2-Micron All Sky Survey (2MASS, \citealt{Skrutskie2006}) whose size was determined
empirically by measuring the radius out to which the wings of the stellar PSF prevented
reliable object detection.
A regression line was then fit to this radius as a function of $K$-band magnitude, and objects
within a radius $318-35.2\ K$ arcseconds (where $K$ is the 2MASS $K$-band magnitude) 
of the 2MASS position were masked out. The final sample area consists of three round areas
with a radius of 1.5 degrees, with the star masks excluded.

\begin{table}
\centering
\begin{tabular}{lllrr}
\hline
\noalign{\vfilneg\vskip -0.2cm}
\hline
Field        &  RA  &   Dec  &    \multicolumn{2}{c}{DeepDrill/SERVS flux ratios}      \\
             &      &        &    3.6 micron   &   4.5 micron          \\
\hline
  ES1        &  00:37:56.2   &  -44:01:52  & 0.952  &   0.959    \\
  XMM-LSS    &  02:22:18.5   &  -04:48:10  & 0.957  &   0.983    \\
  CDFS       &  03:31:50.1   &  -28:07:06  & 0.951  &   0.969    \\
\hline
\noalign{\vfilneg\vskip -0.2cm}
\hline
\end{tabular}
\caption{As the SERVS and DeepDrill data were taken at different epochs during the \spitzer\
warm mission, there is a small difference in their fluxes, which we list here as ratios for 
each band and each field \changed{(for which the central coordinates are provided here too)}.
These \changed{ratios} were set by equalising the overal galaxy numbers density of
the two fields in each band, while retaining a colour distribution matching that of S-COSMOS.
The ratios thus found are consistent with those listed in the survey paper, \citep{Lacy2021}.}
\label{tab:flux-ratios}
\end{table}

Because the \servs\ and \deepdrill\ datasets were taken at different epochs of the 'warm' part
of the \spitzer\ mission, there is a small difference in photometric calibration between
the two datasets.
This is discussed in more detail in the survey paper \citep{Lacy2021}, where the ratio between
the fluxes in each survey is given for each band in each field. Due to the survey geometry it is
important to take this into account as well as we can: the \servs\ fields are rectangular, which were
extended all around by \deepdrill\ to yield a circular field of view with the \servs\ field
in the middle. Any mismatch in flux calibration would result in a different mean source density
for the \servs\ rectangle as compared to the circular surrounding of the \deepdrill\ pointings.
Such a large-scale mismatch produces an obvious artificial bump in the large-scale part of the angular
correlation function, as we don't expect any ring-like large-scale structure in the galaxy density
distribution. We therefore equalise the mean number density of sources for the two
datasets to derive a flux ratio that removes this artificial bump (which would arise from
small differences in photometric calibration between the \servs\ and \deepdrill\ datasets).

One subtlety here is that we need to do this for both Spitzer bands, where different flux
\changed{ratios} for each band changes the colour distribution of the population somewhat.
We therefore constrain the colour distribution to be as close as possible to that derived
for the S-COSMOS survey \citep{Sanders2007}, which was observed in the same \spitzer\ bands as
\servs+\deepdrill, while at the same time matching the mean source number density for both bands,
which is done using number density maps for each Spitzer band. This exercise yields the flux
ratios listed in \autoref{tab:flux-ratios}, which are somewhat more precise but still consistent
with the flux ratios reported in \citep{Lacy2021}, \changed{who derived these from the sources
in the overlap regions of the \servs\ and \deepdrill, i.e. sources found in both surveys}.

Once this flux ratio correction is applied,
we cut our source sample to account for completeness (in both magnitude and colour),
as defined in the survey definition paper \citep{Mauduit2012}: the full sample is
complete down to an AB magnitude of just over 23
in both bands. However, we cut at [3.6]$<$22 and [4.5]$<$22 to allow for colour cuts with
a range of about a magnitude, so that both band will always remain sufficiently complete.
\changed{We also cut at [3.6]$>$18 and [4.5]$>$18, as the contribution of stars increases
significantly at brighter magnitudes \citep{Sanders2007}.}

\begin{figure}
\centering
\includegraphics[width=1.05\columnwidth]{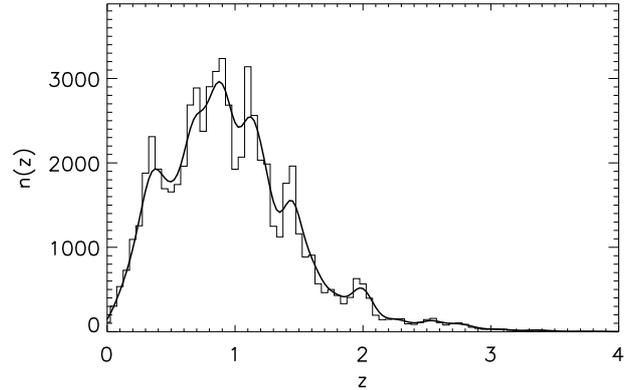}         
\caption{Histogram of the \changed{photometric} redshift distribution for S-COSMOS galaxies
with 18$<$[3.6]$<$22 and 18$<$[4.5]$<$22,
which is the overall flux cut used for our dataset for all clustering measures.
The smooth function is \changed{a Gaussian smoothed (with a width $\sigma_z=0.1$) version of the histogram,
and is subsequently used for estimating the spatial clustering strength (see main text for details).}
}
\label{fig:redshift-distribution} 
\end{figure}

\begin{figure}     
\includegraphics[width=1.01\columnwidth]{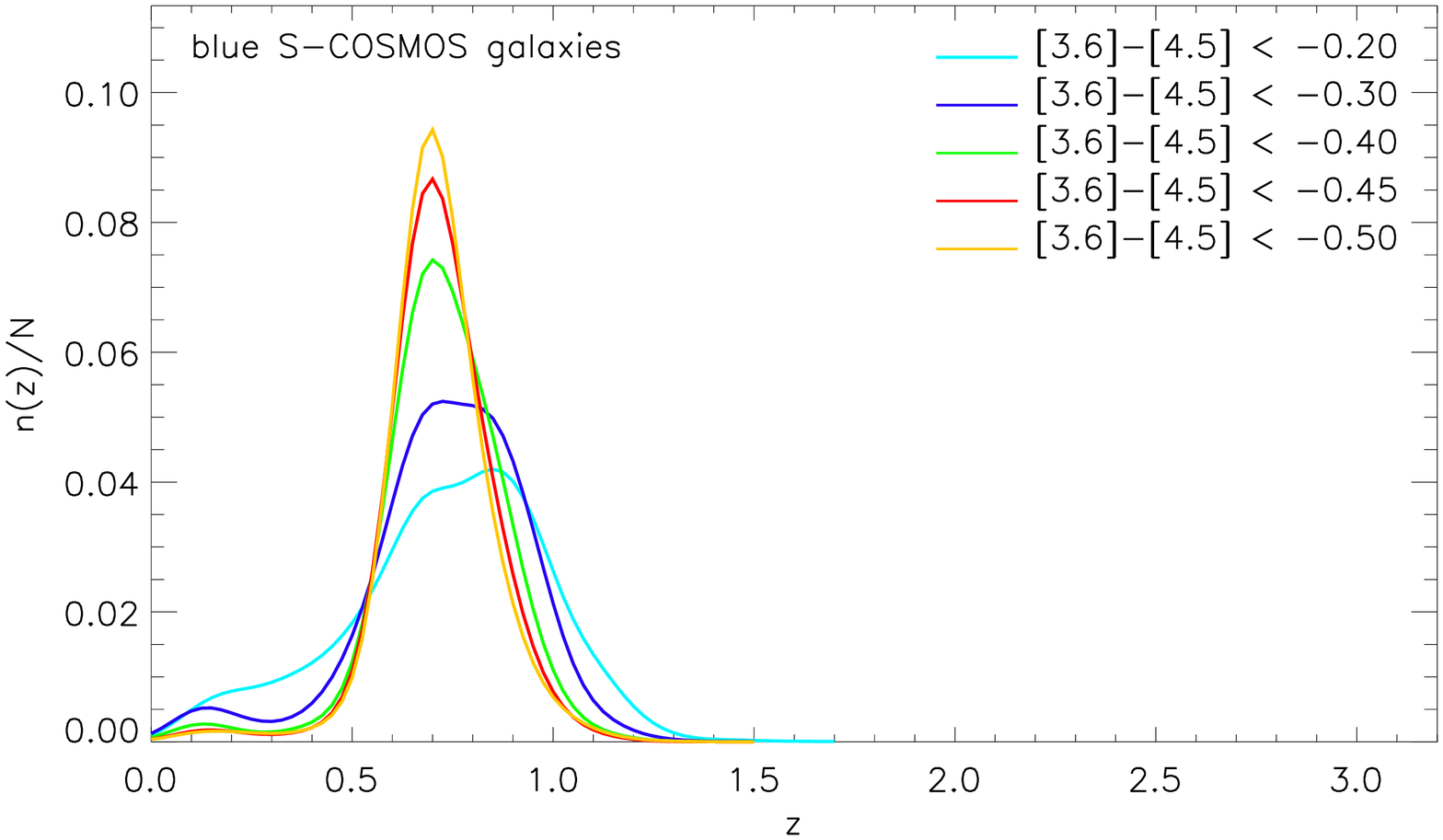}         
\includegraphics[width=1.01\columnwidth]{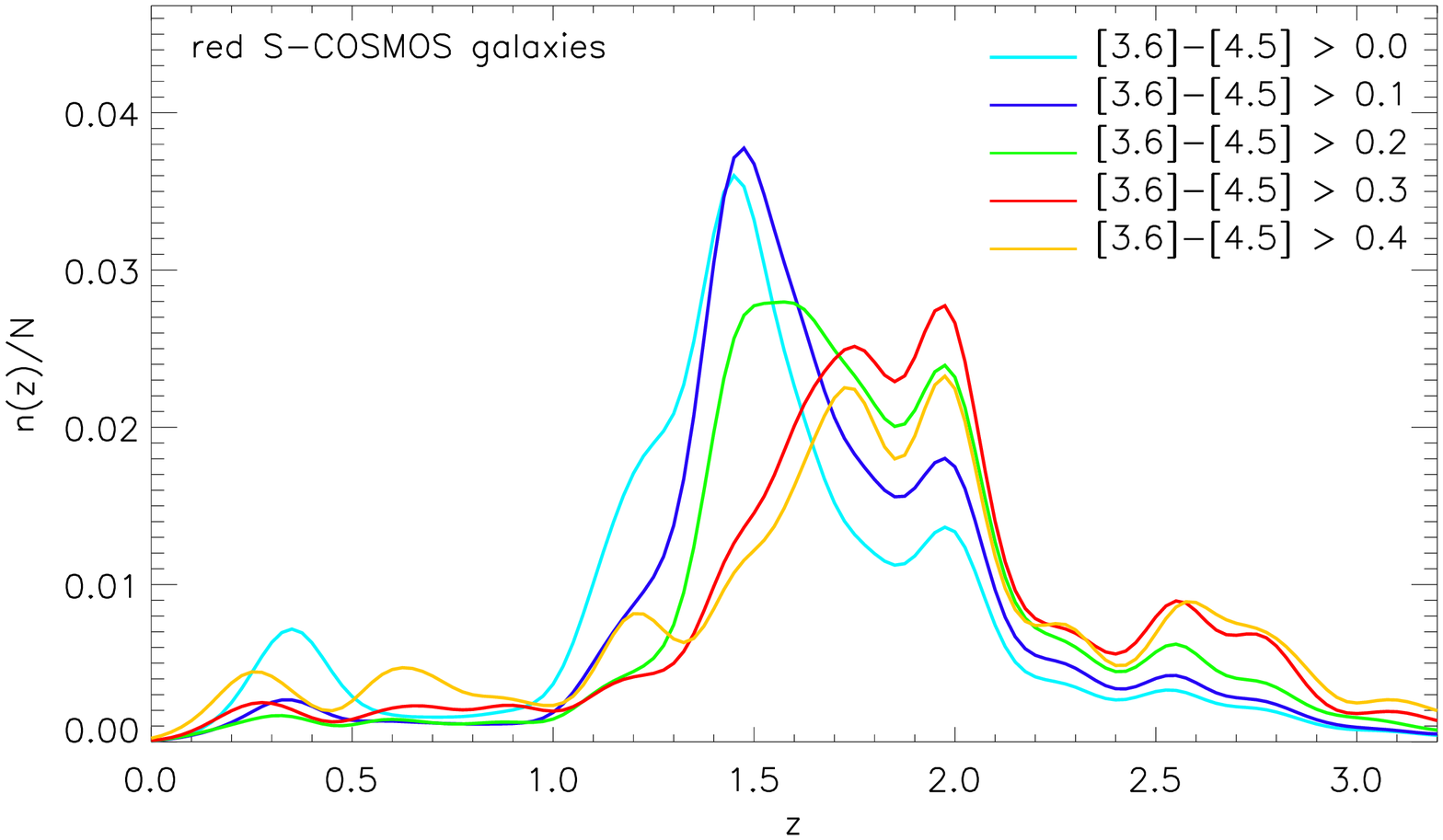}          
\caption{\changed{Photometric} redshift distribution for S-COSMOS galaxies for
different [3.6]-[4.5] colour cuts \changed{and down to our flux cuts}.
These distributions are taken as representative for subsamples
taken from our own survey (which lacks sufficient numbers of \changed{photometric} redshifts)
with the same colour cuts. The top panel is for blue galaxies, the bottom panel red ones.
The aim is to find those colour cuts that yield the narrowest redshift distributions (see text
for details). As the area of the S-COSMOS field is 2 deg$^2$, these distributions are not that
smooth (especially for the red subsamples). }
\label{fig:colour-cuts-zdist} 
\end{figure}

\subsection{Redshift distributions from colour cuts}

Following \citet{Papovich2008}, we employ the [3.6]-[4.5] colour to select subsamples with
a more restricted redshift distribution than the full sample. For angular clustering
measures, a narrow redshift range results in a much stronger clustering signal as it is not
averaged out as much along the line-of-sight. The cleanest subsample is derived
from the bluest galaxies: \citet{Papovich2008} found that cutting at [3.6]-[4.5]$<$-0.6 selects
(blue) galaxies in a relatively narrow redshift range around $z\sim 0.7$. They also found that
cutting at [3.6]-[4.5]$>$-0.1 selects for predominantly high redshift galaxies ($z>1.3$).

In order to establish which redshift distribution results from which cuts, both in colour and magnitude,
we use the 2 sq. degrees S-COSMOS field \citep{Sanders2007} observed in the same \spitzer\ bands as
\servs+\deepdrill, and with a wealth of supplementary data (30 bands and homogeneous coverage), which
yielded a large number of spectroscopic and photometric redshifts \citep{Ilbert2009}. 
\changedtwo{For our purposes the \citep{Sanders2007} catalog is sufficient, as it has a similar depth and
completeness as our sample, but we would like to mention that quite recently a deeper COSMOS
catalog became available (\citealt{Weaver2022}).}
The photometric redshift distribution for the S-COSMOS sample down to our flux limit in the [3.6]
and [4.5] bands is shown in \autoref{fig:redshift-distribution}. We currently
lack sufficient photometric redshifts for our own samples: we only have these for \servs\
(see \citealt{Pforr2019}), and even for these the coverage and depth are unfortunately
fairly inhomogeneous because some data lack the minimum of five bands needed for finding a reliable
photometric redshift.

\begin{figure}
\includegraphics[width=1.01\columnwidth]{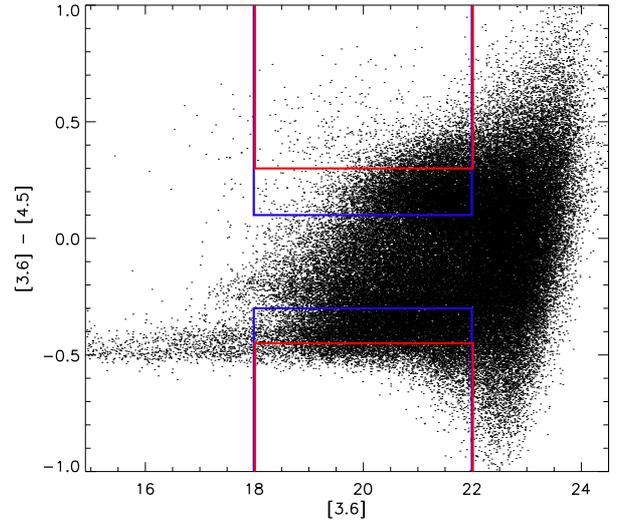} 
\caption{Colour selection and flux cut for the XMM field shown in the colour magnitude
relation of all observed galaxies, including those beyond the flux cut (in both bands).
The coloured lines correspond to the coloured redshift distributions shown in
\autoref{fig:colour-cuts-zdist}, where the top lines, selecting for red galaxies,
correspond to those in the bottom panel of \autoref{fig:colour-cuts-zdist},
and vice-versa for the bottom lines (selecting for blue galaxies).  }
\label{fig:colour-selection} 
\end{figure}

The choice of which colour cuts to use is somewhat arbitrary. The aim is to produce subsamples
with sufficient numbers of galaxies each and a redshift distribution that is as narrow as possible and
without too many outliers (both at lower and higher redshifts). Using the S-COSMOS sample we obtain
redshift distributions for a range of [3.6]-[4.5] colour cuts, with fixed magnitude cuts in both bands
as for our own samples: \changed{18$<$[3.6]$<$22 and 18$<$[4.5]$<$22}.
Two sets, one for negative (blue) cuts and one for positive (red) cuts, of such
(normalised and somewhat smoothed) redshift distributions are shown in \autoref{fig:colour-cuts-zdist}.
The actual number of galaxies left in each subsample is decreasing top-down for each of the panels.

Based on \autoref{fig:colour-cuts-zdist}, we choose two blue and two red cuts for our clustering
analysis. To fully exploit the large sample size while still restricting the redshift range as
compared to the full sample, we first cut moderately at [3.6]-[4.5]$ <$-0.3 and [3.6]-[4.5]$ >$0.1
to form a blue and red subsample respectively (the blue lines in \autoref{fig:colour-cuts-zdist}),
each containing roughly a quarter of all galaxies. The redshift distribution of the blue
subsample relatively clean and still somewhat broad, but not as broad as the [3.6]-[4.5]$ <$-0.2 subsample
(cyan line in \autoref{fig:colour-cuts-zdist}).
The redshift distribution of the red subsample has a clear
peak near $z\approx 1.5$, but with a high redshift tail and a second, smaller peak near $z\approx 2$.
It does lack the significant fraction of low-redshift source that a [3.6]-[4.5]$ >$0 cut includes,
which partly motivates our choice for a cut at [3.6]-[4.5]$ >$0.1 for the red sample.

For the blue part of the sample a narrower redshift distribution can be achieved
by cutting more severely. 
A cut at [3.6]-[4.5]$<$-0.5 (\changed{yellow} line in the top panel of \autoref{fig:colour-cuts-zdist})
produces the highest peak and the lowest fraction of low and high redshift galaxies,
\changed{but yields too small a sample for a robust clustering measurement.
We therefore cut at [3.6]-[4.5]$ <$-0.45 instead (red line in the top panel
of \autoref{fig:colour-cuts-zdist}), thus}
retaining sufficient numbers of galaxies for a clean clustering analysis.
\changed{Note that beyond [3.6]-[4.5]$ <$-0.5 the peak decreases again, and yielding very small
subsample sizes too.}
The peak of these 'very blue' galaxies is not that different from the 'blue' ones.
Such a narrow redshift distribution cannot be achieved in the red part of the sample,
and what is noteworthy here is that the peak of the
redshift distribution moves with the choice of the colour cut.
For a 'very red' subsample we choose to cut at [3.6]-[4.5]$ >$0.3 (red line in the bottom
panel of \autoref{fig:colour-cuts-zdist}), as this gives the highest peak at around $z\approx 2$,
and less foreground contamination as compared to an even more severe cut. However, there is a
significant fraction of high redshift galaxies beyond $z>2.5$ that cannot be avoided
by a [3.6]-[4.5] colour cut alone.

\begin{figure}
\includegraphics[width=1.04\columnwidth]{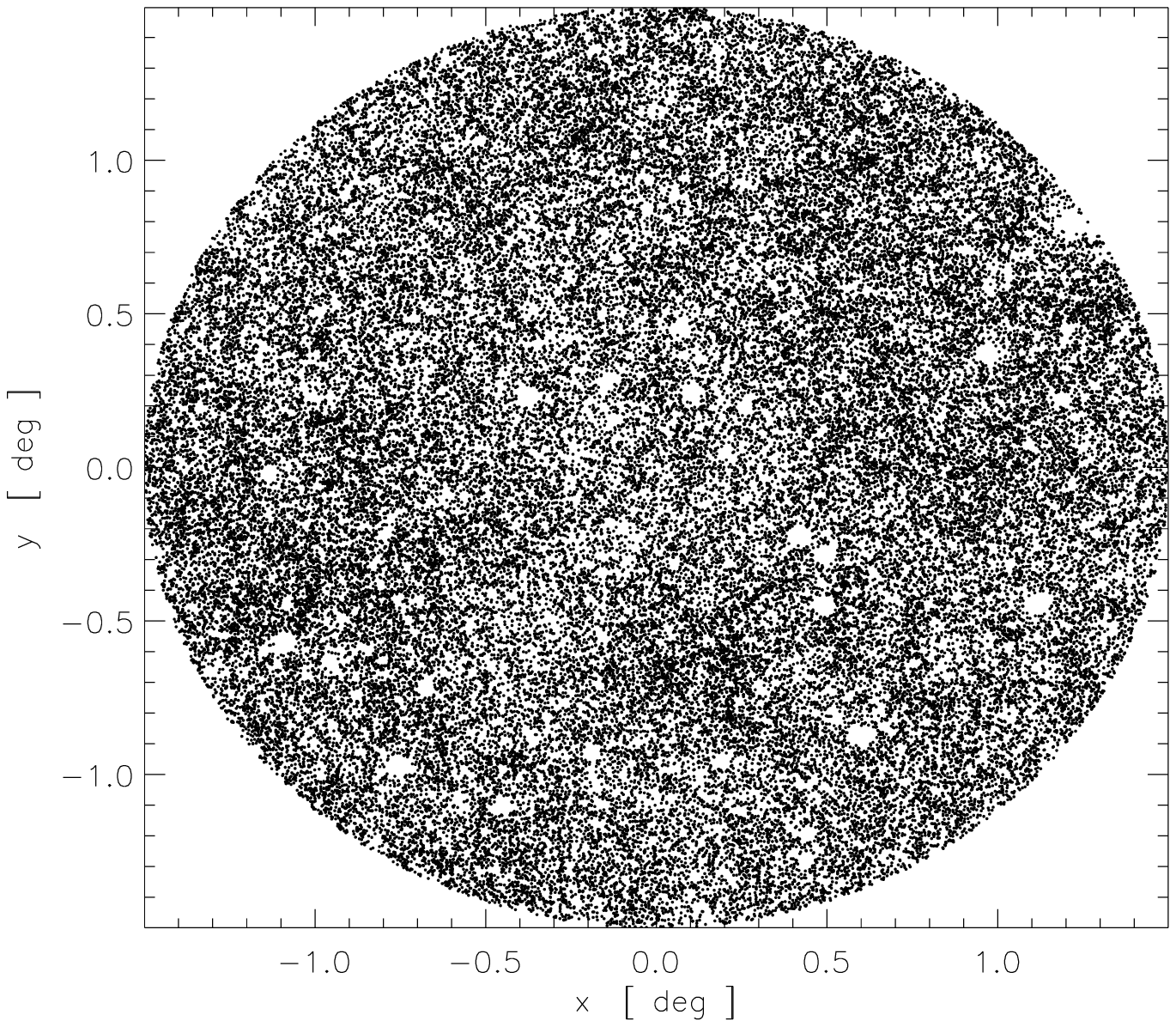}\\
\includegraphics[width=1.04\columnwidth]{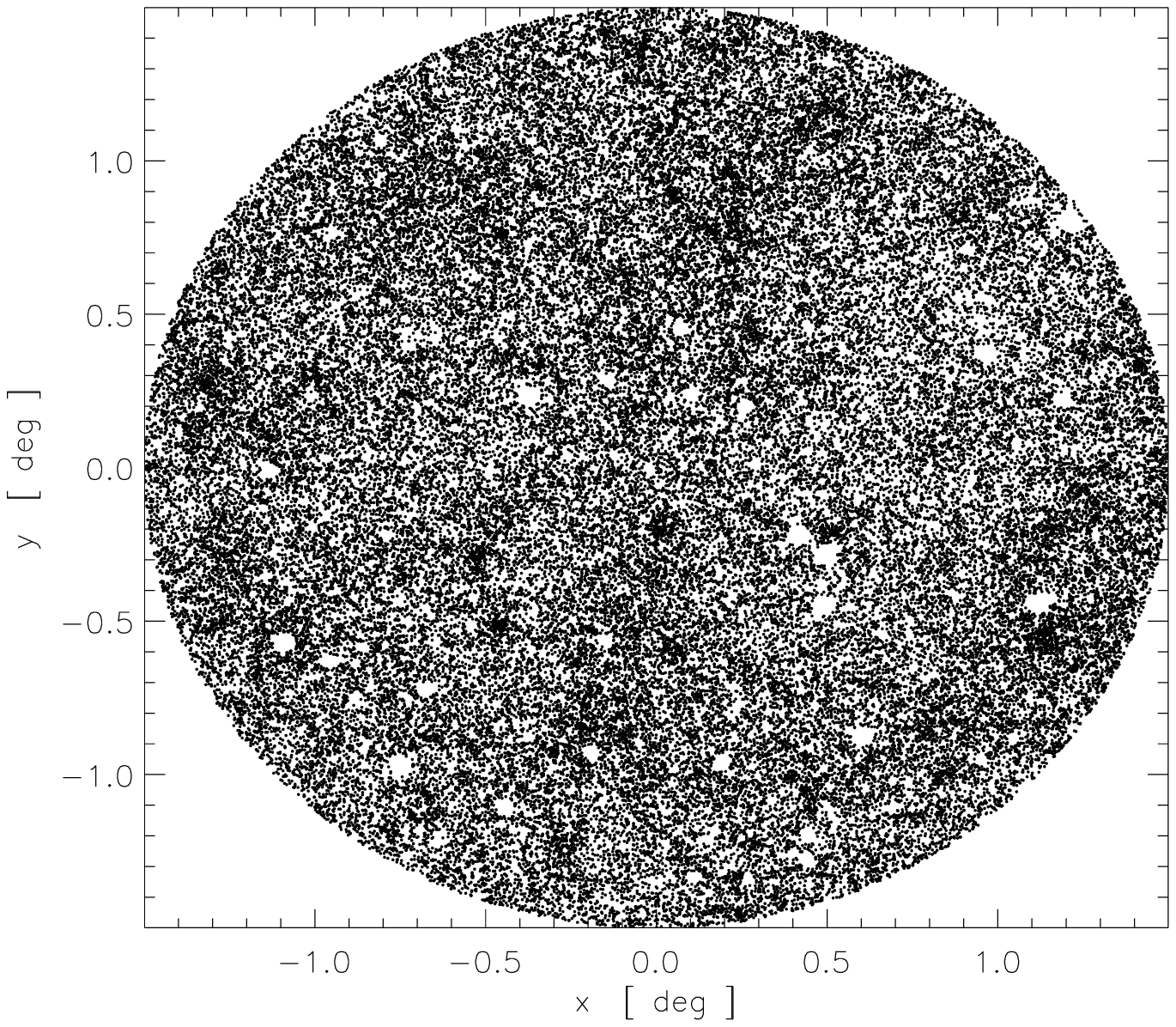} 
\caption{red (top) and blue (bottom) galaxies selected for the CDFS field. These colour
selections, [3.6]-[4.5]$>$0.1 and [3.6]-[4.5]$<$-0.3 respectively, are aimed at retaining a relatively
large number of galaxies for each colour, while still significantly narrowing the redshift
distribution as compared to the full sample. The star masks are clearly visible as white
circular areas.}
\label{fig:cdfs-selected} 
\end{figure}

The chosen cuts are shown in \autoref{fig:colour-selection}, which is the colour-magnitude diagram
for the XMM field. The red
and blue galaxy distributions resulting from these cuts are shown in \autoref{fig:cdfs-selected}.
Holes appear where bright star masks have been applied in the observed distribution.
The same masks are also applied to the random catalogues used to estimate the clustering strength.
The resulting subsamples for each field analysed in this paper are listed in the first three parts of
\autoref{tab:corr-results}, where the second column displays the number of galaxies remaining
after each selection. The fourth part of the table display the numbers for the three fields taken
together, which is what we use for the combined clustering estimates, helping to overcome field-to-field
variations. The \changed{bottom} part of the table concerns the simulated galaxy populations,
discussed next.

\begin{figure}
\includegraphics[width=1.04\columnwidth]{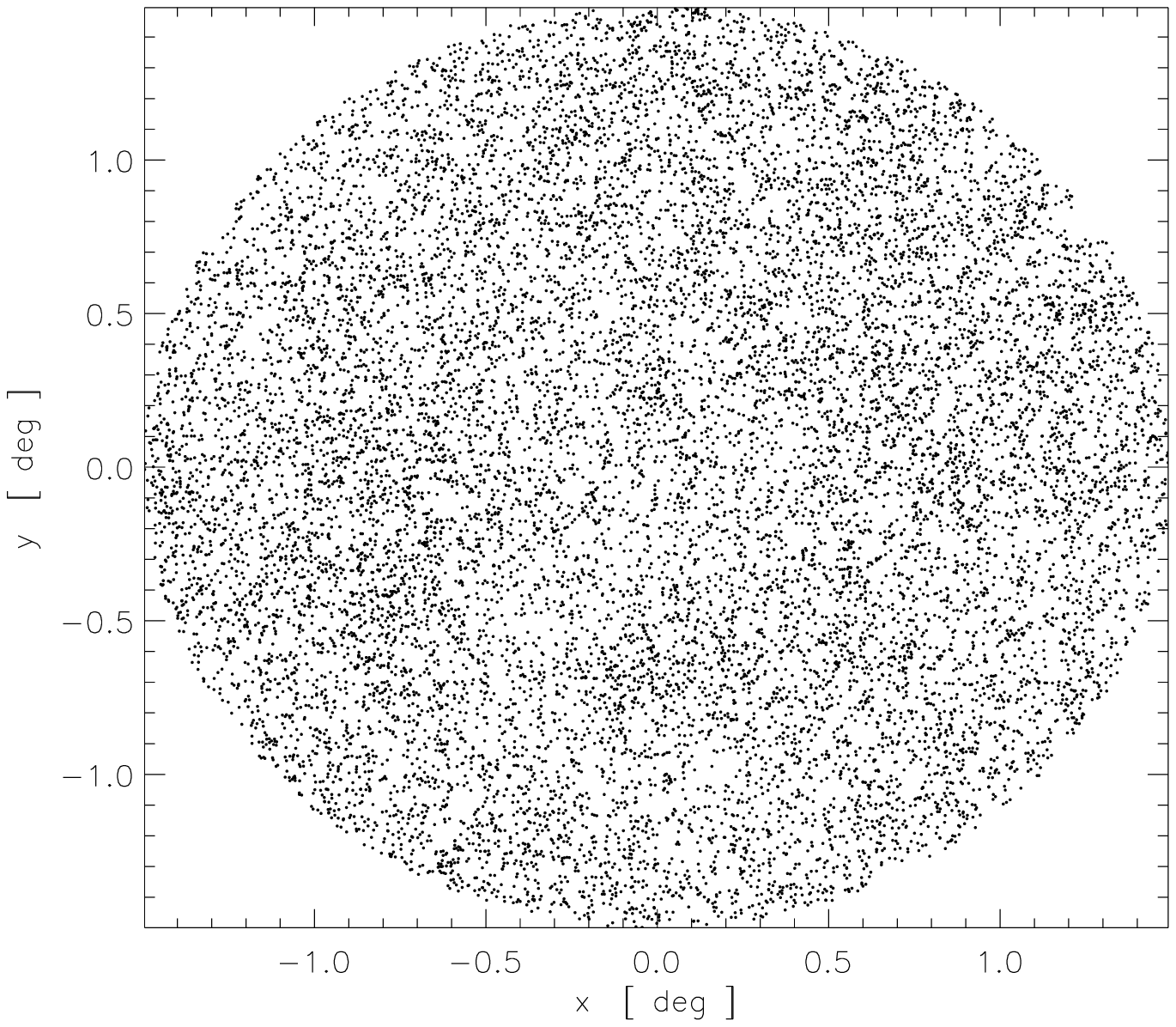}\\
\includegraphics[width=1.04\columnwidth]{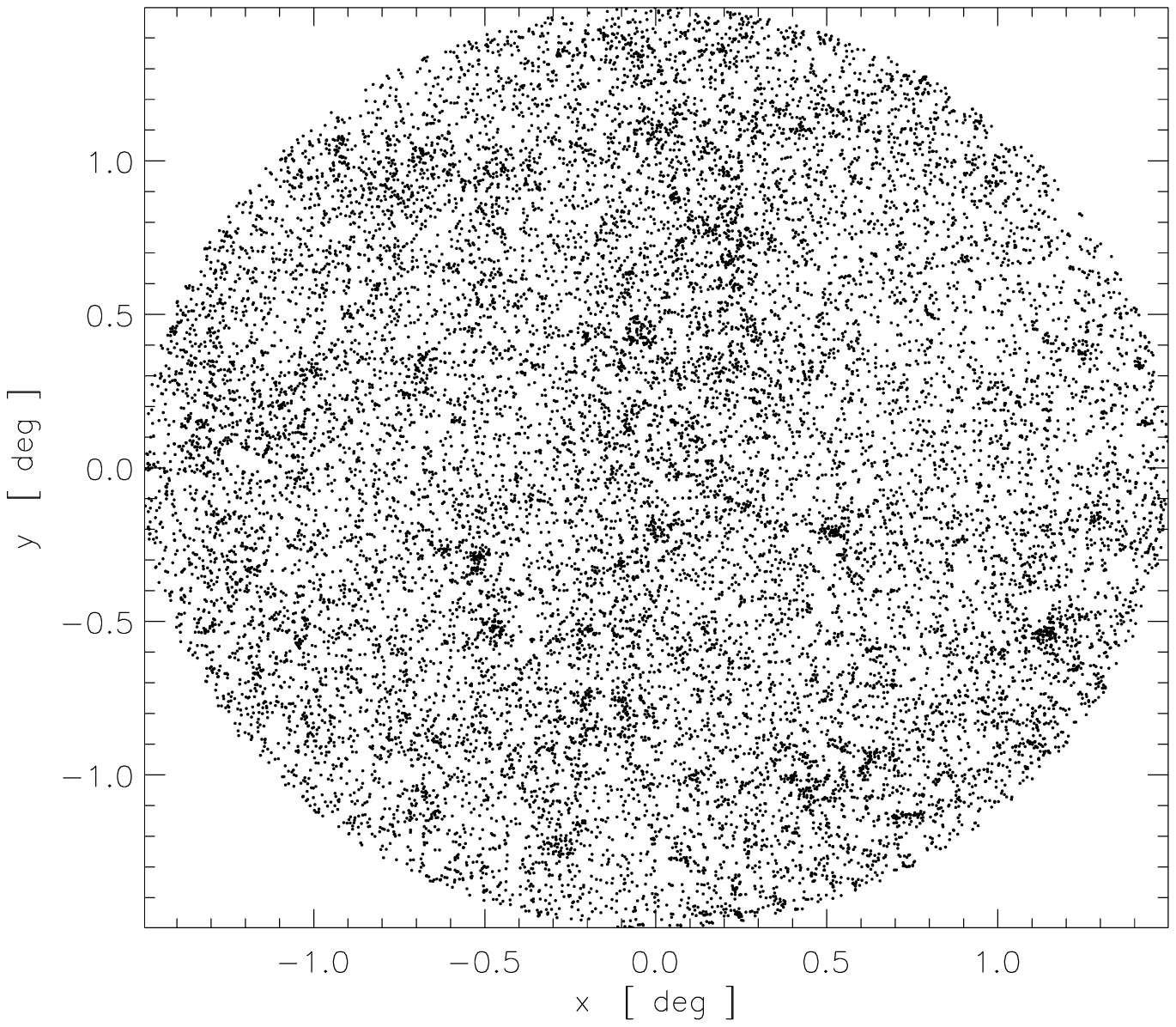} 
\caption{'Very red' (top) and 'very blue' (bottom) galaxies selected for the CDFS field. The
aim of 'very red' selection ([3.6]-[4.5]$>$0.3) is to select the highest redshift distribution,
with a similar dispersion as the blue and red population shown in \autoref{fig:cdfs-selected}, 
whereas the 'very blue' cut ([3.6]-[4.5]$<$-0.45) yields 
the narrowest redshift range with a clear peak $z\sim 0.7$. The clumpiness of the latter is easy
to spot by eye.}
\label{fig:cdfs-selected2} 
\end{figure}

\subsection{Simulated data used}

We employ the SHARK semi-analytical model of galaxy formation (\citealt{Lagos2018},2019), 
in which the intrinsic SEDs are modelled based on the galaxy's star formation and gas
metallicity history, to then be attenuated based on the galaxy's dust surface density
and employing the radiative transfer results of EAGLE from \cite{Trayford2020}.
Re-emission in the IR is done using the templates of \cite{Dale2014}.
We refer the reader to \cite{Lagos2019} and \cite{Robotham2020}
for more details on how SEDs are produced in SHARK. The outcome of this process is a set of 
mock datasets which are at least as complete as the observational data, in both \spitzer\
bands. We cut three simulated light-cones from the complete 107.9 deg$^2$ SHARK volume
with the same field of view and depth as for the three observed \servs +\deepdrill\ fields.

\subsubsection{Mimicking photometric accuracy}

For the flux range we consider, the photometry of the \servs +\deepdrill\ data has a fractional
uncertainty $\Delta f_{\rm [3.6]} / f_{\rm [3.6]}$ that varies with flux roughly as $f_{\rm [3.6]}^{-0.5}$,
as derived from the photometric catalogue of \cite{Lacy2021}. This uncertainty with respect to
the true flux is important for the position of a galaxy in the colour-magnitude diagram. There is a
density gradient in this diagram on the lines representing our flux and colour cuts (the coloured
line in \autoref{fig:colour-selection}). As the photometric uncertainty diffuses positions in the colour-magnitude
diagram, this gradient causes more galaxies to move to lower density parts of the diagram
(i.e. more extreme colours) than vice-versa, and affects the subsamples selected by the flux and/or colour cuts.

The mock galaxies have no photometric uncertainty, so this diffusion does not take place, and fewer mock galaxies
will scatter towards the more extreme colours as compared to the observed \servs + \deepdrill\ galaxies.
In order to mimic the photometric uncertainty that is present in the observational data, and its consequences for
subsample selection, a random offset for each mock galaxy is drawn from a normal distribution with the same width
as the uncertainty in the data as derived from the source finder used by \cite{Lacy2021},
ranging from (on average) less than a per cent at the bright end, to more than 5 per cent
at our flux limit of [3.6]$=22$.

\subsubsection{Incorporating source blending}

The mock light-cones are built from large cosmological volumes, which when projected
(in a random direction) will have sources overlapping on the sky. In the [3.6] band where we
estimate clustering, the point spread function (PSF) is 1.8 arcsec, so sources separated by less than that
in projection on the sky will be blended. These are mostly pairs, but also triples and some quadruples.
Those sources that are blended in this way are turned into single sources, with their fluxes summed
for both the [3.6] and [4.5] band. This is not immediately visible in the autocorrelation
function at the smallest scales, as we only estimate this function from around 5 arcsec onwards.
The blending does increase the flux for about 10 per cent of the sources, although this is
mostly by a relatively small amount in that many of these blend with fairly faint sources,
including those beyond our \changed{faint-end} flux cuts (at $AB=22$ for both bands).
The sample size decreases with sources above the flux cut being blended, but increases with
sources getting brighter through blending that then end up above the flux cut. The blending
somewhat changes the subsamples derived from selection in luminosity bins. Sources that blend also
change the colour of the resulting blended source, affecting any subsamples derived from colour cuts.

\begin{table}
\centering
\begin{tabular}{lrr}
\hline
\noalign{\vfilneg\vskip -0.2cm}
\hline
AB Mag bin   &      \multicolumn{2}{c}{Completeness fraction}      \\
             &      3.6 micron   &   4.5 micron                    \\
\hline									   
  18-19      &    1.0    &   1.0      \\   
  19-20      &    0.97   &   0.97     \\   
  20-21      &    0.945  &   0.94     \\   
  21-22      &    0.89   &   0.915    \\   
\hline  
  18-22      &    0.93   &   0.94     \\   
\hline
\noalign{\vfilneg\vskip -0.2cm}
\hline
\end{tabular}
\caption{Completeness as derived from Tables 8 and 9 of \citep{Lacy2021}. The completeness
fraction for the full sample (denoted as 'all galaxies') considered for our clustering analysis,
i.e. those in the magnitude interval $18<$AB<$22$, is the weighted mean of the fractions per bin.}
\label{tab:completeness}
\end{table}

\subsubsection{Matching number counts and completeness}

Comparing the number counts for the \servs+\deepdrill\ sample and the SHARK light-cones,
\cite{Lacy2021} found that these agree relatively well at faint magnitudes ($AB > 22$), but not
at brighter magnitudes: this may in part be driven by the lack of modelling of the AGN contribution
to the galaxy emission in SHARK.
At intermediate magnitudes, $\sim$ 18 -- 21, the model light-cone predicts a lower number of galaxies:
the models are somewhat too faint ($\sim$0.6 mag), with a small offset of 0.05 mag in colour.
As this is the range of magnitudes for which we estimate the autocorrelation function, we now
consider two sets of mocks: one set in which only the photometric uncertainty and source blending are 
mimicked, as described above, leaving the physics of the models as generated by SHARK untouched, and a second
set in which we additionally match for sample size through flux brightening (as the mock counts
are too low), which can be seen as compensating for the lack of AGN modelling in SHARK, amongst others.

\begin{figure}
\includegraphics[width=0.95\columnwidth]{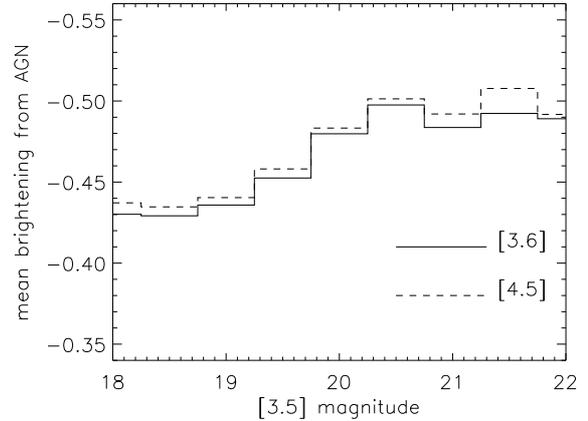}
\caption{Mean brightening (in magnitude) as a function of [3.6] when an AGN component is included in
the SED fitting for the DEVILS sample (see main text and \citealt{Thorne2022} for details), for each of the
two \spitzer\ bands considered in this paper. At the faint end the offset is half a magnitude, decreasing
somewhat for the brighter galaxies.}
\label{fig:agns}
\end{figure}

\changed{The contribution of AGN to the overal galaxy SEDs, including the near-infrared bands,
was recently estimated for two large samples by \cite{Thorne2022}. This was achieved by comparing SED
fits (using ProSpect: \citealt{Robotham2020}) with and without an AGN component. The \spitzer\ fluxes of the galaxies
derived from the global best fit SED can then be compared to those where the best fit AGN template has been
removed, yielding a flux ratio between these for each of the [3.6] and [4.5] bands.
In \autoref{fig:agns} we show these flux ratios as a magnitude difference (brightening) as a function of the
[3.6] magnitude for each of these two bands, 
for the deepest (and largest) of the two samples studied by \cite{Thorne2022}: DEVILS (\citealt{Davies2018}).
For [3.6]$>$20 the mean offset for each band is about half a magnitude. For the brighter galaxies this goes
down a bit to a brightening of -0.43 in magnitude.
}

For our second set of mocks we brighten the predicted fluxes with a single offset for each
band to get sample sizes that match the observed ones for {\it all} samples and subsamples considered.
Given the uncertainties, it is not realistic to get all simulated and observed sample sizes to match,
so the aim is to get a best match for most (sub)samples. We also need to deal with completeness, as the
observed sample will not be complete down to the flux limit, but the mock sample is (as it is much deeper).
The completeness fractions of \servs+\deepdrill\ for [3.6] and [4.5] magnitude intervals are listed
in Tables 8 and 9 in \cite{Lacy2021}. We map these fractions to a bin size of one magnitude,
which we use later on, and work out the overall completeness for our flux limits
\changed{18$<$[3.6]$<$22 and 18$<$[4.5]$<$22} by taking the weighted mean of these bins
(weighted by number counts).
These are listed in \autoref{tab:completeness} for each [3.6] and [4.5] catalogue individually.
As we select in both bands, completeness will be somewhat worse than either fraction listed (0.93 and 0.94
for [3.6] and [4.5] respectively) because of the variation in colour: a source missing in one band might
not be missed in the other, and vice versa, but it could be missing in both too. This means that the
completeness for our joint [3.6] and [4.5] cut will be below 0.93, but most likely not below $\sim$0.90.
As the simulated light-cones are several magnitudes deeper than the observed data, any mock 
survey drawn from SHARK down to our shallower flux limit should be practically complete. This implies
that for the nearly \changed{870000} \servs+\deepdrill\ sources found for our flux limits
\changed{18$<$[3.6]$<$22 and 18$<$[4.5]$<$22}
there should be between around \changed{935000} and \changed{967000} mock galaxies from SHARK
(roughly corresponding to completeness fractions of 0.93 and 0.90, respectively).

For these completeness fractions we find that brightening the mock galaxies by -0.565 mag at 3.6 micron
and -0.61 at 4.5 micron gives the best matching sample sizes for the three fields: the combination of
mimicking photometric accuracy and source blending, and applying these completeness corrections,
the three fields we selected from the full SHARK volume comprise a total of \changed{952970} galaxies,
corresponding to a joint [3.6] and [4.5] completeness fraction for the observational data of 0.916.
Besides yielding a reasonable sample size, this brightening of the mock galaxies is consistent
with the offsets found by \cite{Lacy2021}, including the small colour offset they found.
The latter is required to get the ratios of simulated red and blue galaxies to match
the ratios for the corresponding observed subsamples. This then constitutes the second set of mocks,
besides the first set of mocks for which the offsets for the two fluxes (-0.565 mag at 3.6 micron
and -0.61 mag at 4.5 micron) are {\it not} applied, and therefore has number counts in both bands that
are too low, and colours that are somewhat off.

\subsubsection{Resulting mock samples}

After mimicking photometric accuracy and source blending, we generated two sets of three mock fields,
one without any flux corrections, and one where each band is brightened by a fixed offset (see 
previous subsection). We then treat these mocks in exactly the same way as the observed fields.
For the first set of mock fields the counts and colours do not match well to the observed ones,
as discussed above, and as can be seen in \autoref{tab:corr-results} (the rows denoted 'all mock fields').
For the second (brightened) set of mocks, applying the same colour cuts to the mock data as for the
simulated data, for the three fields we retain \changed{225683} blue galaxies ([3.6]-[4.5]$<$-0.3),
and \changed{218231} red ([3.6]-[4.5]$>$0.1) ones (shown in the bottom part of \autoref{tab:corr-results}).
For the more extreme cuts we get \changed{42242} 'very blue' ([3.6]-[4.5]$<$-0.45) galaxies,
and \changed{46077} 'very red' ([3.6]-[4.5]$>$0.3) ones. There is some variation from the observed
counts for each of the colour cuts, but this cannot be changed by a colour offset, as there are too
many 'very blue' mock galaxies, too few blue ones, and more or less the right amount of red and
'very red' mock galaxies. As the choice of the colour cuts remains somewhat arbitrary, this is
good enough for our purposes.

\begin{table}
\centering
\begin{tabular}{llrrr}
\hline
\noalign{\vfilneg\vskip -0.2cm}
\hline
Subsample    & Colour cut     & median $z$  &  median sfr   &  median M$_*$    \\
             &  [3.6]-[4.5]   &             &  M$_\odot$/y   &   $10^{10}$M$_\odot$   \\
\hline
all          &    --    &   1.08  &  4.7   &  1.79   \\ 
very blue    &  < -0.45 &   0.68  &  1.2   &  4.57   \\ 
blue         &  < -0.3  &   0.72  &  2.5   &  1.23   \\ 
red          & > 0.1    &   1.77  &  12.3  &  2.63   \\ 
very red     &  > 0.3   &   1.89  &  28.2  &  5.62   \\ 
\hline
\noalign{\vfilneg\vskip -0.2cm}
\hline
\end{tabular}
\caption{\changedtwo{Typical properties for the galaxy population in our five subsamples,
as derived from the (brightened) mock catalogues.
The table lists median values for redshift, star formation rate, and stellar mass. The latter
two are expected to be somewhat larger for the corresponding observed samples, as in the step
of brightening the mocks these properties were not changed.}}
\label{tab:properties}
\end{table}

\begin{figure}       
\includegraphics[width=1.03\columnwidth]{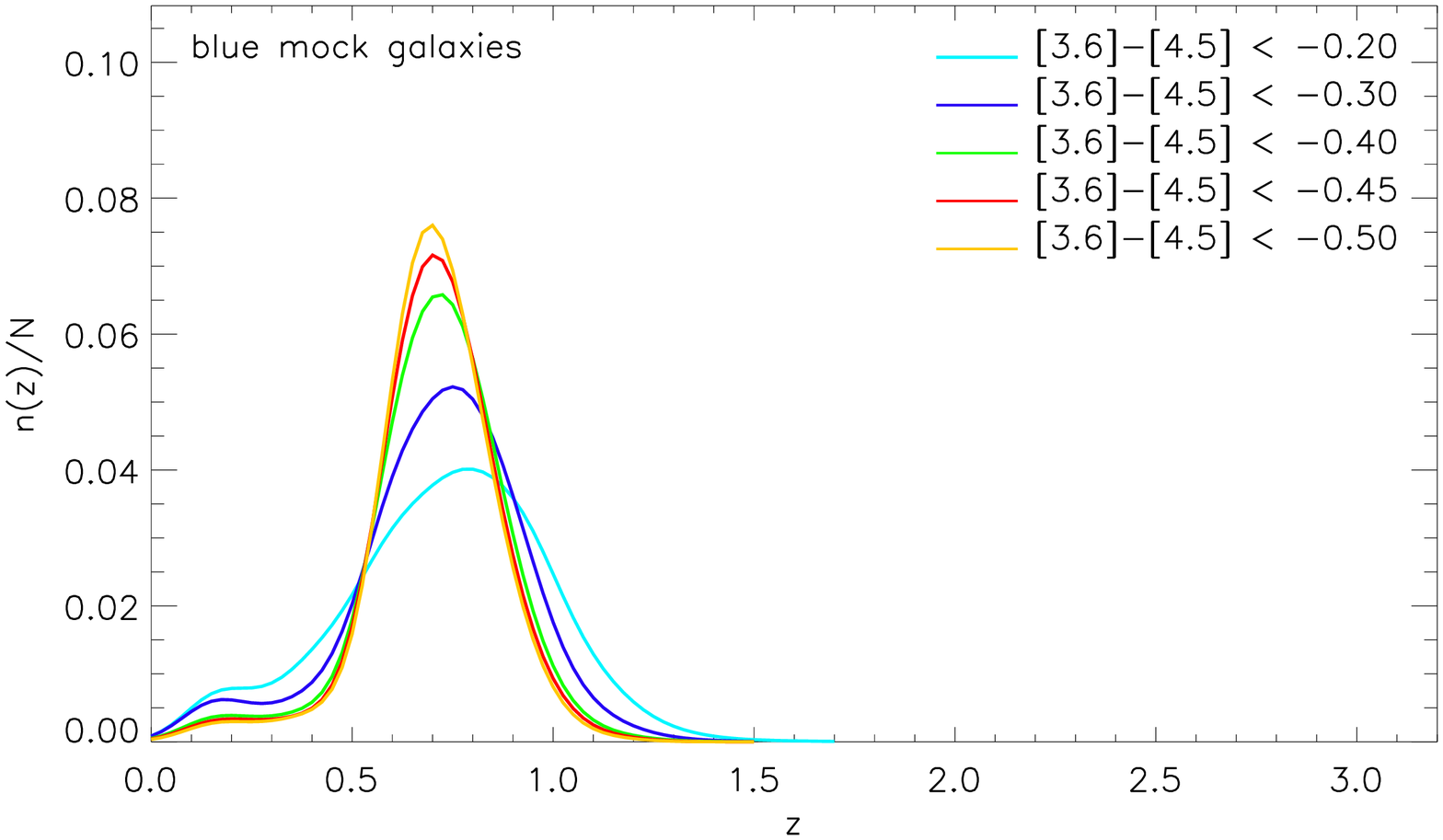}         
\includegraphics[width=1.03\columnwidth]{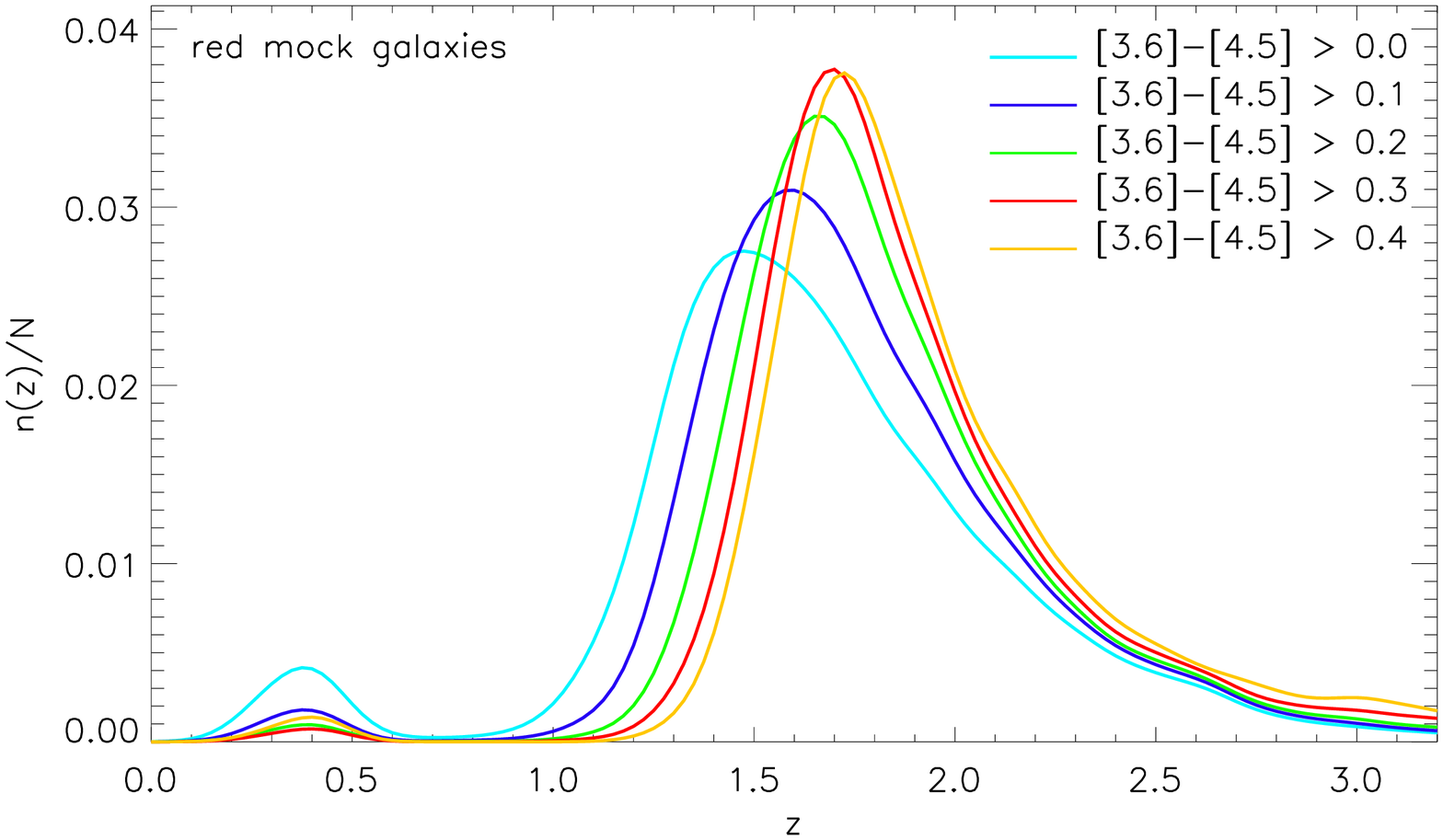}          
\caption{Redshift distribution of the SHARK mock galaxies for the same [3.6]-[4.5] colour cuts
as for observed galaxies in the S-COSMOS field, as shown in \autoref{fig:colour-cuts-zdist}.
The top panel is for blue mock galaxies, the bottom panel red ones.
Please note that the area of the S-COSMOS field is 2 deg$^2$, whereas the total area of the mocks
is the same as for \servs +\deepdrill, i.e. just over 20 deg$^2$. This much larger area yields smoother
redshift distributions for the SHARK mock galaxies as compared to the S-COSMOS ones, especially
for the red subsamples.}
\label{fig:colour-cuts-zdist-mocks} 
\end{figure}

In the following we will mostly look at the brightened mock samples (the second set), but also 
look at estimates for the first set where fluxes (and colours) are not changed at all.
Using the actual redshifts of the mock galaxies from the SHARK lightcones we derive the redshift
distribution of the colour-selected (brightened) mock subsamples, as we have done for the S-COSMOS field
(\autoref{fig:colour-cuts-zdist}), with the same colour and flux cuts. The distributions are shown
in \autoref{fig:colour-cuts-zdist-mocks}. What is immediately clear is that the distributions
for the blue S-COSMOS and SHARK subsamples are very similar, for all cuts, whereas the distributions
for the red ones are not: the SHARK redshift distribution for the red galaxies is quite smooth,
whereas the corresponding S-COSMOS one is not, and contains more high-redshift ($>2$) galaxies.
The main reason for this is field size: the area covered by S-COSMOS is 2 deg$^2$, whereas the
three SHARK mocks cover just over 20 deg$^2$, by construction (to mimic the area of the
\servs+\deepdrill\ field), to the same depth. This much larger volume clearly smoothens the
redshift distribution of the red subsamples.

\changedtwo{To get an idea what type of galaxies are selected by the colour cuts, we obtained from
the (brightened) mock samples the median values of the redshift, stellar mass and star formation
rate ('sfr' for short) for each of the populations selected. These are listed in \autoref{tab:properties}.
As the mocks are too faint overall, the stellar mass and star formation rates, which were not changed in
the step of brightening the mocks, are to
be viewed as lower limits (either or both are likely to be somewhat larger in value for the observed data).
However, the relative differences between the (sub)samples should remain similar for the mocks and the
observed datasets, allowing us to look for trends.
The colour cut correlates with both the median star formation rate and median redshift of the subsample,
but as these populations vary significantly in size this is hard to interpret physically: at best
the colour cut can be seen as a proxy for these median values. The sample sizes for the red and blue
subsamples as well as the two more severely cut samples are similar, so the 'red' populations 
have a higher median stellar mass than the corresponding 'blue' ones.}

\section{Methods}

We combine an estimate for the angular clustering of our galaxies with a
redshift distribution obtained for the S-COSMOS sample \citep{Sanders2007} using the same
selection criteria used for our samples. This provides, through inversion of the Limber equation,
a measure for spatial clustering. We perform this for various [3.6]-[4.5] colour cuts in each
of our three fields, and for the three fields combined.
We also estimate clustering for the full sample, which is less informative as the
redshift distribution is much broader than for the colour-selected samples,
but this is a useful measure for reference to previous work and to estimates for the various
subsamples we consider.

\subsection{Estimating correlation functions}\label{estimating}

We employ the standard estimator \citep{Landy1993} for measuring angular
correlations $w(\theta)$:  $w_{LS}=(DD-2DR+RR)/RR$ ,
where $DD$, $DR$ and $RR$ are the (normalized) galaxy-galaxy,
galaxy-random and random-random pair counts at separation $\theta$.
The source catalogue extracted from the observed maps provide the
galaxy positions, and we employ a ten times more abundant random catalogue
that Poisson samples the same survey region.
For the estimate of $w(\theta)$ and its errors we use the Jackknife technique
(e.g. \citealt{Wall2003,Norberg2009}), employing 16 pie slices of the circular fields
and estimating errors from the Jackknife sampling variance.

The estimator is fitted by its expected value
\begin{equation}
1+\langle w_{LS} \rangle = [1+w(\theta)]/(1+w_\Omega)\ ,
\end{equation}
where the ''integral constraint'' $w_\Omega$ is the integral of the model
function for the two-point correlation function over the survey area.

We consider a two-parameter fit for the generic power-law $w(\theta) = (\theta/A)^{-\delta}$.
Our samples and subsamples are large enough to not have to restrict ourselves to a
one-parameter fixed-slope power-law function.
The fitting technique is the one used by \citet{vanKampen2005}, but adapted for the survey
geometry of \deepdrill\ and multiple fields.
We employ non-linear $\chi^2$-fitting using the Levenberg-Marquardt method \citep{Press1988},
which allows us to easily take into account the multiplicative integral constraint,
but also produces the covariance matrix of the fitted parameters which provides
a good estimate for their uncertainties. The integral constraint
is not a free fitting parameter, but a function of the clustering amplitude $A$
and power-law slope $\delta$, and fitted that way.
For each fit the $\chi^2$ probability $Q$ is calculated using the
incomplete gamma function, and any fits with $Q<0.1$ are discarded. With the large
number of sources available here, this does not actually happen even for our smallest
subsample.

\begin{figure}
\centering
\begin{tabular}{cc}
\includegraphics[width=0.87\columnwidth]{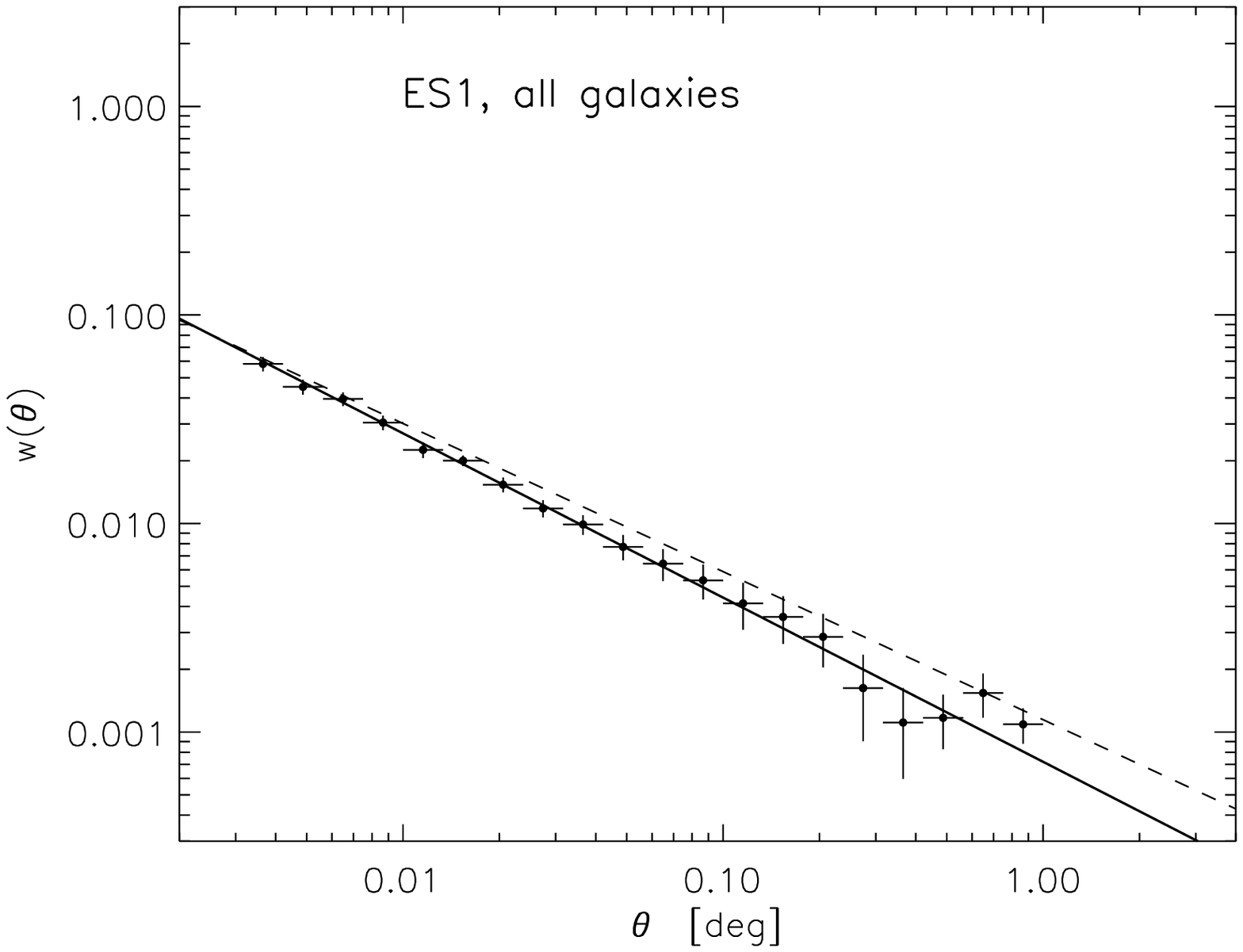} \\
\noalign{\vfilneg\vskip -0.4cm}
\includegraphics[width=0.87\columnwidth]{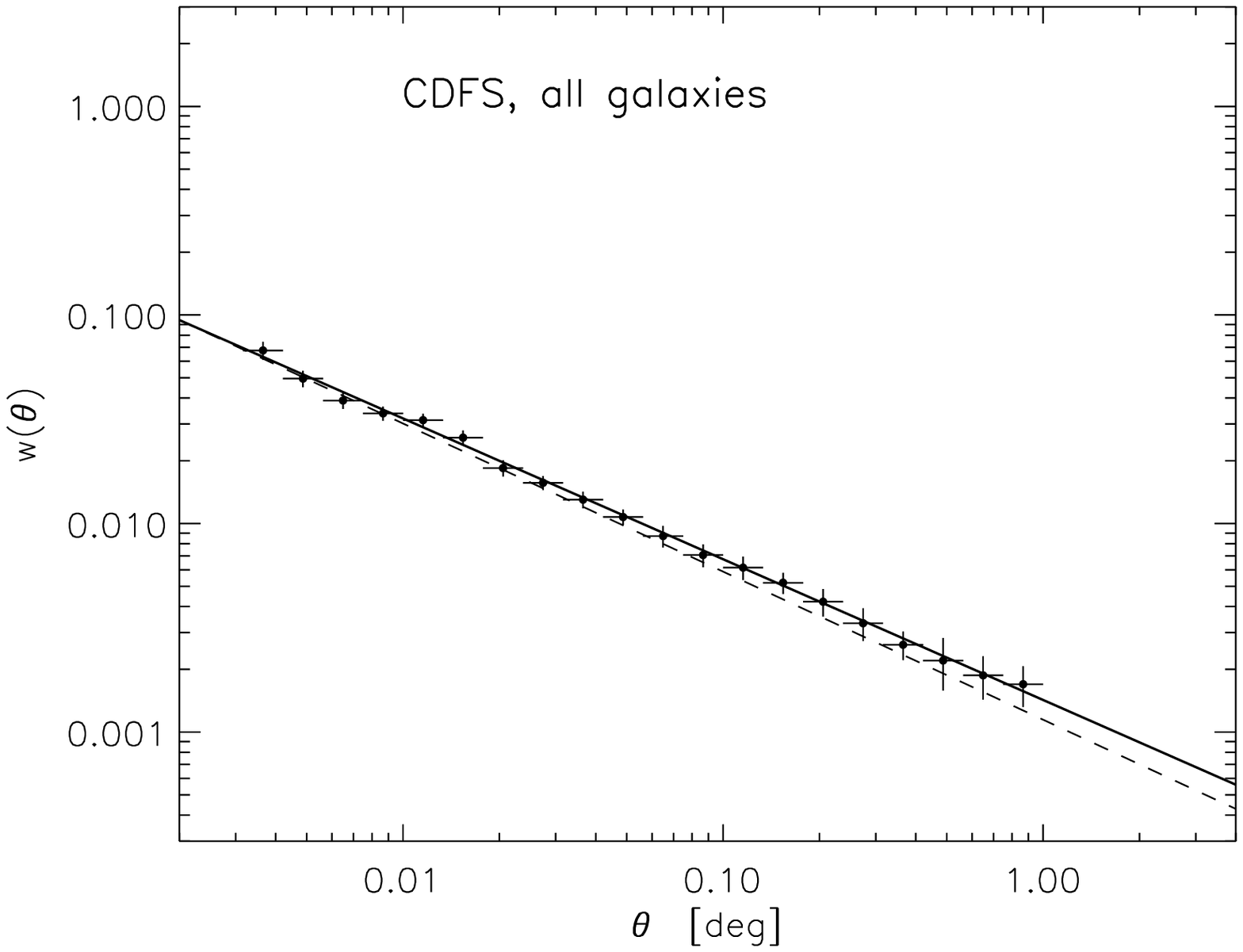} \\
\noalign{\vfilneg\vskip -0.4cm}
\includegraphics[width=0.87\columnwidth]{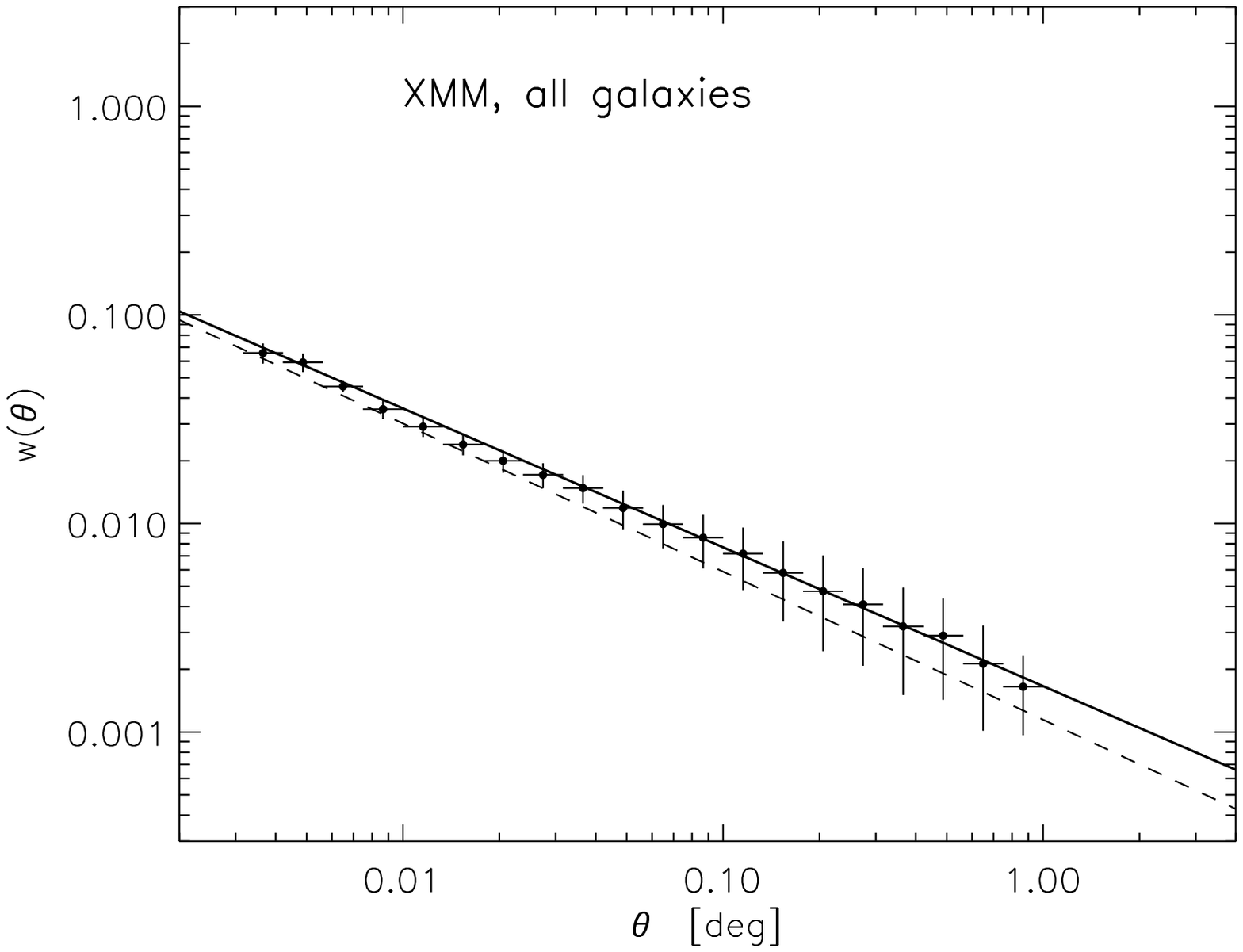} \\
\noalign{\vfilneg\vskip -0.4cm}
\includegraphics[width=0.87\columnwidth]{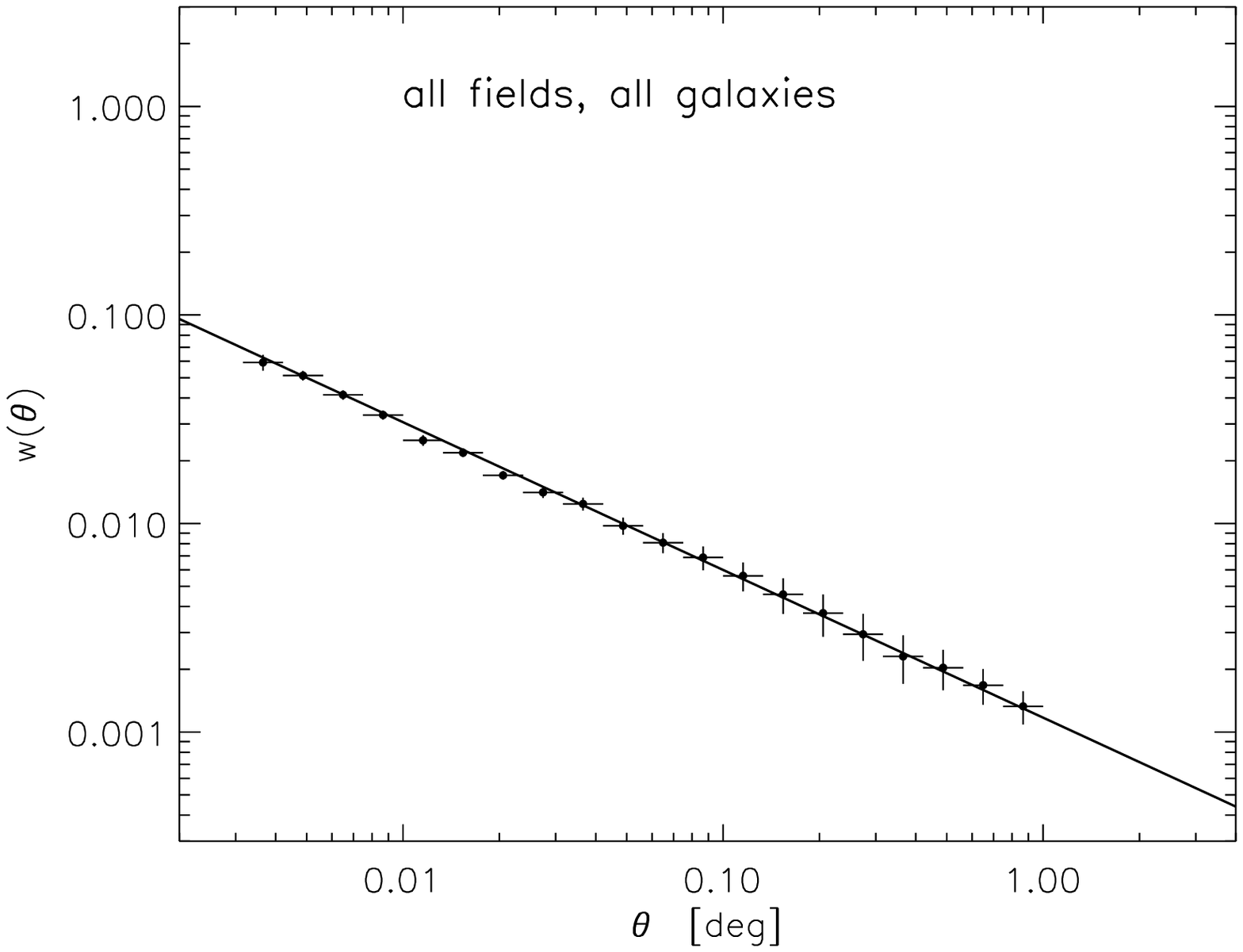} \\
\end{tabular}
\caption{Angular clustering for each of the fields separately and for all fields together
(bottom panel), for all galaxies below our flux cut 18$<$[3.6]$<$22
and 18$<$[4.5]$<$22. Solid lines indicate the power-law functions fitted to these data.
To aid visual comparison, the dashed line in the top three panels corresponds to the solid line in
the bottom panel (for all three fields).}  
\label{fig:angclus-all}
\end{figure}

\subsection{Spatial clustering}

We lack spectroscopic redshifts for most (around 98 percent) of our sources, so we cannot
measure spatial clustering directly. Also, we only have photometric redshifts for around half the 
sources (the original \servs\ fields), with inhomogeneous coverage and depth \citep{Pforr2019}.
However, we can employ a similar sample from the literature for which sufficient numbers of
photometric (and spectroscopic) redshifts, with homogeneous in coverage and depth, are available and
to which we can apply the same magnitude and colour cuts as for our own samples. The S-COSMOS
data provides such a sample \citep{Ilbert2009}, even though the total area of 2 deg$^2$ is 10
times smaller than that of our \servs+\deepdrill\ sample, resulting in redshift distributions
that are not very smooth (which they would be for 10 times the S-COSMOS area).

Assuming such redshift distributions are representative for our own dataset, Limber's
equation \cite{Limber1953} provides an estimate for the spatial clustering length $r_0$
from the angular clustering function (eg. \citealt{Phillips1978}).
For the Limber equation inversion we employ the code used by \cite{Farrah2006},
\changed{using the (Gaussian smoothed with width $\sigma_v=0.1$) measured redshift
distribution from S-COSMOS,} corresponding to the same selection (cuts in
magnitude and colour) as for our own (sub)samples. 

The redshift distribution for the full S-COSMOS sample down to our flux limits of
18$<$[3.6]$<$22 and 18$<$[4.5]$<$22 is shown
in \autoref{fig:redshift-distribution},
whereas the corresponding redshift distributions for the colour cuts are shown in 
\autoref{fig:colour-cuts-zdist}, where we only consider two colour cuts in red and blue each.

\begin{figure*}
\centering
\begin{tabular}{cc}
\includegraphics[width=0.9\columnwidth]{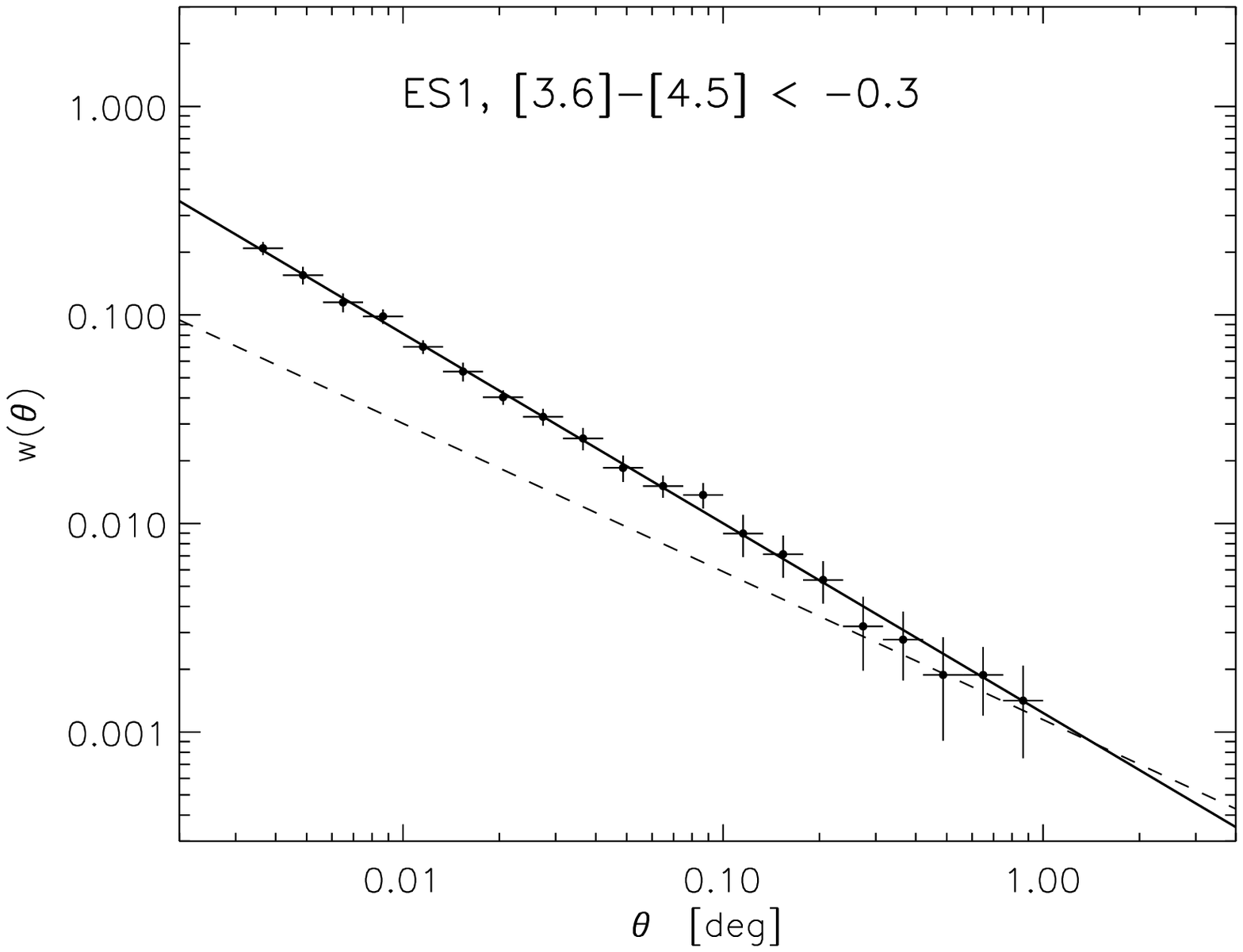} &
\includegraphics[width=0.9\columnwidth]{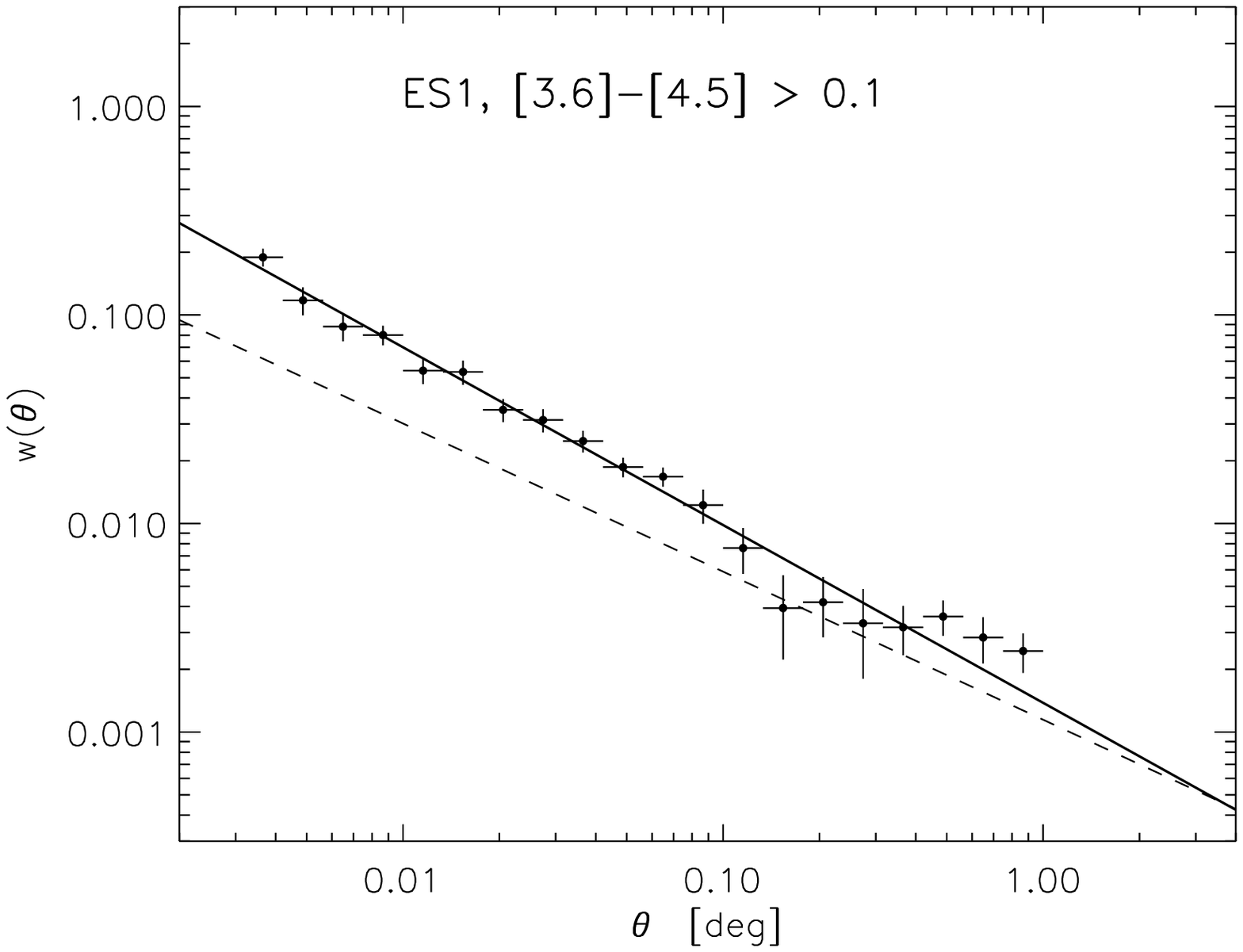} \\
\noalign{\vfilneg\vskip -0.4cm}
\includegraphics[width=0.9\columnwidth]{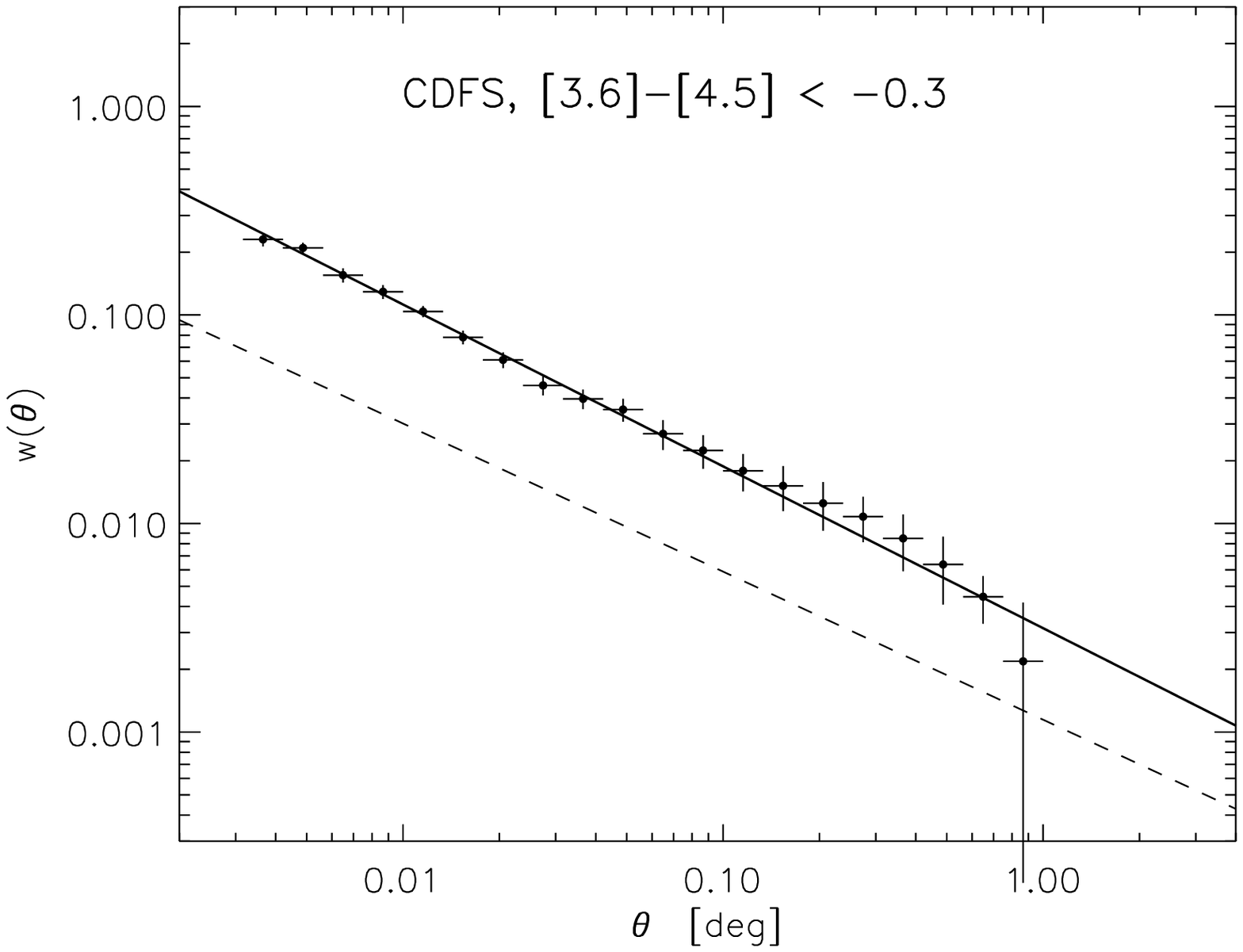} &
\includegraphics[width=0.9\columnwidth]{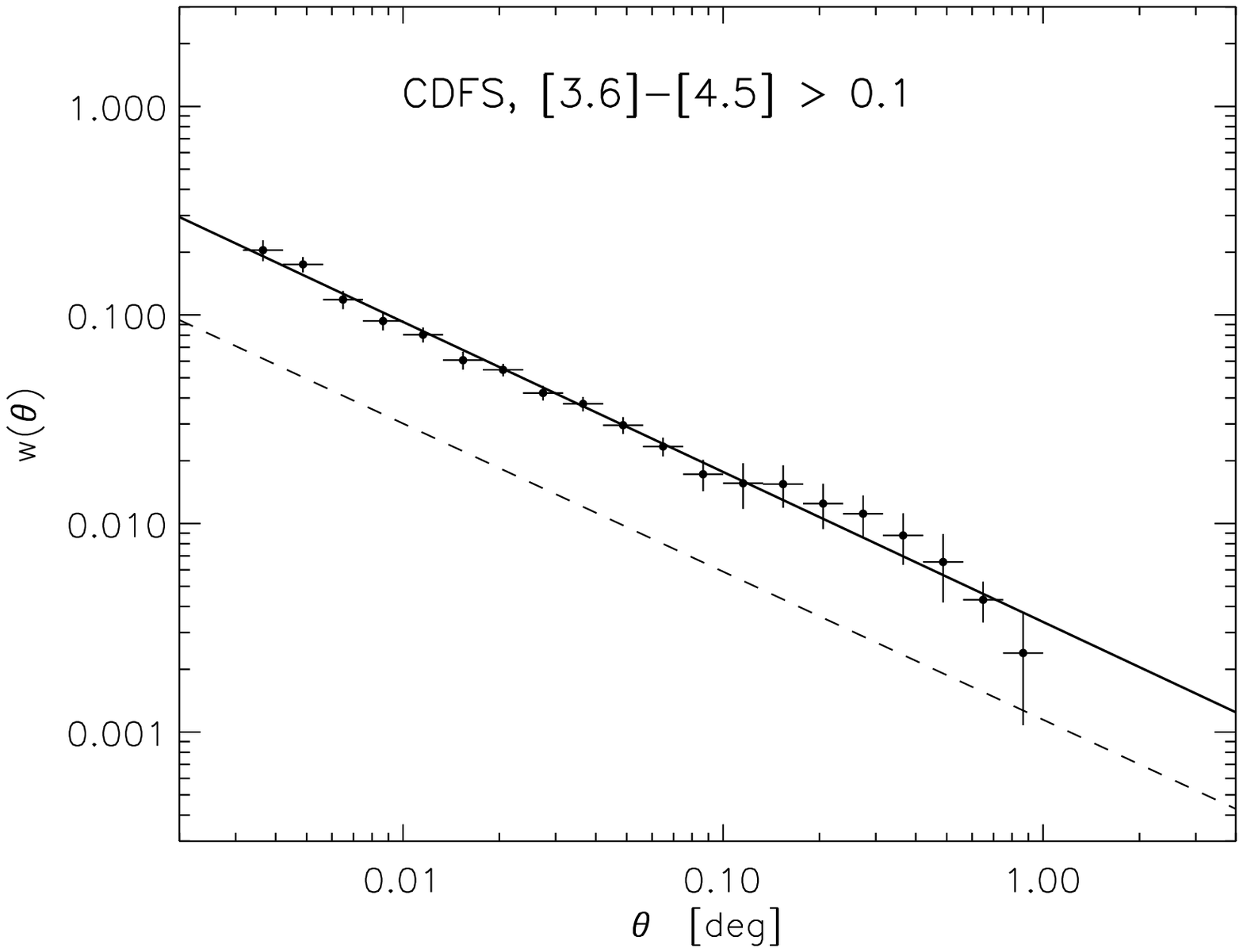} \\
\noalign{\vfilneg\vskip -0.4cm}
\includegraphics[width=0.9\columnwidth]{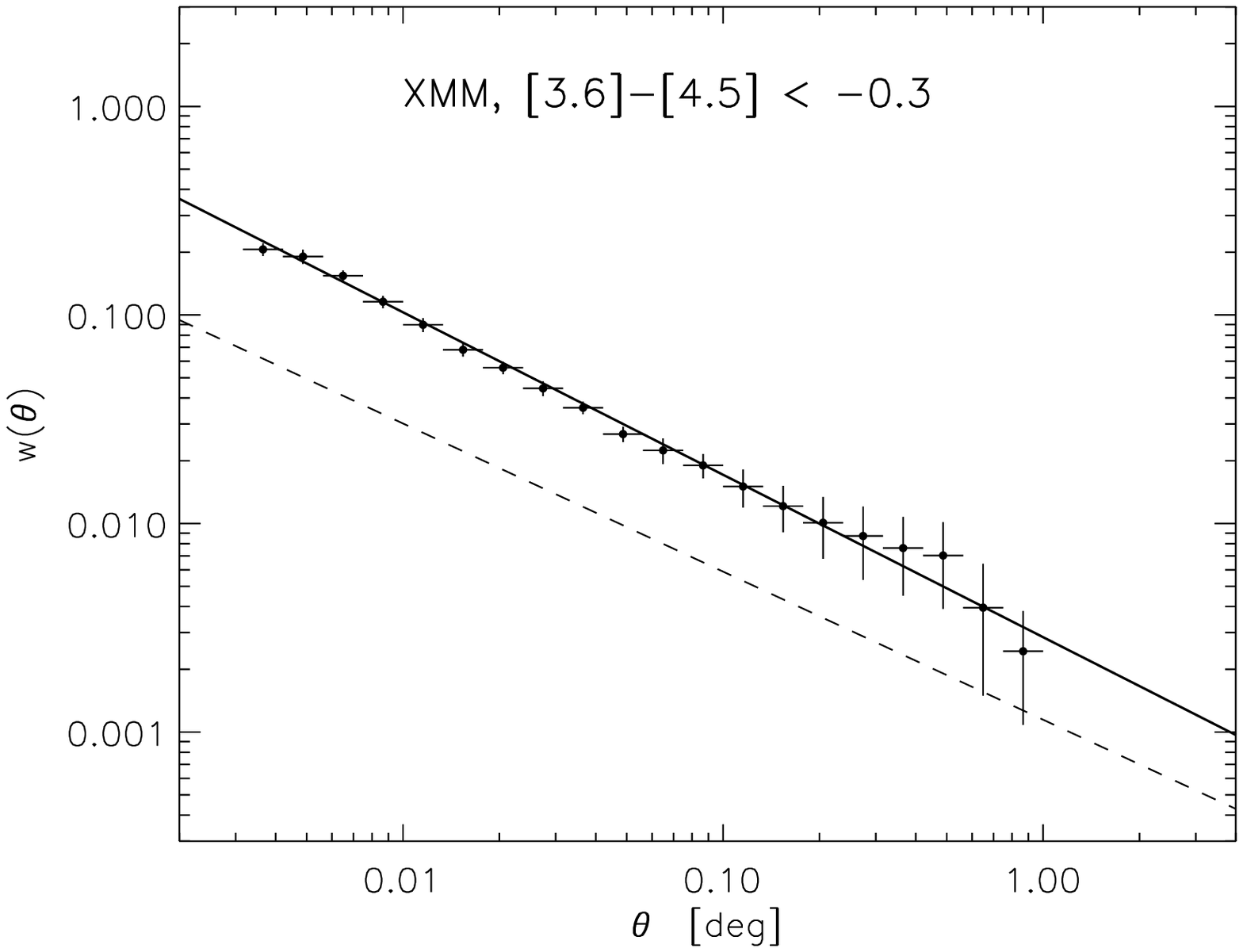} &
\includegraphics[width=0.9\columnwidth]{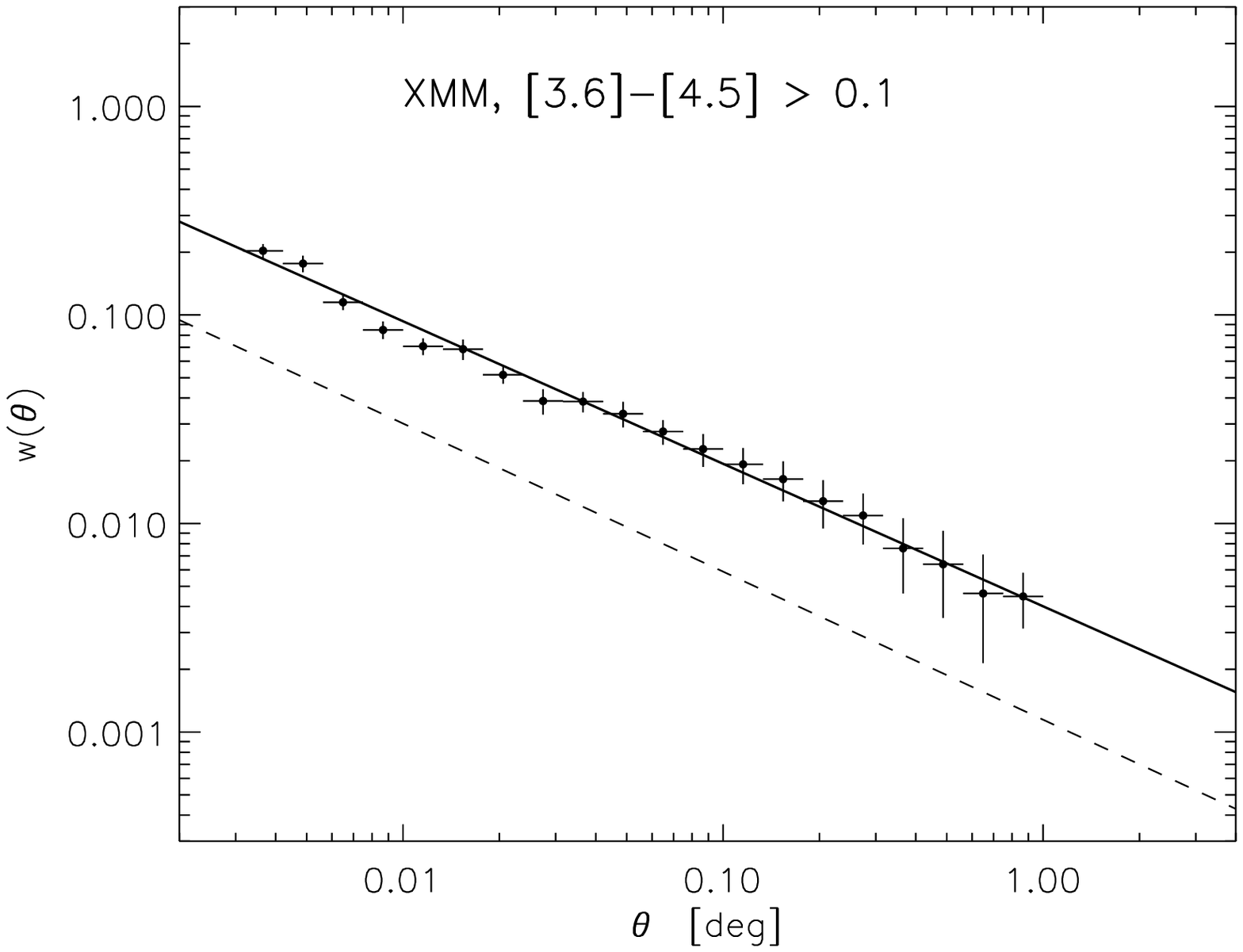} \\
\noalign{\vfilneg\vskip -0.4cm}
\includegraphics[width=0.9\columnwidth]{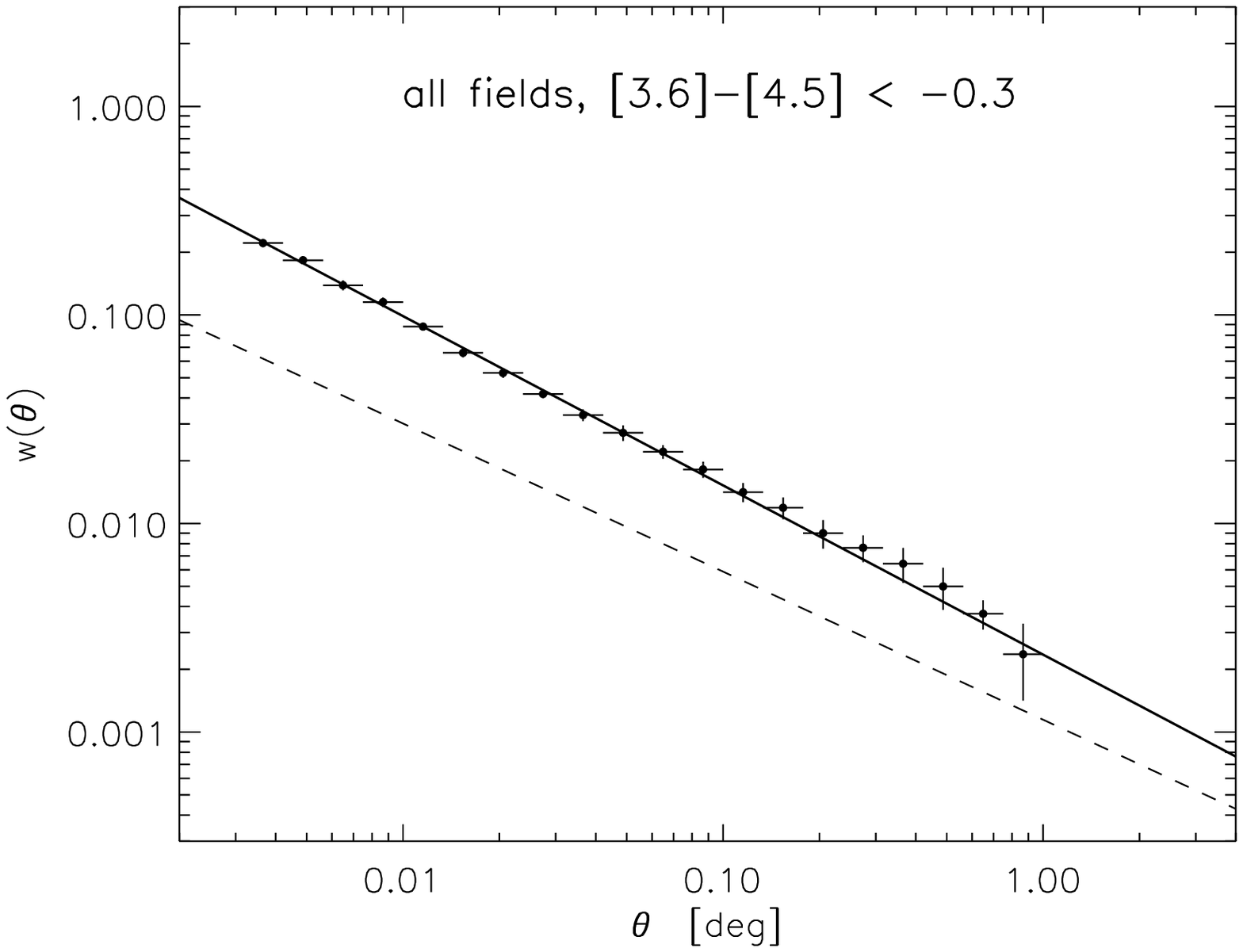} & 
\includegraphics[width=0.9\columnwidth]{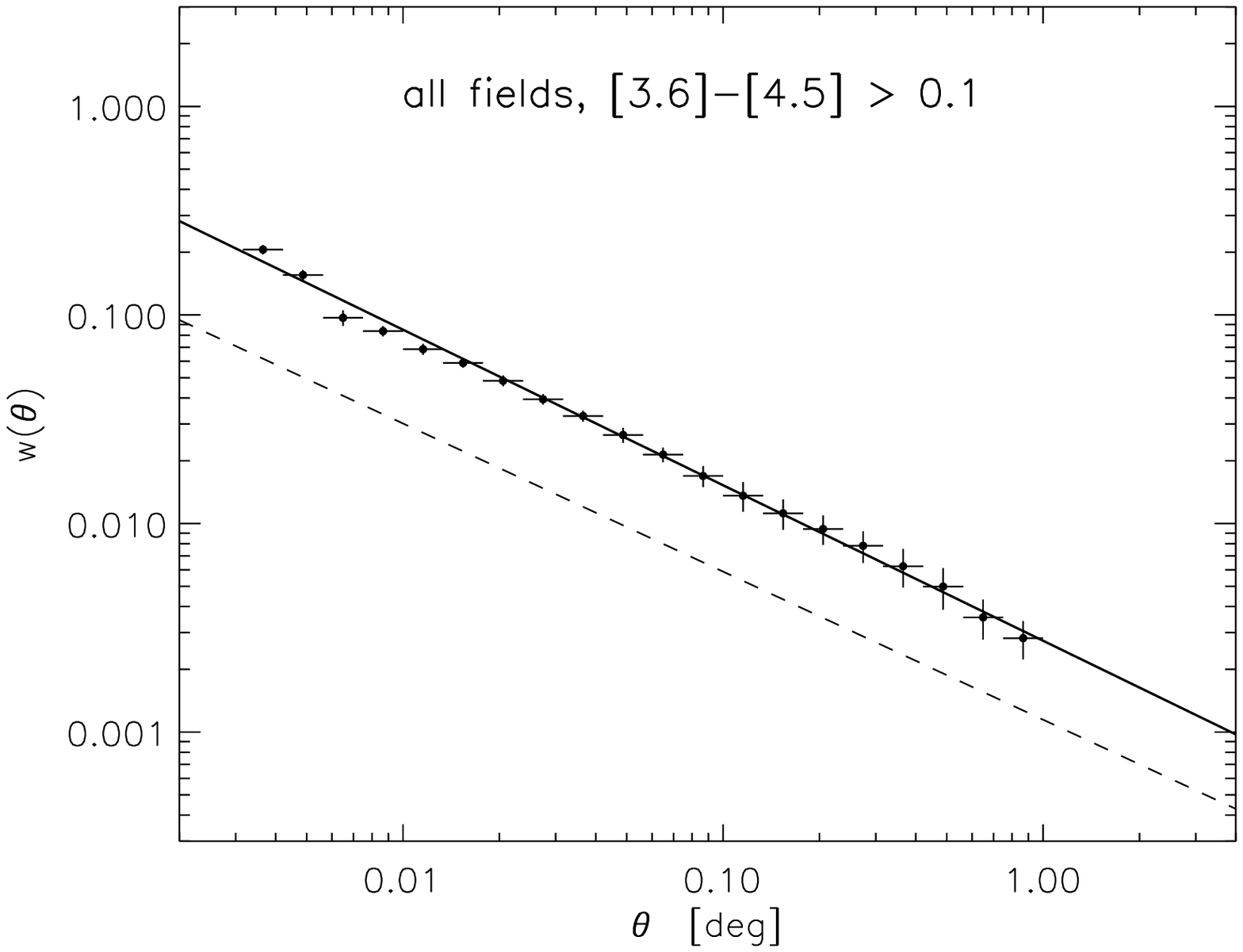}   
\end{tabular}
\caption{Angular clustering for ES1, CDFS, XMM, and for all fields together
(bottom panels), for blue (left panels) and red galaxies (right panels),
i.e. [3.6]-[4.5]$ >$0.1 and [3.6]-[4.5]$ <$-0.3, respectively.
Solid lines indicate the power-law functions fitted to these data.
To aid visual comparison, the dashed line in each panel corresponds
to the solid line in the bottom panel of \autoref{fig:angclus-all}, which
is for all galaxies in all fields.}
{\label{fig:angclus-rb1}
}\end{figure*}

\begin{figure*}
\centering
\begin{tabular}{cc}
\includegraphics[width=0.9\columnwidth]{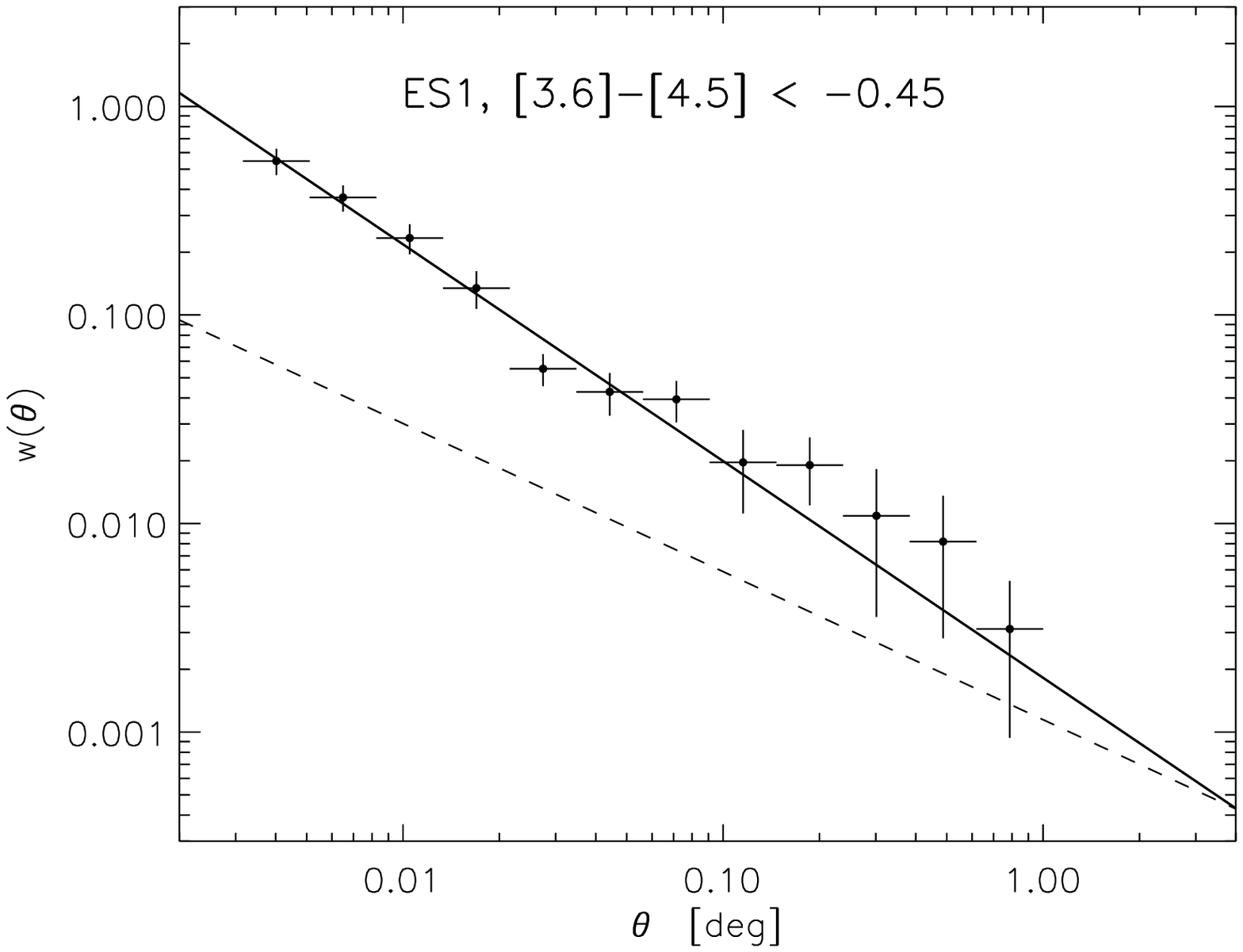} &
\includegraphics[width=0.9\columnwidth]{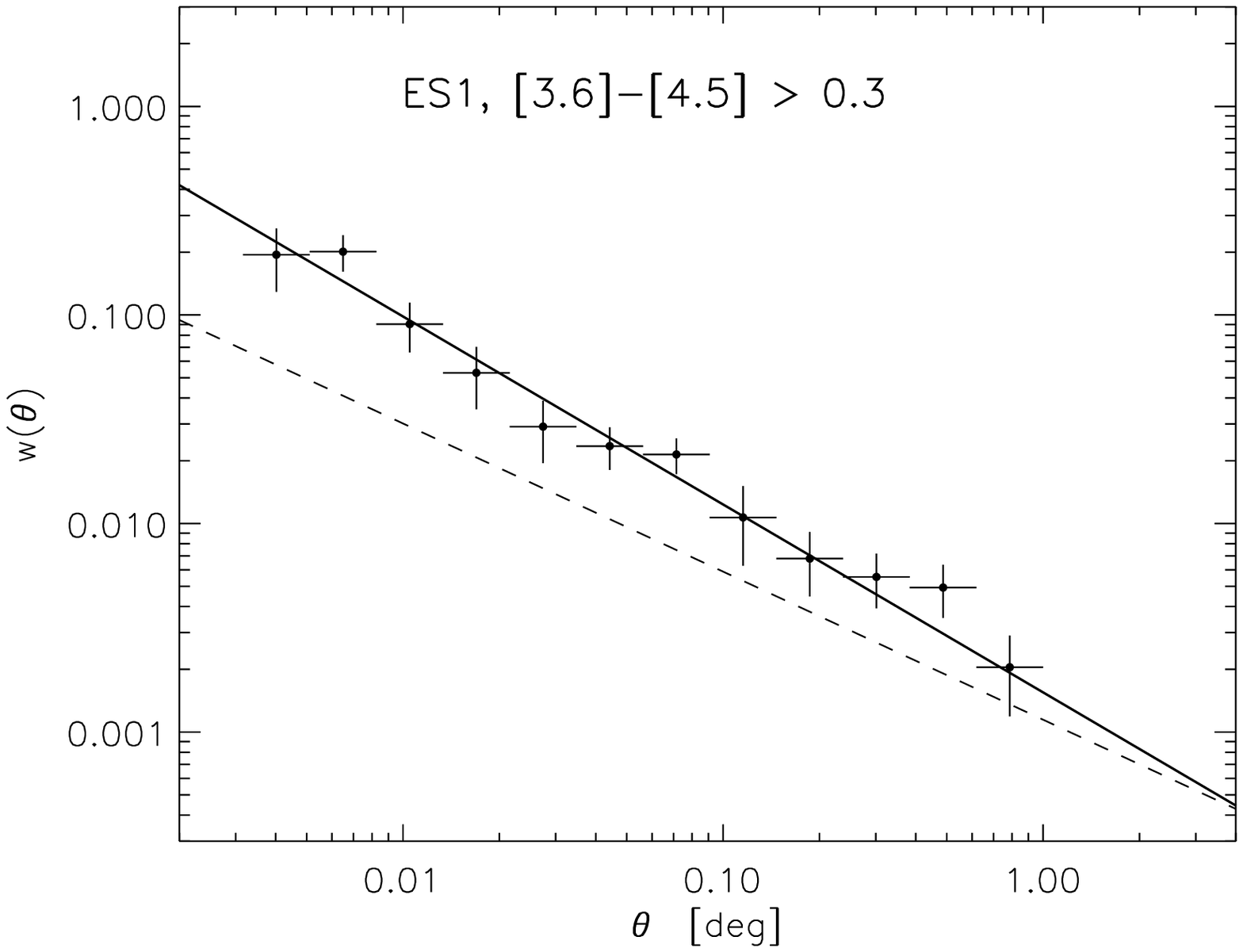} \\
\noalign{\vfilneg\vskip -0.4cm}
\includegraphics[width=0.9\columnwidth]{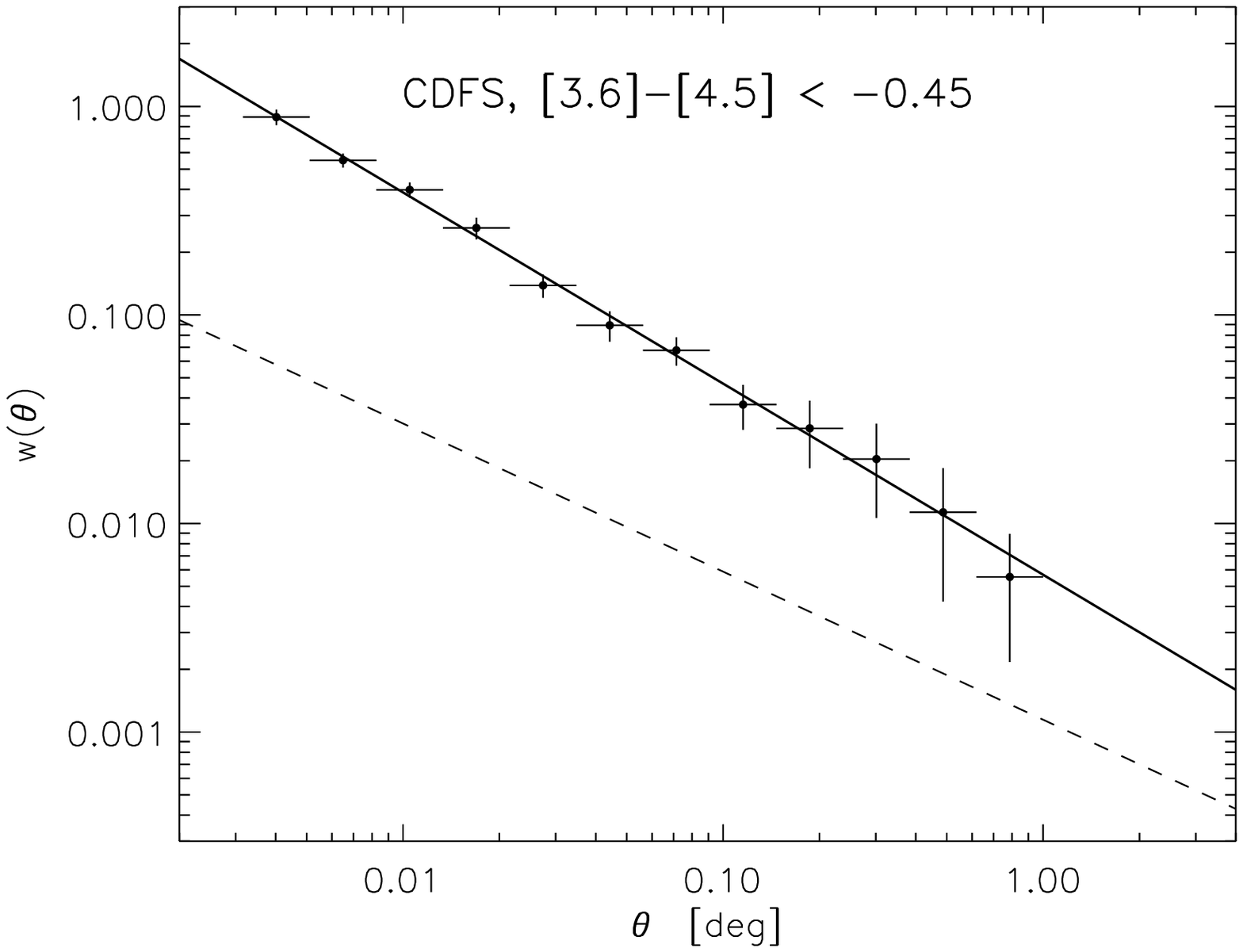} &
\includegraphics[width=0.9\columnwidth]{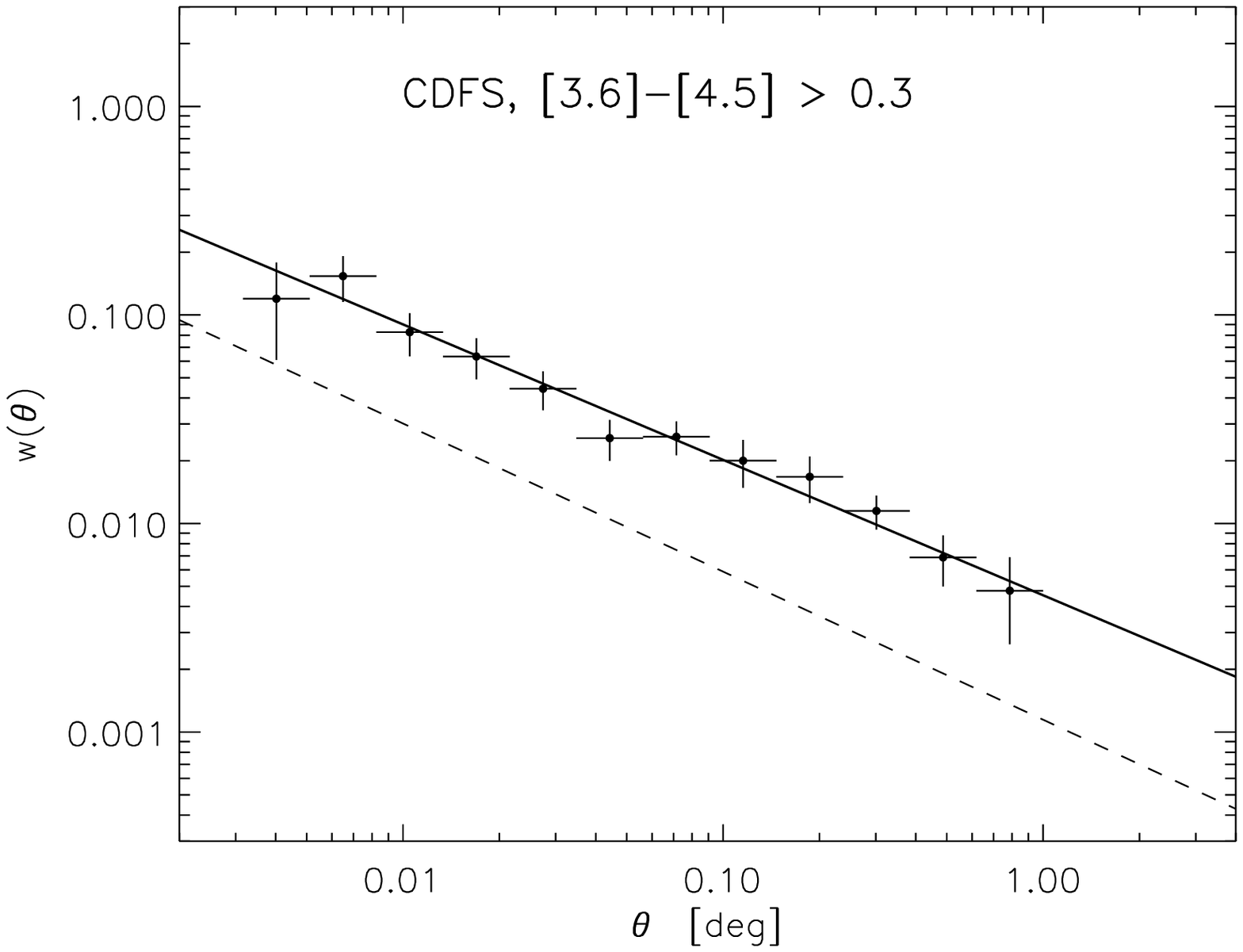} \\
\noalign{\vfilneg\vskip -0.4cm}
\includegraphics[width=0.9\columnwidth]{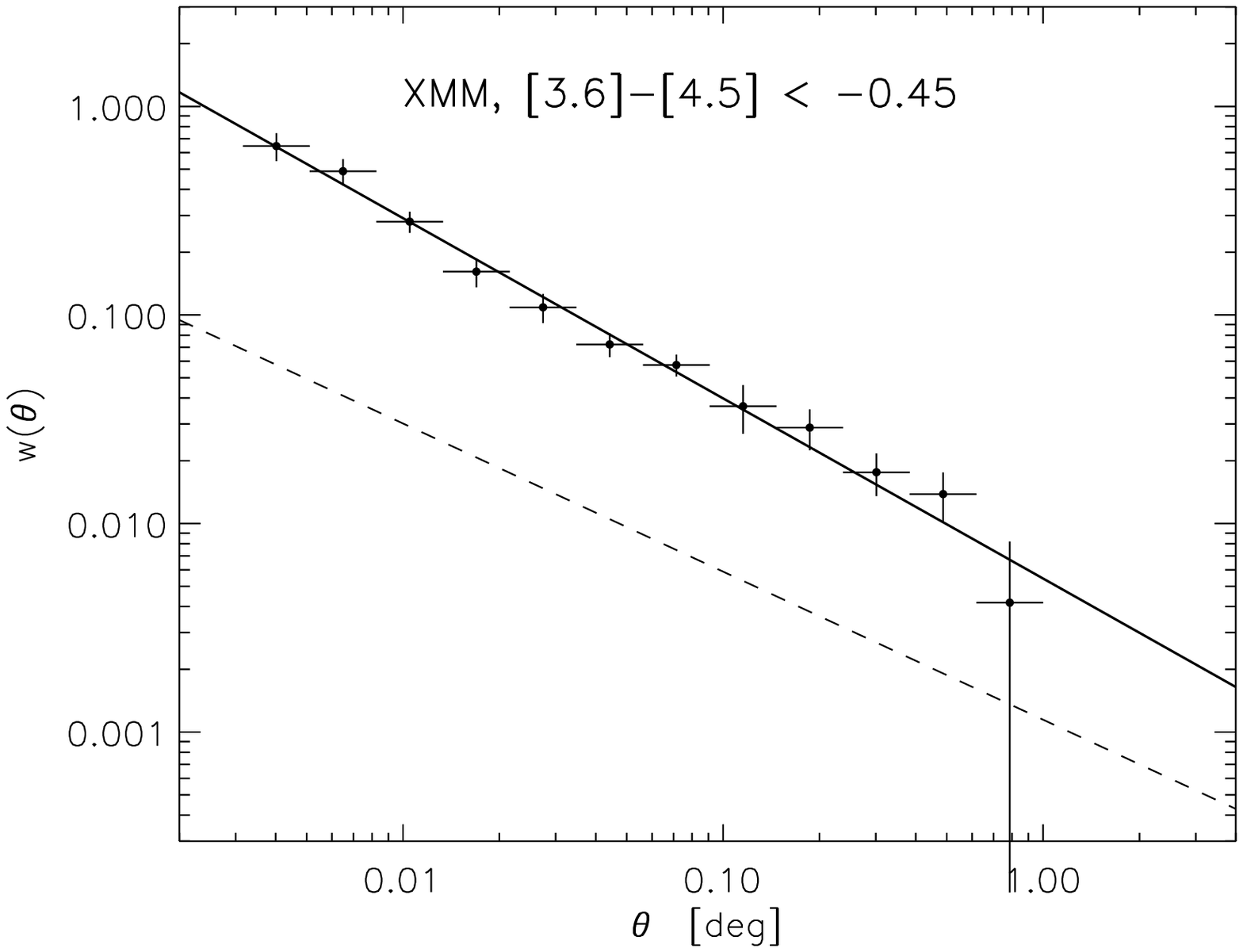} &
\includegraphics[width=0.9\columnwidth]{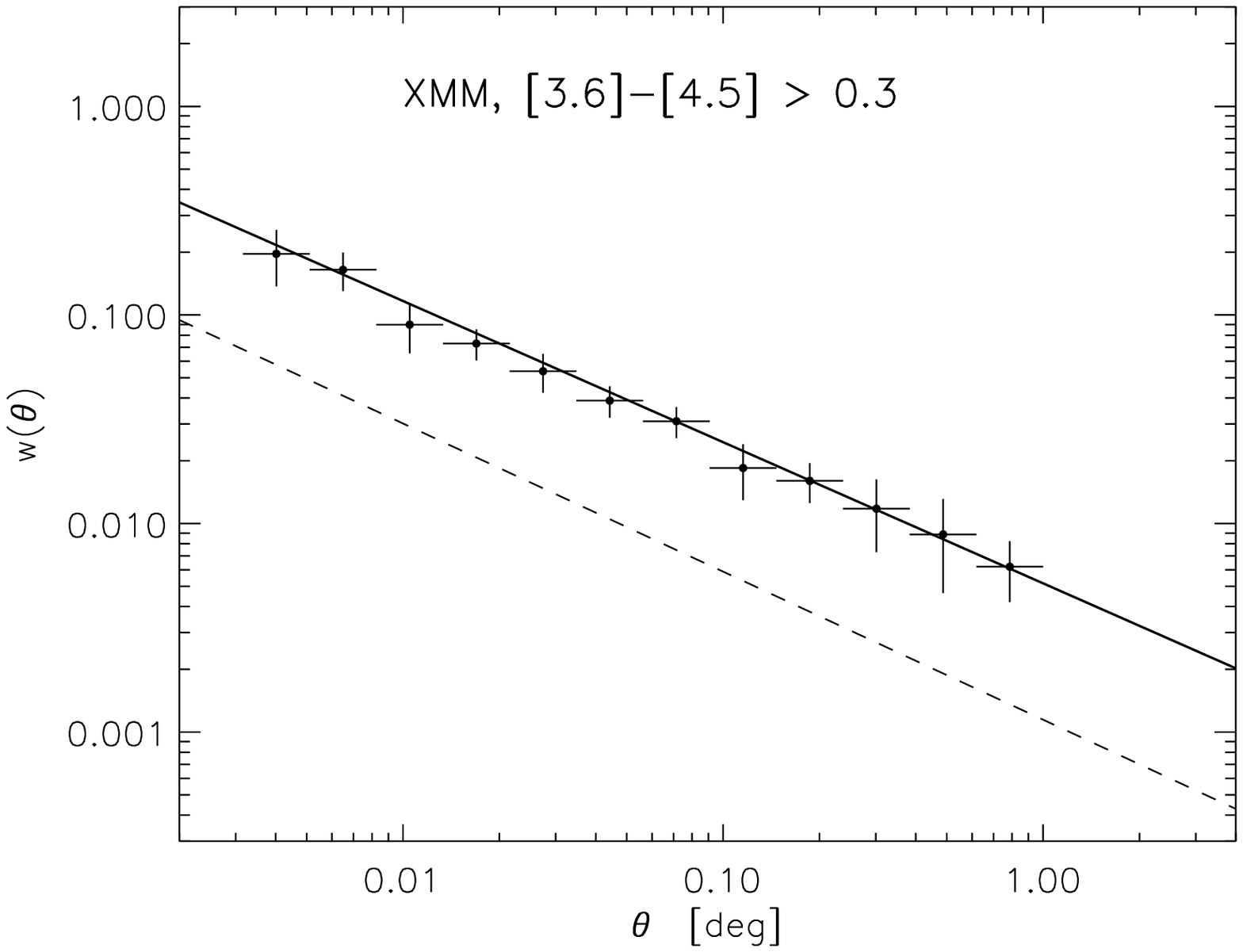} \\
\noalign{\vfilneg\vskip -0.4cm}
\includegraphics[width=0.9\columnwidth]{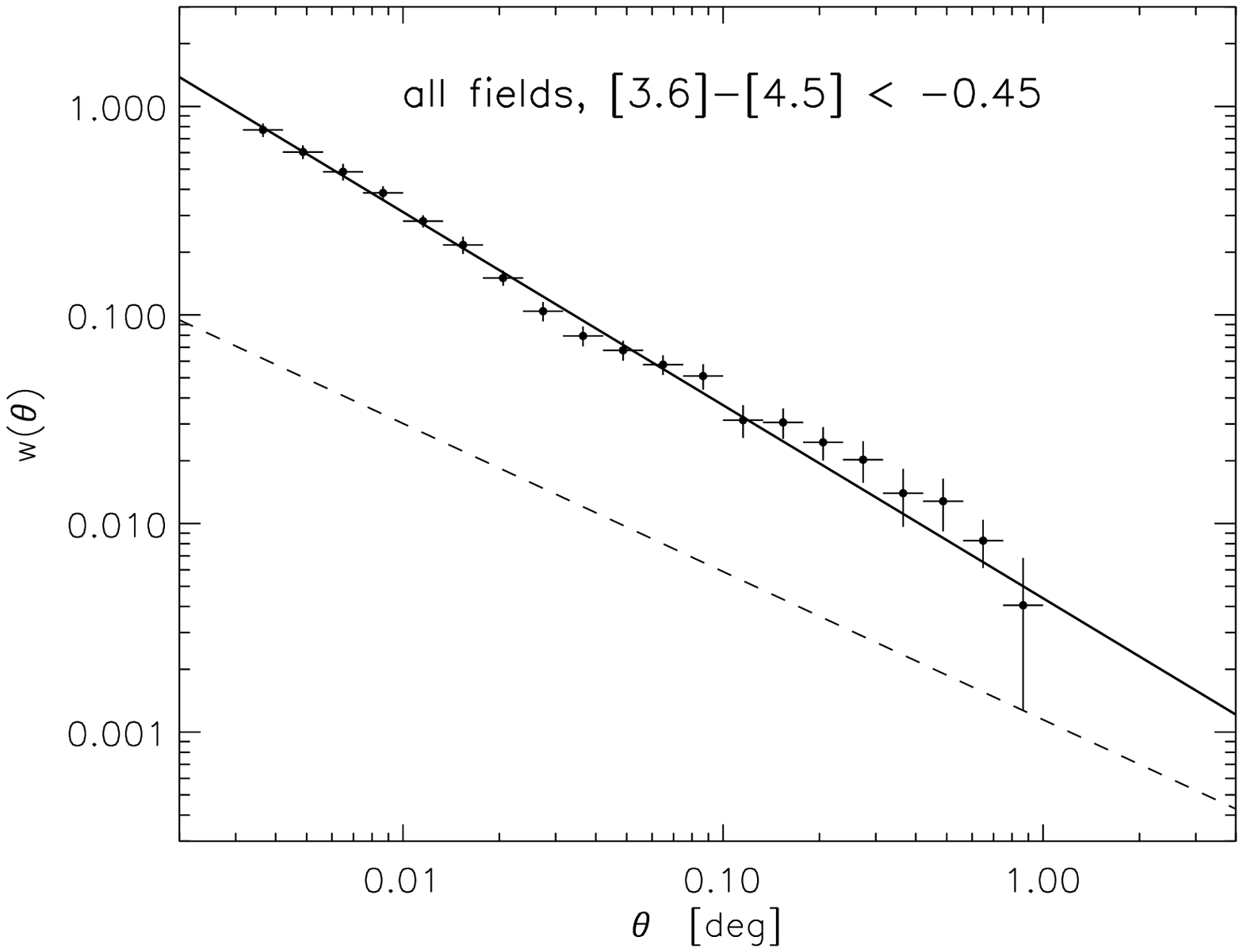} & 
\includegraphics[width=0.9\columnwidth]{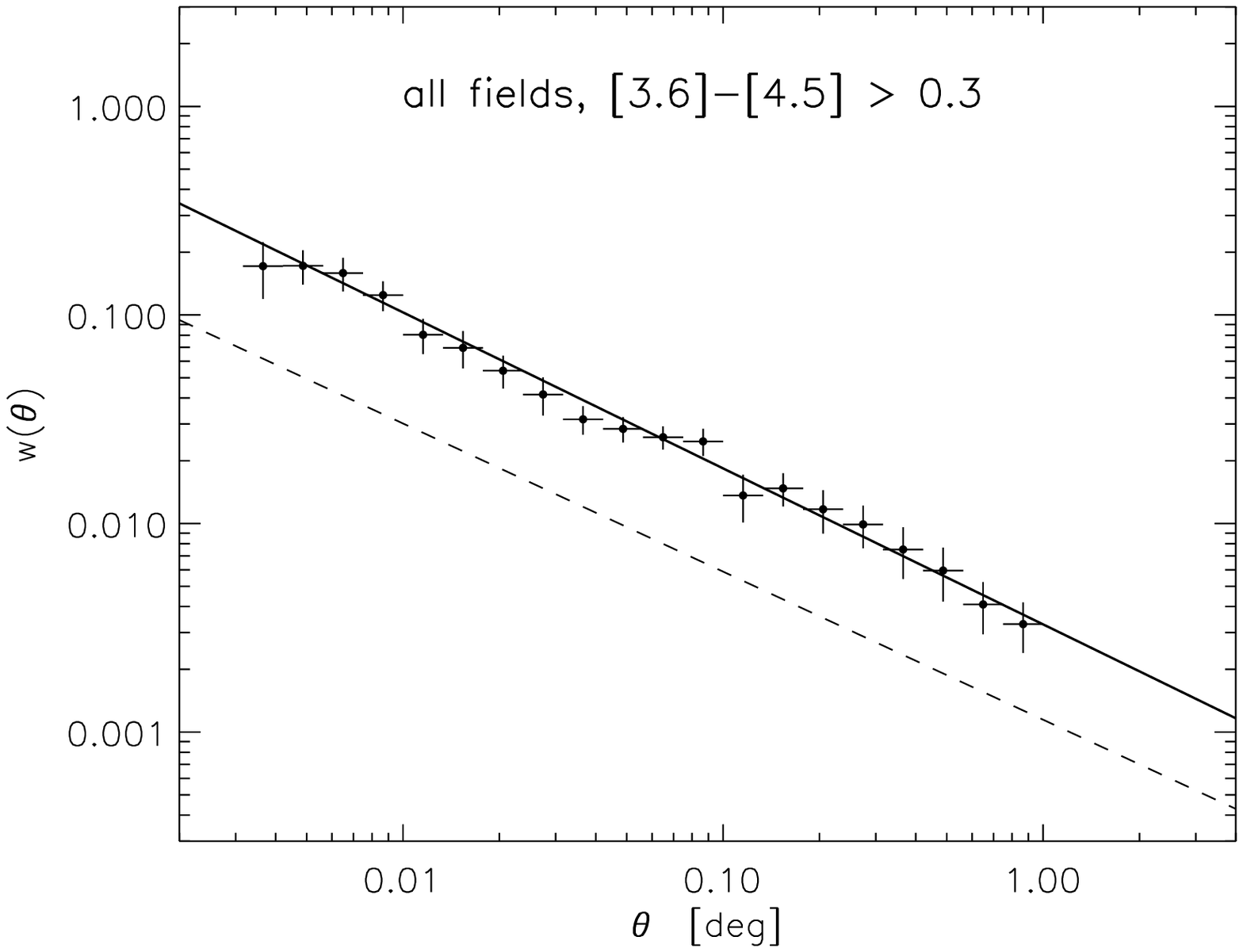}   
\end{tabular}
\caption{Angular clustering for ES1, CDFS, XMM, and for all fields together
(bottom panels), for the 'very blue' and 'very red' galaxy populations, 
i.e. [3.6]-[4.5]$ >$0.3 and [3.6]-[4.5]$ <$-0.45, respectively.
Solid lines indicate the power-law functions fitted to these data.
To aid visual comparison, the dashed line in each panels corresponds
to the solid line in the bottom panel of \autoref{fig:angclus-all} (which
is for all galaxies in all fields).}   
\label{fig:angclus-rb2}
\end{figure*}

\begin{table*}
\centering
  \begin{tabular}{lrrrr}
  \hline
  \noalign{\vfilneg\vskip -0.2cm}
  \hline
Sample & N &  A           &   $\delta$\hskip 0.5cm  &     $r_0$            \\
       &   &  [arcsec]    &                         &   [$h^{-1}$Mpc]            \\
\hline
ES1, all galaxies  &  289256 &  0.37 $\pm$  0.08 &  0.79 $\pm$  0.04 &  5.08 $\pm$  1.13 \\
ES1, [3.6]-[4.5] < -0.45 &   16352 &  8.32 $\pm$  1.13 &  1.04 $\pm$  0.08 &  9.38 $\pm$  1.44 \\
ES1, [3.6]-[4.5] < -0.3 &   81600 &  2.28 $\pm$  0.26 &  0.91 $\pm$  0.03 &  6.40 $\pm$  0.76 \\
ES1, [3.6]-[4.5] > 0.1 &   69507 &  1.59 $\pm$  0.28 &  0.85 $\pm$  0.04 &  7.79 $\pm$  1.41 \\
ES1, [3.6]-[4.5] > 0.3 &   13301 &  2.74 $\pm$  1.03 &  0.90 $\pm$  0.10 & 10.64 $\pm$  4.16 \\
\hline
CDFS, all galaxies  &  291734 &  0.22 $\pm$  0.06 &  0.67 $\pm$  0.04 &  4.87 $\pm$  1.37 \\
CDFS, [3.6]-[4.5] < -0.45 &   17501 & 12.77 $\pm$  0.87 &  0.92 $\pm$  0.05 & 11.71 $\pm$  1.02 \\
CDFS, [3.6]-[4.5] < -0.3 &   84930 &  2.15 $\pm$  0.27 &  0.78 $\pm$  0.04 &  6.75 $\pm$  0.91 \\
CDFS, [3.6]-[4.5] > 0.1 &   73846 &  1.32 $\pm$  0.26 &  0.72 $\pm$  0.04 &  8.01 $\pm$  1.61 \\
CDFS, [3.6]-[4.5] > 0.3 &   14668 &  0.88 $\pm$  0.68 &  0.65 $\pm$  0.12 &  8.08 $\pm$  6.37 \\
\hline
XMM, all galaxies  &  288597 &  0.24 $\pm$  0.09 &  0.67 $\pm$  0.06 &  5.14 $\pm$  2.04 \\
XMM, [3.6]-[4.5] < -0.45 &   16619 &  8.63 $\pm$  1.27 &  0.86 $\pm$  0.06 &  9.75 $\pm$  1.60 \\
XMM, [3.6]-[4.5] < -0.3 &   81216 &  1.94 $\pm$  0.23 &  0.78 $\pm$  0.03 &  6.44 $\pm$  0.80 \\
XMM, [3.6]-[4.5] > 0.1 &   73495 &  1.12 $\pm$  0.24 &  0.68 $\pm$  0.04 &  7.73 $\pm$  1.72 \\
XMM, [3.6]-[4.5] > 0.3 &   14524 &  1.51 $\pm$  0.77 &  0.68 $\pm$  0.10 &  9.73 $\pm$  5.19 \\
\hline
all fields, all galaxies  &  869587 &  0.26 $\pm$  0.05 &  0.71 $\pm$  0.03 &  4.96 $\pm$  0.99 \\
all fields, [3.6]-[4.5] < -0.45 &   50472 & 10.22 $\pm$  0.55 &  0.93 $\pm$  0.03 & 10.52 $\pm$  0.67 \\
all fields, [3.6]-[4.5] < -0.3 &  247746 &  2.07 $\pm$  0.15 &  0.81 $\pm$  0.02 &  6.50 $\pm$  0.51 \\
all fields, [3.6]-[4.5] > 0.1 &  216848 &  1.32 $\pm$  0.15 &  0.75 $\pm$  0.03 &  7.83 $\pm$  0.95 \\
all fields, [3.6]-[4.5] > 0.3 &   42493 &  1.72 $\pm$  0.48 &  0.75 $\pm$  0.06 &  9.68 $\pm$  2.77 \\
\hline
all mock fields, all galaxies  &  650496 &  1.23 $\pm$  0.08 &  0.91 $\pm$  0.02 &  7.65 $\pm$  0.52 \\
all mock fields, [3.6]-[4.5] < -0.45 &   66333 &  8.66 $\pm$  0.35 &  1.01 $\pm$  0.02 &  9.62 $\pm$  0.45 \\
all mock fields, [3.6]-[4.5] < -0.3 &  216273 &  3.60 $\pm$  0.15 &  0.95 $\pm$  0.02 &  7.82 $\pm$  0.36 \\
all mock fields, [3.6]-[4.5] > 0.1 &   97562 &  4.40 $\pm$  0.32 &  0.99 $\pm$  0.03 & 11.64 $\pm$  0.91 \\
all mock fields, [3.6]-[4.5] > 0.3 &    9614 &  4.32 $\pm$  2.28 &  0.90 $\pm$  0.16 & 13.19 $\pm$  7.33 \\
\hline
brightened mocks, all galaxies  &  946217 &  0.74 $\pm$  0.06 &  0.85 $\pm$  0.02 &  6.40 $\pm$  0.57 \\
brightened mocks, [3.6]-[4.5] < -0.45 &   52441 &  7.22 $\pm$  0.44 &  0.98 $\pm$  0.03 &  8.82 $\pm$  0.59 \\
brightened mocks, [3.6]-[4.5] < -0.3 &  221476 &  2.92 $\pm$  0.17 &  0.88 $\pm$  0.02 &  7.30 $\pm$  0.46 \\
brightened mocks, [3.6]-[4.5] > 0.1 &  220590 &  2.39 $\pm$  0.16 &  0.93 $\pm$  0.02 &  8.91 $\pm$  0.61 \\
brightened mocks, [3.6]-[4.5] > 0.3 &   42589 &  2.59 $\pm$  0.56 &  0.87 $\pm$  0.06 & 10.55 $\pm$  2.39 \\
\hline
\noalign{\vfilneg\vskip -0.2cm}
\hline
\end{tabular}
\caption{Clustering measures for all samples considered, observational as well as simulated
(see main text for details) for the two-parameter fits. This table also lists the actual sizes of all samples
used, as well as the derived spatial clustering strength using the Limber equation inversion technique.}
\label{tab:corr-results}
\end{table*}

\section{Results}

We now employ the methods described above to yield clustering estimates down to our selected
flux limit and for various redshift distributions selected through the [3.6]-[4.5] colour cuts.
Because the three fields have the same depth, with no systematic differences
between the fields, we also combine the estimates for each of the fields to a single estimate
for the survey as a whole. We do not use the full coverage of \servs+\deepdrill: we leave out
areas near the border of each field and around bright stars (see section 2.1), resulting in a
clean set of three round fields, totalling just over 20 sq.\ degrees. Corresponding mock fields
of the same field size, depth and geometry, including bright star removal, are analyzed in the
same fashion as the observed fields.

\subsection{Estimates for all galaxies}

For the three \servs+\deepdrill\ fields, which each measure 3 degrees in diameter, we first estimate the
angular correlation function for all galaxies in the \changed{magnitude intervals
18$<$[3.6]$<$22 and 18$<$[4.5]$<$22}.
With no colour selection, we probe the full depth of the sample from the local Universe out to
$z\approx 2-3$, as can be seen in \autoref{fig:redshift-distribution}. This shows the redshift distribution
for the full S-COSMOS sample \changed{(with the same magnitude selection as for the observed samples)}.
So we are clearly not estimating the local auto-correlation function, but the angular
correlation over a wide redshift range. This results in relatively low amplitudes for the estimated angular
correlation functions, as can be seen in \autoref{fig:angclus-all}, with fairly large uncertainties
(derived from the Jackknife technique - see Section 3.1). The uncertainty is noticeably smaller for the
joint estimate for the three fields taken together. Power-law functions fit well to each of the estimates
(the reduced $\chi^2$ values for the fits are within the range expected for the number of bins used),
with fitting parameters as listed in \autoref{tab:corr-results}. Despite the low amplitudes for the
angular correlation functions, the well-defined redshift distribution allows us to estimate the spatial 
correlation length $r_0$ fairly well, especially for the estimate for all three fields together:
$r_0 = 4.96 \pm 0.99\ h^{-1}$Mpc, i.e. an uncertainty of around 20 percent. This estimate mostly serves as
a reference for the other estimates which are derived below, for more specific subsamples which sample
a more restricted redshift range.

Also using \spitzer\ data, \cite{Waddington2007} employed the SWIRE survey \cite{Lonsdale2003} to estimate
clustering at 3.6 micron as well, with a survey size of 8 deg$^2$. This survey is not as wide and as deep
\servs+\deepdrill\ , and less complete near our [3.6]=22 limit. Our magnitude limit corresponds to 5.6 $\mu$Jy,
which is close to the one for a sample down to 6.3 $\mu$Jy that they consider. For this sample
they find a mean redshift of 0.9, which is very close to our mean of 0.89,
and a spatial clustering length $r_0 = 3.18 \pm 0.94\ h^{-1}$Mpc. This is smaller than what we find, but within
the uncertainties still, and therefore consistent. The reason for their smaller clustering strength could
be related to their larger incompleteness near the faint magnitude limit.

\subsection{Estimates for colour selected subsamples}

Colour cuts allow us to measure the angular correlation function for narrower redshift distributions,
yielding larger amplitudes (as there is less averaging out along the line of sight), and cleaner
measurements. The colour cuts also select for more specific populations of galaxies, in terms of type,
star formation rate, and other properties.
For each of the three fields, and for each of the four colour cuts, we estimate the 
galaxy correlation function and fitted a two-parameter power-law function,
as for the full sample. Again we use the S-COSMOS sample to derive the redshift distribution
correspond to a given colour cut, which is then used in the Limber equation inversion technique
(see Section 3.2).

Looking at the moderate colour cuts first, we select a blue and red subsample cutting 
at [3.6]-[4.5]$<$-0.3 and [3.6]-[4.5]$>$0.1 respectively. Their redshift distributions are much
narrower than for the total sample,
and should provide stronger clustering estimates which are also more informative.

We select two more redshift distributions through more severe colour cuts,
yielding much smaller but also much bluer and redder subsamples. As described and argued
in Section 2.2, we cut at [3.6]-[4.5]$<$-0.45 and [3.6]-[4.5]$>$0.3 to achieve this.
The same selection for the S-COSMOS sample gives us to the two corresponding redshift
distributions.
The 'very blue' cut produces the narrowest redshift distribution of all four cuts, with a
clear peak around $z\approx 0.7$, which has less than half the width of the blue one.
In comparison, the 'very red' cut mostly moves the red redshift distribution to higher redshifts,
and produces a less pronounced peak and a slightly broader distribution as well.

For each of the four subsamples we show our estimates for the galaxy correlation function
in \autoref{fig:angclus-rb1} for the blue and red subsamples, and in \autoref{fig:angclus-rb2}
for the more extreme cuts (where we use less bins for the fitting).
For each cut, the top three panels in each figure (left- and right-hand
side) show the galaxy correlation functions obtained for each \servs+\deepdrill\ field
individually, whereas the bottom panels show the combine estimate using all three fields for each
subsample. The latter clearly produce the most accurate estimates (smallest error bars), and also
the cleanest fits. The variance between the fields can also be seen: the 'very blue' subsamples
for each field probably show the most consistency.
In all panels the dotted line represents the measured galaxy correlation function for the
full sample, as already shown in the bottom panel of \autoref{fig:angclus-all}
(which has amplitude $A=0.18\pm0.04$ arcsec and slope $\delta=0.67 \pm 0.03$), but repeated here
for reference.

A power-law shape is observed for each correlation function, with small deviations
for some (the largest of these are for the ES1 field). A power-law is robustly fitted in each case,
using the methods described in \autoref{estimating}, with reduced $\chi^2$ values for the fits within
the expected range for the number of bins \changed{used for the fit}.
The resulting fitted parameters are listed in \autoref{tab:corr-results}.
Fractional uncertainties for these are smaller than for the full sample,
as can be expected. However, the differences are larger for the fitted amplitudes than
for the fitted slopes. The reason for this is the uncertainty in the integral constraint
for the individual fields, which is still significant because of the limited size
of the fields. Here again the advantage of using three independent fields is apparent,
as this alleviates this uncertainty somewhat.

\begin{figure*}
\centering
\begin{tabular}{cc}
\includegraphics[width=0.9\columnwidth]{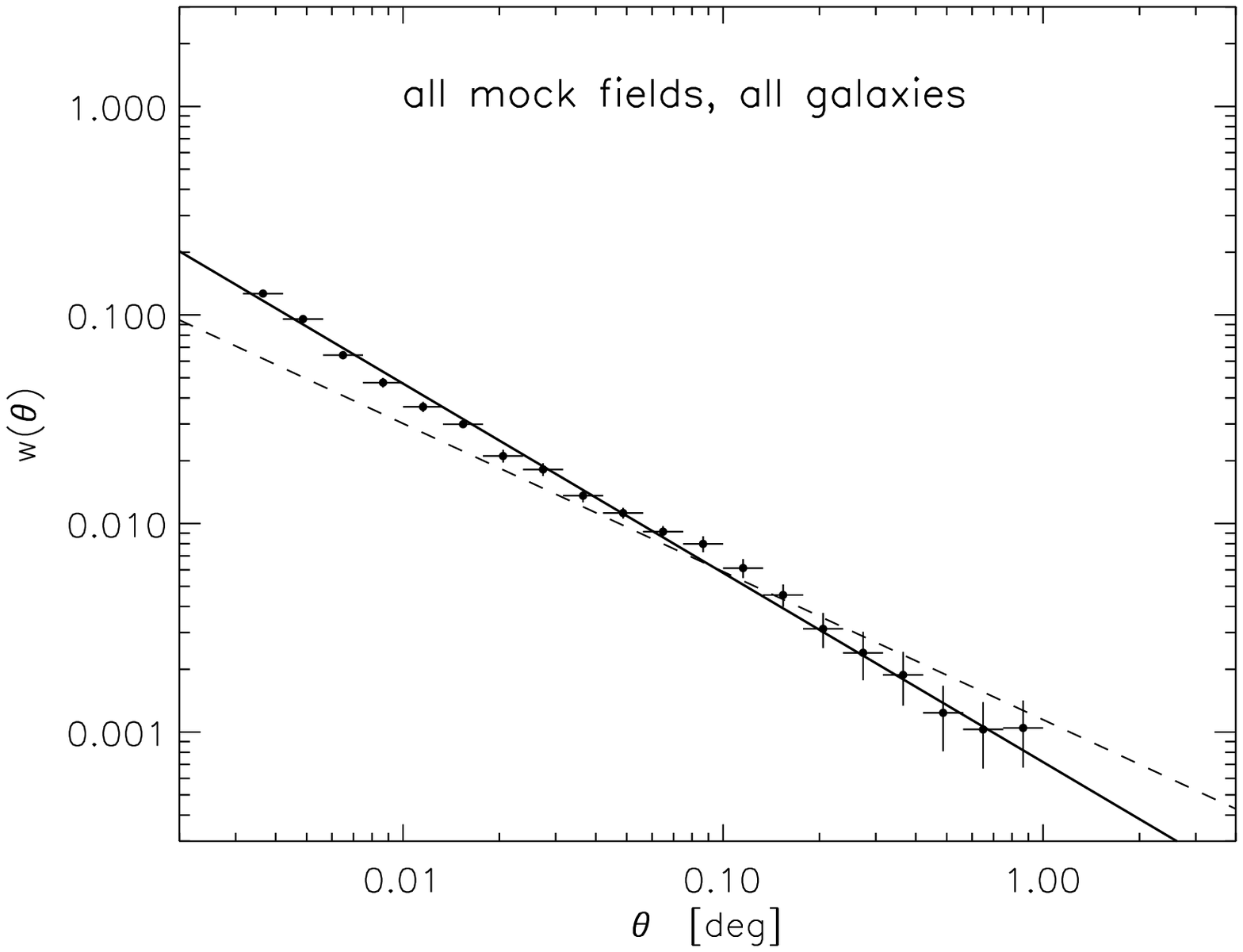} &
\includegraphics[width=0.9\columnwidth]{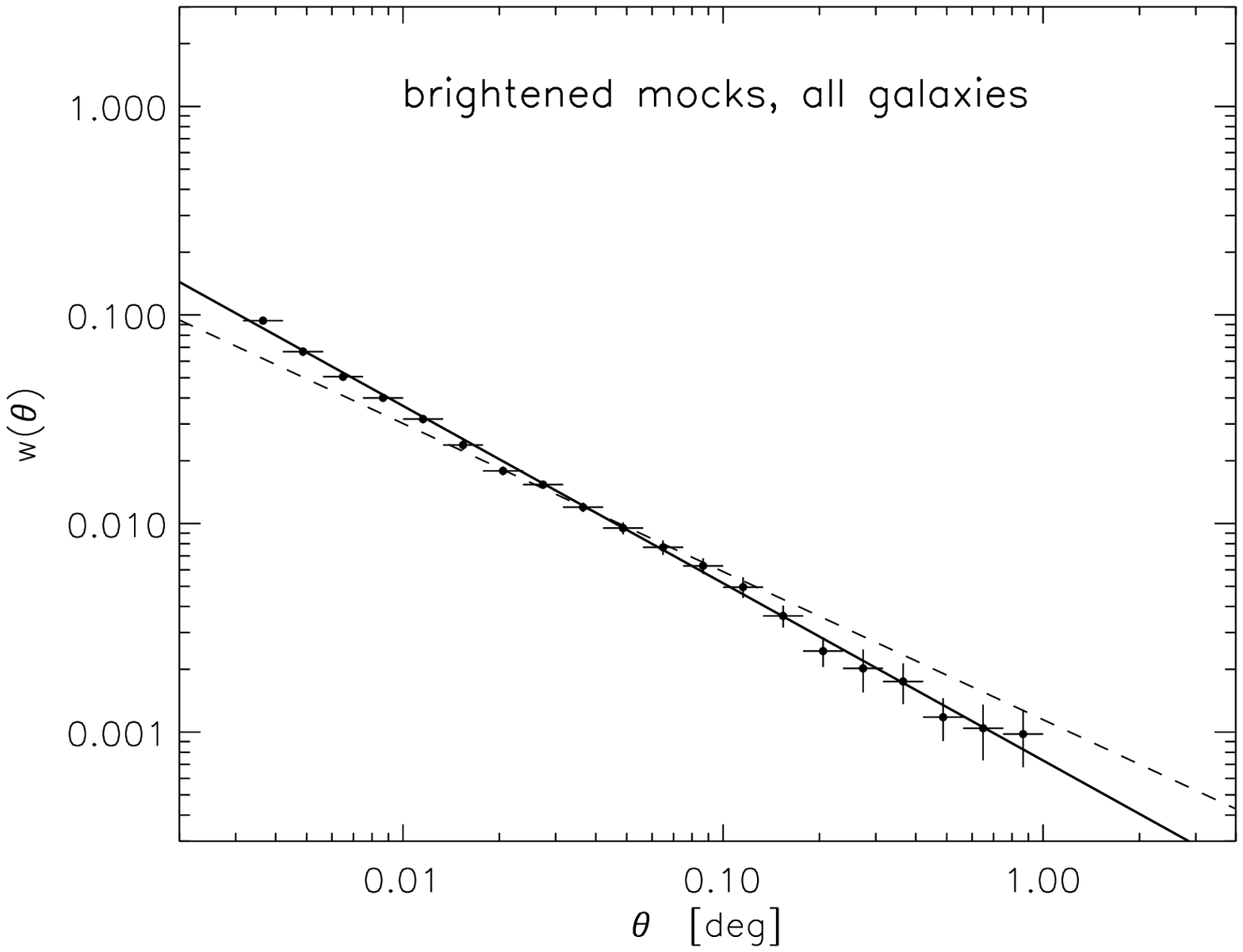} \\
\end{tabular}
\caption{Angular clustering for the two sets of SHARK samples, for the three mock fields combined,
for 'all' mock galaxies (that is, for 18$<$[3.6]$_{\rm AB}<$22 and 18$<$[4.5]$_{\rm AB}<$22).
The left-and panel is for the mocks without flux corrections to match sample sizes, the right-hand panel
is for the mocks that were corrected, where the two bands have been brightened by somewhat different
amounts (see main text).
Both sets of mocks do mimic photometric accuracy and source blending: the only difference between the two
sets are the flux offsets. The dashed line denotes the power-law fit to the observed angular correlation
function for the same flux cut. The simulated galaxies are more strongly clustered, with a steeper slope
than the observed population, most significantly for the mocks that have not been corrected (brightened)
to match sample size.} 
\label{fig:angclus-shark-all}
\end{figure*}

The uncertainties in the estimates for the clustering strengths using the three fields
combined are also smaller than for the individual fields. It is worth checking whether
the spread in $r_0$ found for the individual fields is consistent with the uncertainty in $r_0$
for the combined field: inspecting the obtained angular galaxy clustering functions shown
in \autoref{fig:angclus-all}, \autoref{fig:angclus-rb1} and \autoref{fig:angclus-rb2}
and the corresponding fitting parameters and derived $r_0$ in \autoref{tab:corr-results},
this is obvious for the full sample, but also for the blue and red samples.
For the 'very red' sample, the variance over the three individual fields is 1.25 $h^{-1}$Mpc,
which is still less than the formal error on $r_0$ for the combined fields.
However, for the 'very blue', the same variance in $r_0$ is 1.3 $h^{-1}$Mpc, which is twice
the formal error on $r_0$ for the combined estimate. The reason
for this is likely that even though the 'very blue' population has the most restricted redshift
range, this is also a fixed redshift range
around $z\approx 0.7$, and therefore sensitive to specific structures like galaxy clusters,
filaments and voids in each of the fields, which are not likely to be synchronous over
all fields: galaxy clusters, for example, have a range of formation and evolution histories.
This can only be alleviated by a survey over an even larger area. However, for our other subsamples
we find that we do not suffer from cosmic variance, and our estimates for the clustering strength
$r_0$ can be considered robust.

\cite{Waddington2007} find the same increasing trend of spatial clustering length with luminosity,
going up to values around $r_0 = 6 h^{-1}$Mpc for their brightest flux cuts (see their Table 1),
which are again somewhat below what we find for our brightest magnitude bins (see Section 4.4
and \autoref{tab:lumdep-results}), noting that they considered flux cuts, not flux bins.

\begin{figure*}
\centering
\begin{tabular}{cc}
\includegraphics[width=0.85\columnwidth]{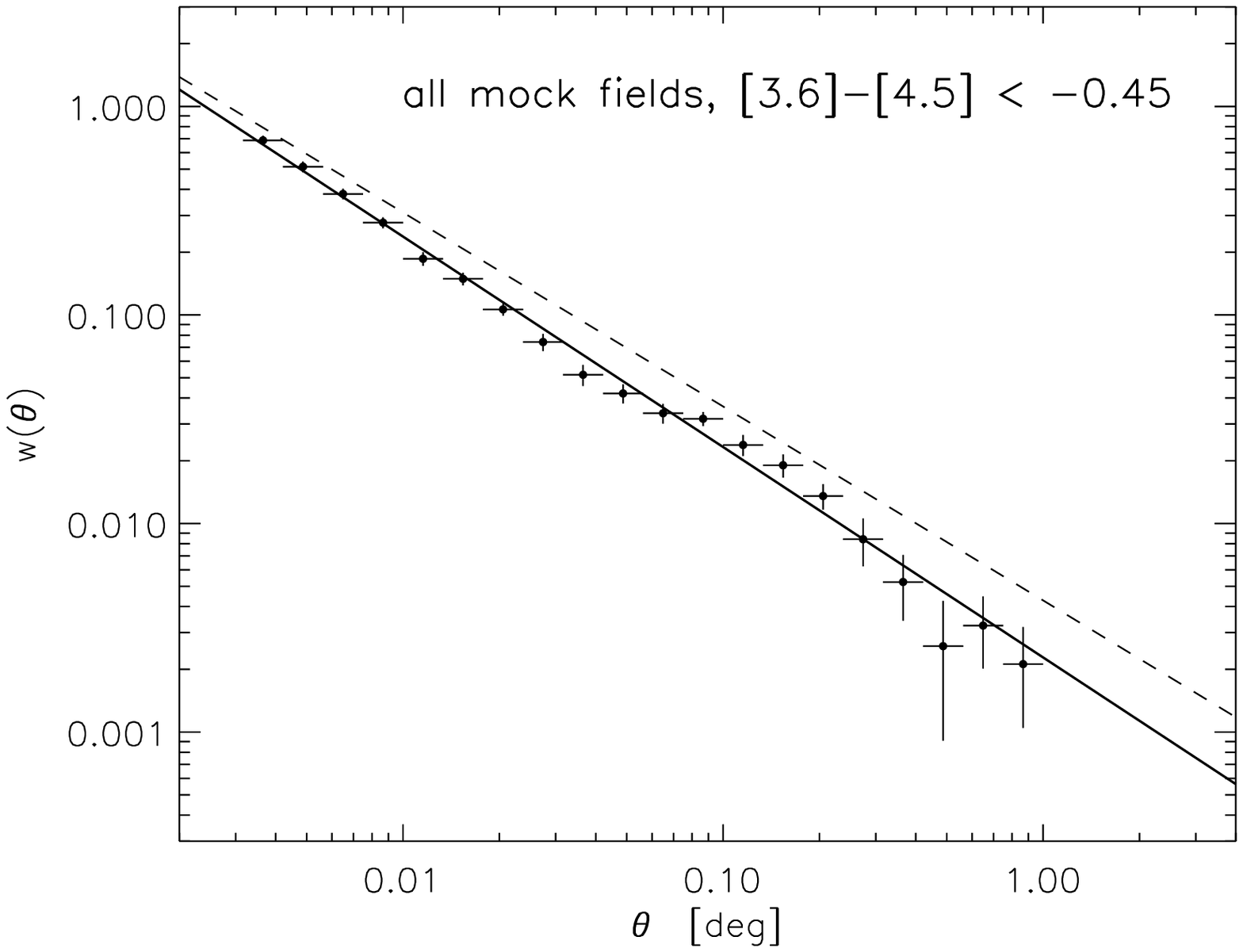} &
\includegraphics[width=0.85\columnwidth]{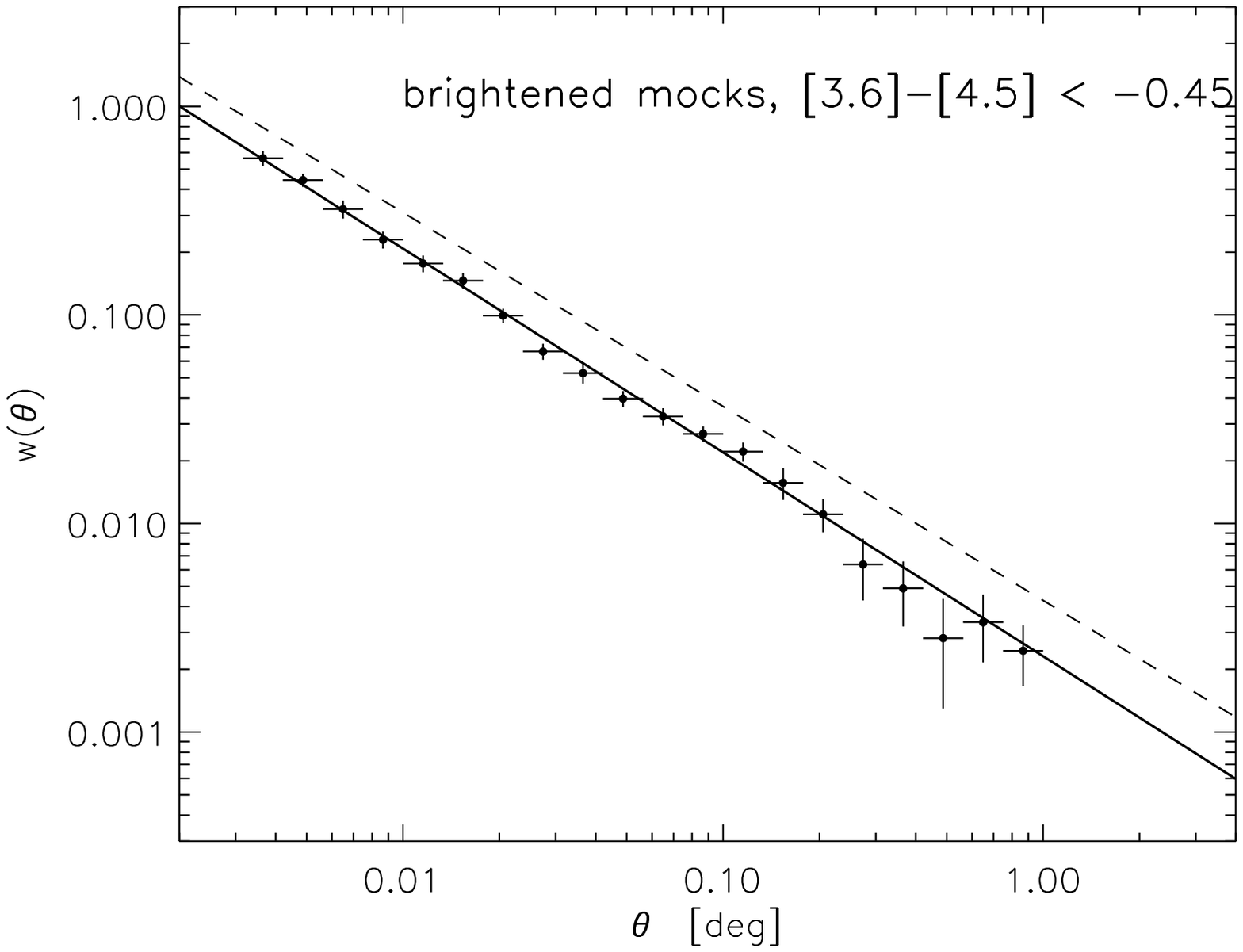} \\
\noalign{\vfilneg\vskip -0.4cm}
\includegraphics[width=0.85\columnwidth]{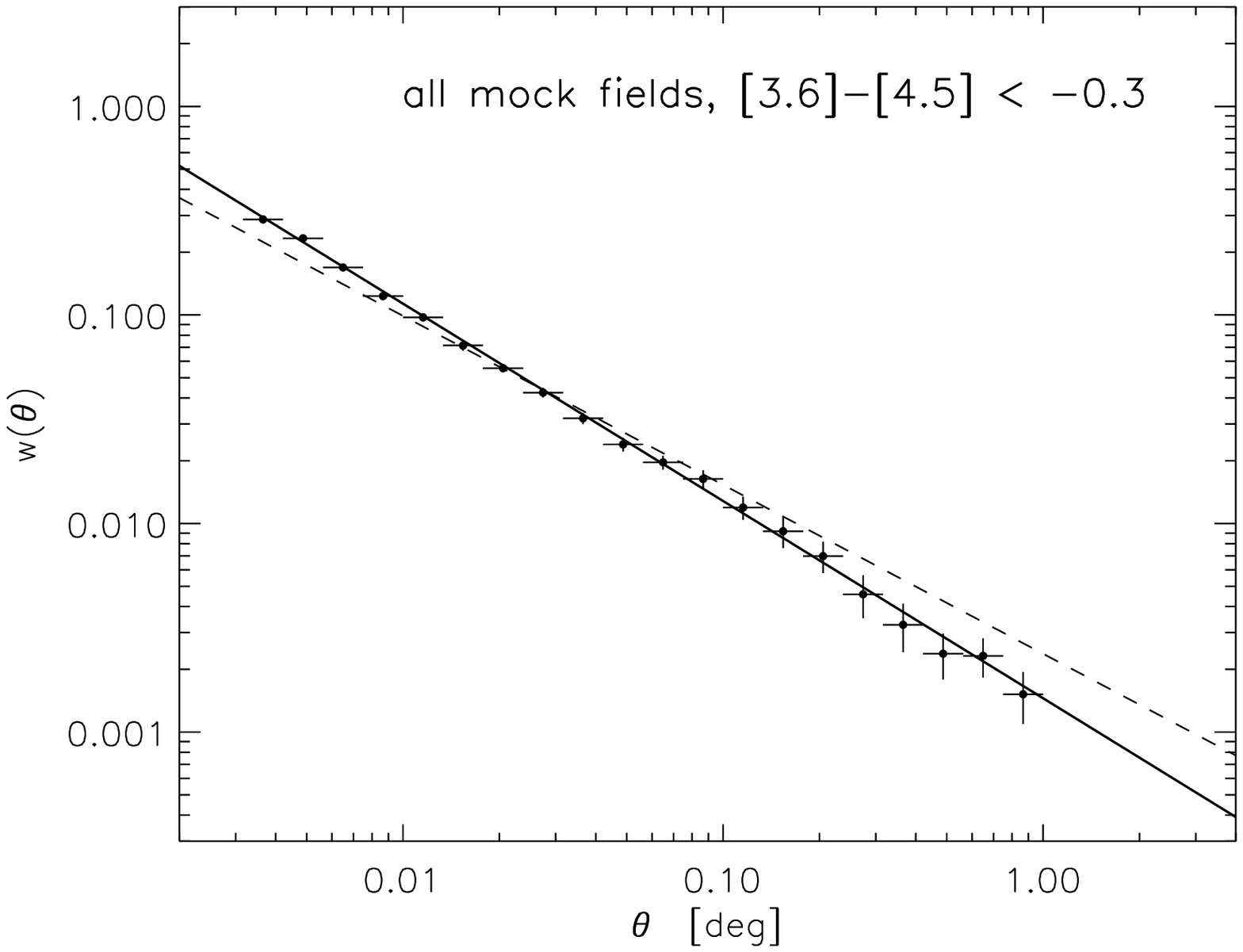} &
\includegraphics[width=0.85\columnwidth]{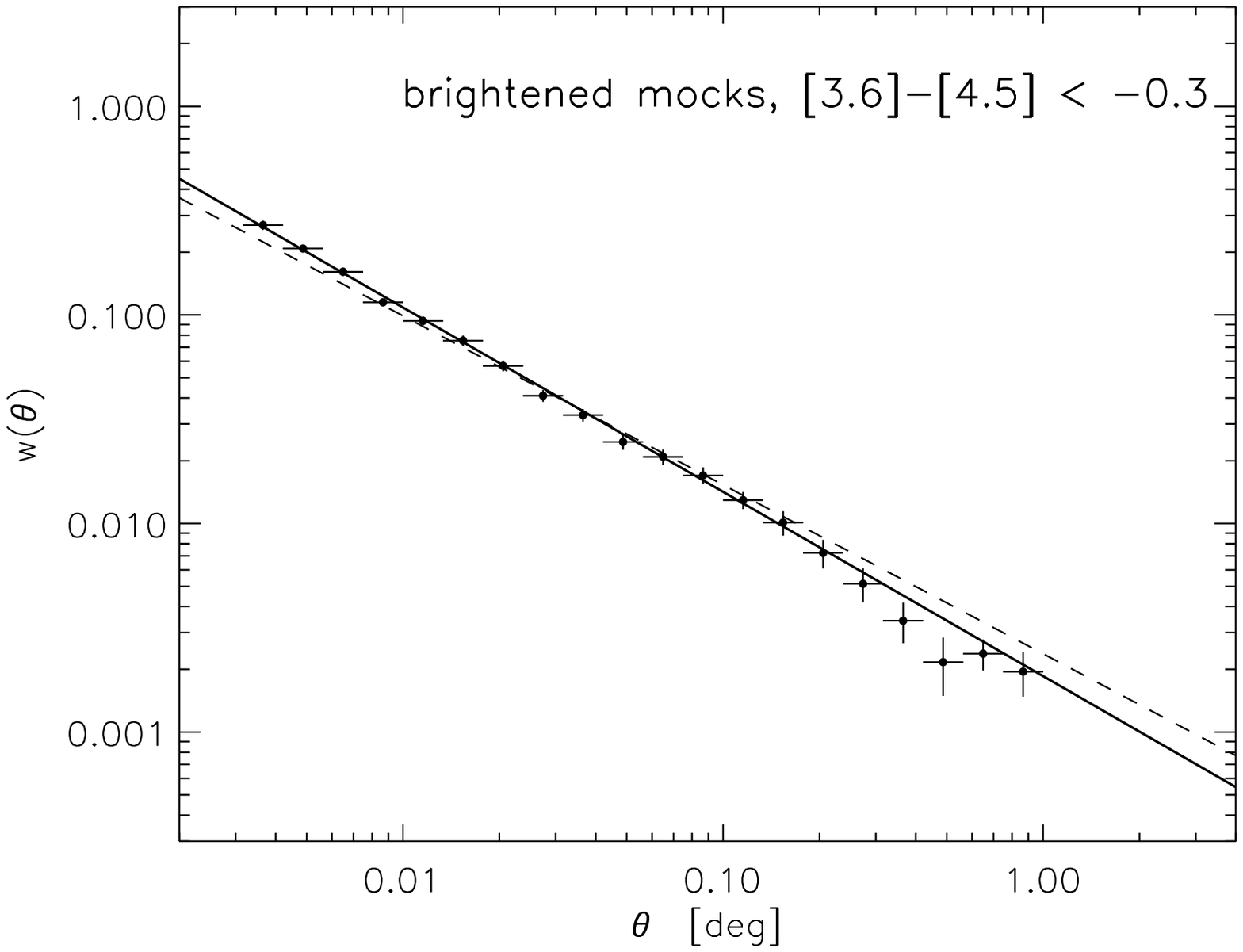} \\
\noalign{\vfilneg\vskip -0.4cm}
\includegraphics[width=0.85\columnwidth]{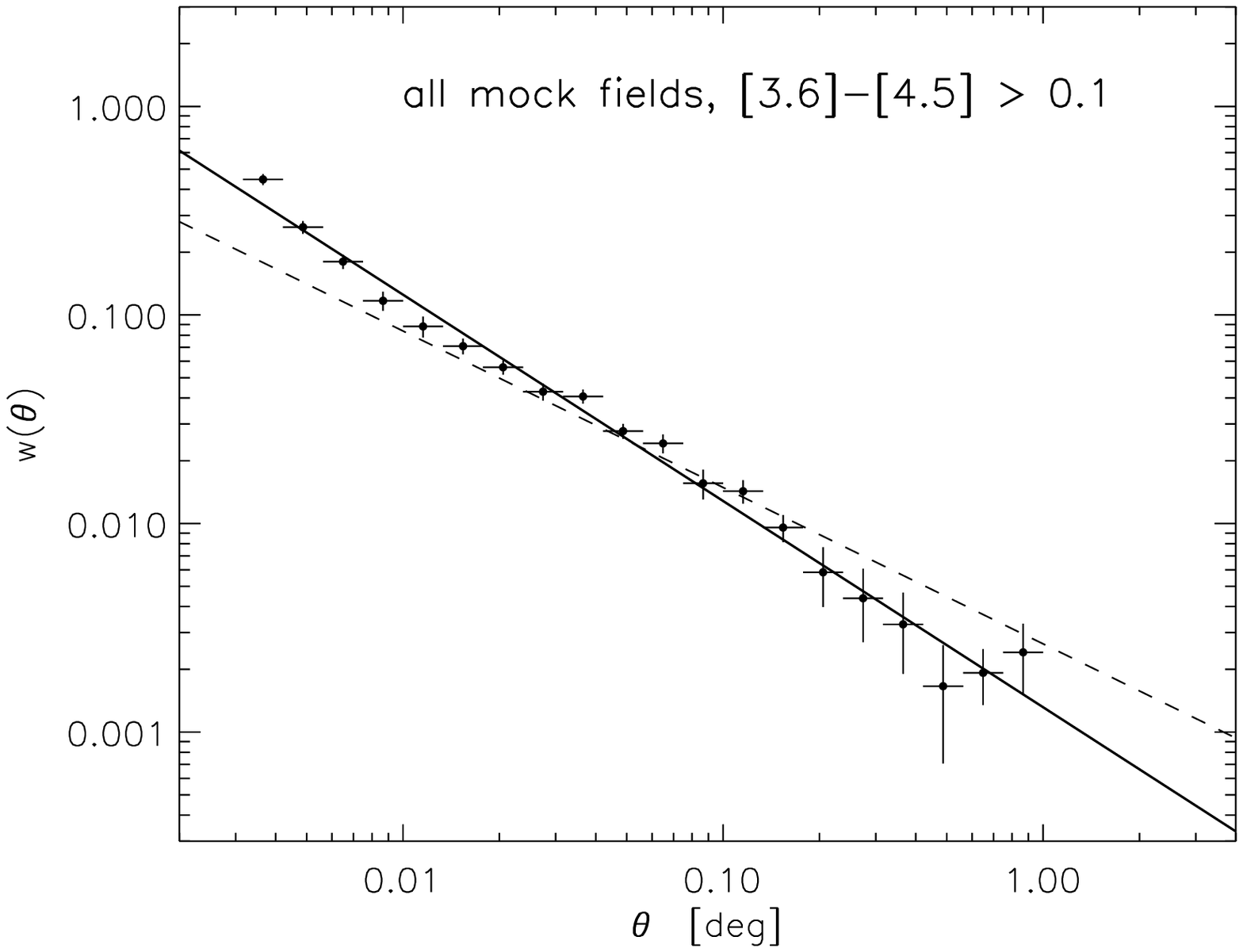} &
\includegraphics[width=0.85\columnwidth]{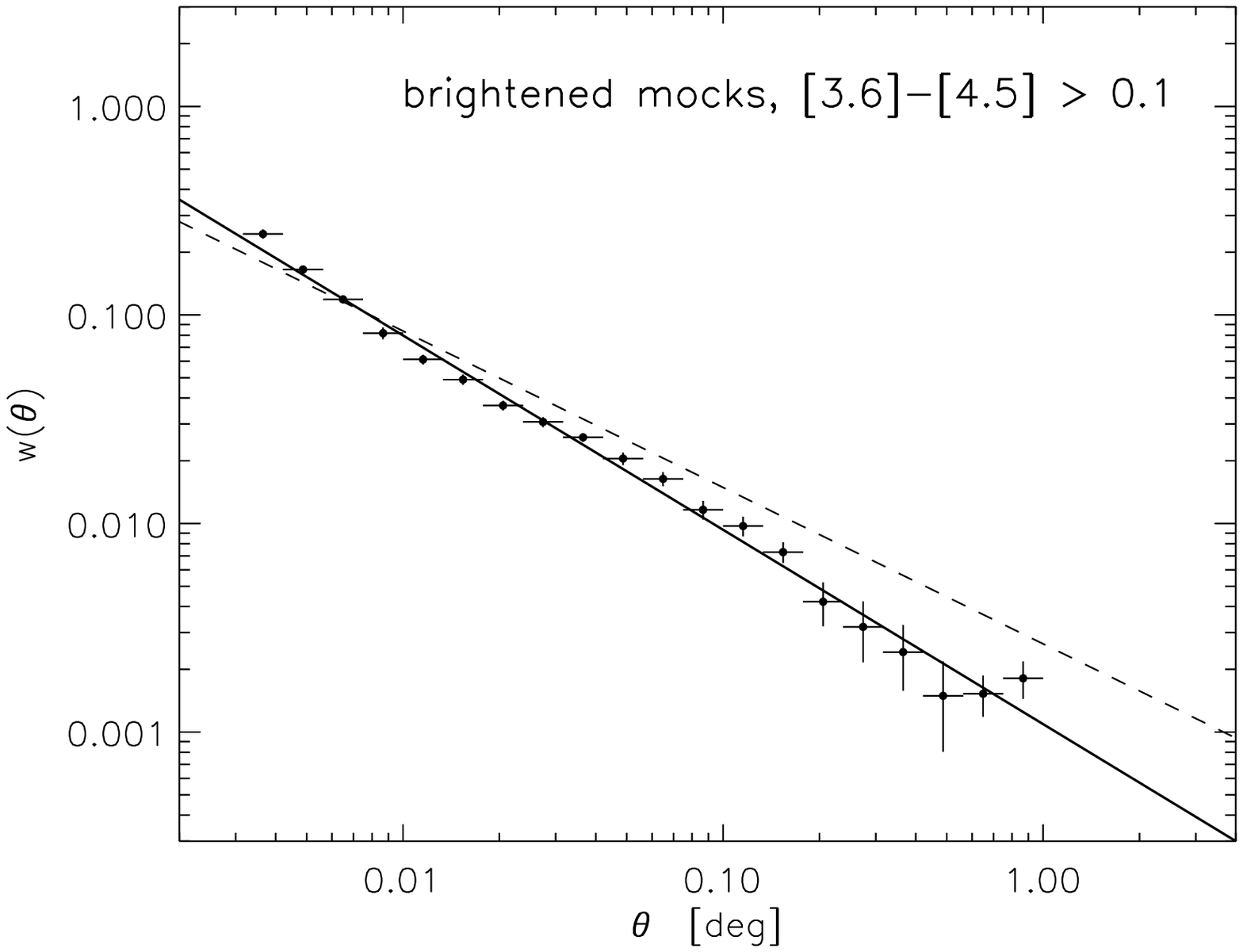} \\
\noalign{\vfilneg\vskip -0.4cm}
\includegraphics[width=0.85\columnwidth]{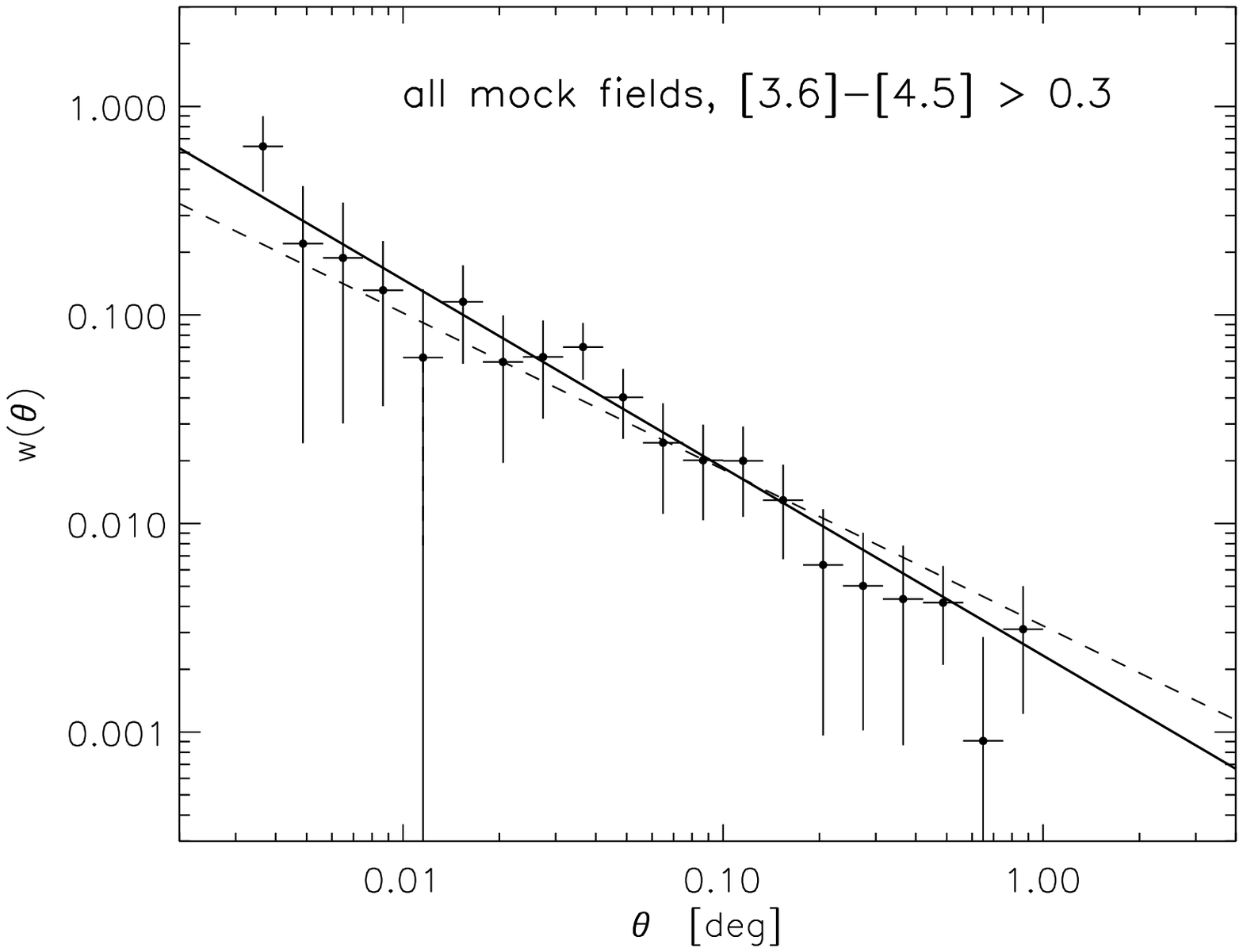} &
\includegraphics[width=0.85\columnwidth]{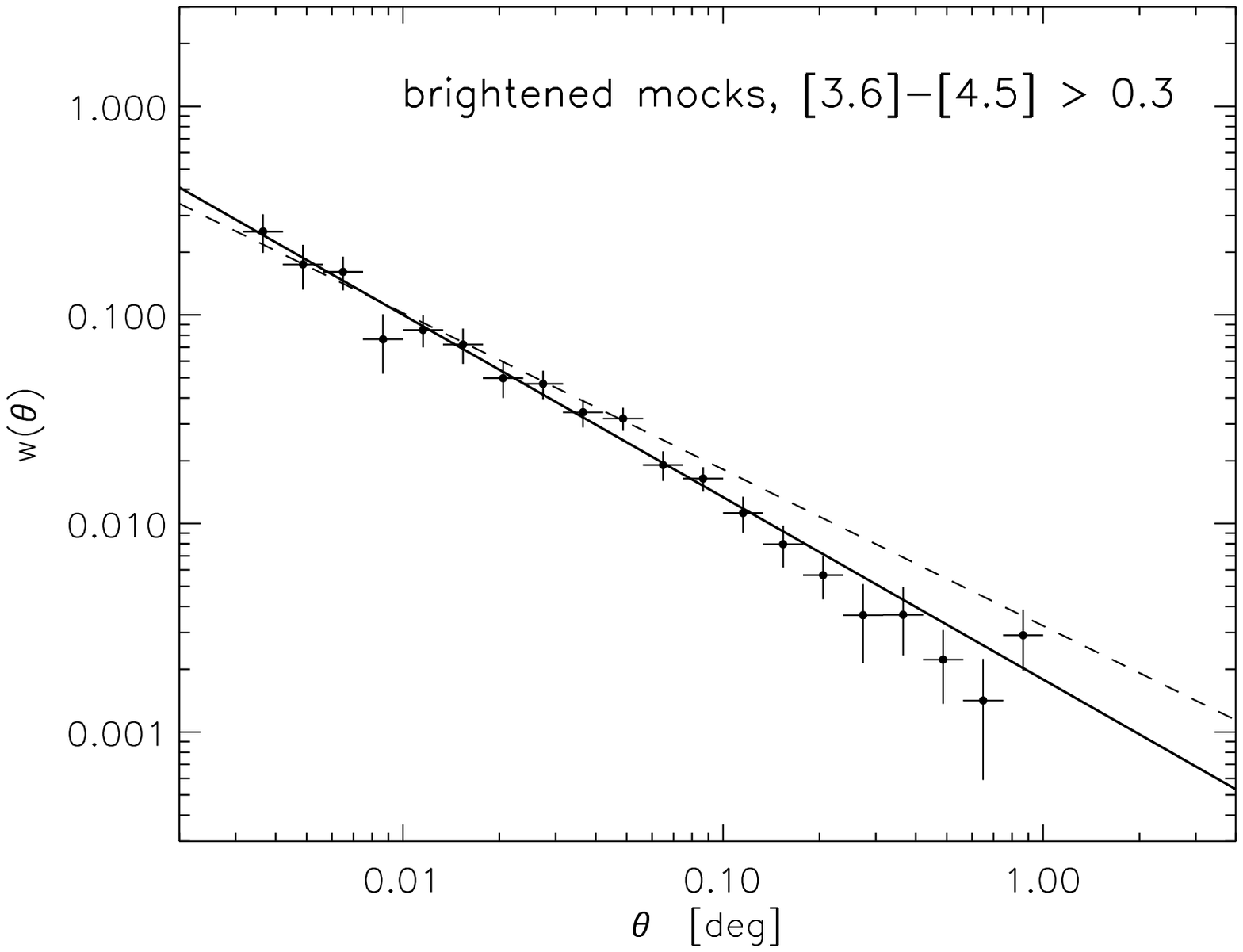} \\
\end{tabular}
\caption{The left-hand side panels show the angular clustering estimates for the first set of SHARK
mock samples with photometric accuracy and source blending (1.8 arcsec radius) taken into account,
for three mock fields combined. Estimates are presented for the same colour cuts used
for the observed samples.
The dashed line in each panel shows the power-law fit to the corresponding observed angular
correlation function with the same colour cut. As for the full sample shown in \autoref{fig:angclus-shark-all},
the simulated galaxy population is more strongly clustered than the observed one,
with a steeper slope, except for the 'very blue' population.
The right-hand side panels are for the second set of mocks, with brightened fluxes
for each of the two bands (a fixed offset) to match the number counts and completeness
of the corresponding observed subsamples. The simulated galaxy population is still
more strongly clustered than the observed one (except for the 'very blue' subsample),
but the clustering amplitudes and slopes are smaller than for the first set of mocks.
The spatial clustering lengths derived for these estimated correlation functions are smaller too
and closer to the observed values, as seen in \autoref{tab:corr-results}.
} 
\label{fig:angclus-shark-colours}
\end{figure*}

\subsection{Clustering predictions for the SHARK models}

The clustering analysis for the two sets of SHARK mock samples and subsamples was performed in exactly
the same way as for the \servs\ and \deepdrill\ data, with similar magnitude and colour cuts, to allow
for a meaningful comparison. We therefore selected three random fields from the full SHARK volume
that are not adjacent to each other (as is the case for the observational data), with the same area as
the three observed datasets from \servs+\deepdrill. Star masks were applied too, employing the same
three sets of masks that were used for the observations, in order to mimic any (small) effect
these might have. 

For both sets of mocks, the resulting estimates for the angular correlation functions for all galaxies
down to our flux limit are shown in \autoref{fig:angclus-shark-all}.
The left-hand side panel shows the estimates for the first set of mocks which only
mimic photometric accuracy and source blending, but no changes to the underlying galaxy formation model,
whereas the right-hand side panel shows the estimates for the brightened mock samples of set two
(see subsection 2.3.3). The variation between the three mock fields is similar to
that between the three observed fields. The power-law fit for all observed galaxies in all fields
(as shown before, in the bottom panel of \autoref{fig:angclus-all}) is plotted for reference as a
single dashed line.

The angular clustering estimates for the same four colour cuts that we employed for the observational data
are shown in \autoref{fig:angclus-shark-colours} for the first set (without brightening) on the left-hand side
and for the second set (with brightening) on the right-hand side. In these figures
only the joint estimates for the three mock fields are shown, and the dashed lines denote the corresponding
observational estimates for the \servs+\deepdrill\ subsamples derived from the same colour cut.
What is apparent is that the slope is somewhat steeper for all mock (sub)samples.
However, the difference in slope between models and data is more pronounced for the first set of mocks
(no brightening), especially for the red subsamples. The flux brightening (the second set of mocks)
does bring the mock samples towards a better match to the estimates for the corresponding observed samples.

All fitted parameters for the models are listed in the bottom
part of \autoref{tab:corr-results}. The spatial clustering length is comparable but mostly
somewhat larger than for the observed samples, except for the 'very blue' brightened mock subsample.
\changed{This is also shown graphically in \autoref{fig:summary}.} 
Overall the clustering lengths for the brightened mock samples are not significantly different from the observed
ones: only for the 'very blue' subsample we find that the difference between the spatial clustering lengths of
the brightened mock and observed subsamples is larger than the formal uncertainty of each, but not by much.
The clustering lengths for the first set of mock samples (with no brightening) are typically too large,
\changed{except for the 'very blue' subsample, noting that both the counts and the colours for the
first set of mocks do not match the observed ones well.}

\begin{table*}
\centering
  \begin{tabular}{llrrrr}
  \hline
  \noalign{\vfilneg\vskip -0.2cm}
  \hline
Colour and sample & Luminosity   & N &  A           &   $\delta$\hskip 0.5cm  &     $r_0$             \\
    selection     &     bin      &   &  [arcsec]    &                         &   [$h^{-1}$Mpc]       \\
\hline   
all, observed  & 22 > [3.6] > 18 &  869587 &  0.26 $\pm$  0.05 &  0.71 $\pm$  0.03 &  4.96 $\pm$  0.99 \\
all, observed  & 22 > [3.6] > 21 &  368889 &  0.17 $\pm$  0.05 &  0.70 $\pm$  0.04 &  4.50 $\pm$  1.49 \\
all, observed  & 21 > [3.6] > 20 &  303735 &  0.32 $\pm$  0.08 &  0.70 $\pm$  0.04 &  5.17 $\pm$  1.35 \\
all, observed  & 20 > [3.6] > 19 &  153987 &  1.29 $\pm$  0.19 &  0.80 $\pm$  0.03 &  7.20 $\pm$  1.12 \\
all, observed  & 19 > [3.6] > 18 &   42976 &  3.49 $\pm$  0.56 &  0.94 $\pm$  0.05 &  7.76 $\pm$  1.31 \\
\hline   
all, brightened mocks & 22 > [3.6] > 18 &  946217 &  0.74 $\pm$  0.06 &  0.85 $\pm$  0.02 &  6.40 $\pm$  0.57 \\
all, brightened mocks & 22 > [3.6] > 21 &  421926 &  0.32 $\pm$  0.07 &  0.76 $\pm$  0.03 &  5.47 $\pm$  1.22 \\
all, brightened mocks & 21 > [3.6] > 20 &  317094 &  0.73 $\pm$  0.09 &  0.81 $\pm$  0.03 &  6.32 $\pm$  0.84 \\
all, brightened mocks & 20 > [3.6] > 19 &  155286 &  2.10 $\pm$  0.20 &  0.94 $\pm$  0.03 &  7.97 $\pm$  0.79 \\
all, brightened mocks & 19 > [3.6] > 18 &   51911 &  5.39 $\pm$  0.46 &  1.01 $\pm$  0.04 &  9.22 $\pm$  0.85 \\
\hline   
blue, observed & 22 > [3.6] > 18 &  247746 &  2.07 $\pm$  0.15 &  0.81 $\pm$  0.02 &  6.50 $\pm$  0.51 \\
blue, observed & 22 > [3.6] > 21 &   62096 &  1.13 $\pm$  0.31 &  0.77 $\pm$  0.05 &  5.52 $\pm$  1.56 \\
blue, observed & 21 > [3.6] > 20 &   94439 &  1.70 $\pm$  0.25 &  0.76 $\pm$  0.03 &  5.93 $\pm$  0.92 \\
blue, observed & 20 > [3.6] > 19 &   68051 &  3.24 $\pm$  0.37 &  0.80 $\pm$  0.03 &  7.53 $\pm$  0.92 \\
blue, observed & 19 > [3.6] > 18 &   23160 &  3.33 $\pm$  0.98 &  0.99 $\pm$  0.09 &  7.21 $\pm$  2.23 \\
\hline   
blue, brightened mocks & 22 > [3.6] > 18 &  221476 &  2.92 $\pm$  0.17 &  0.88 $\pm$  0.02 &  7.30 $\pm$  0.46 \\
blue, brightened mocks & 22 > [3.6] > 21 &   55389 &  0.65 $\pm$  0.24 &  0.70 $\pm$  0.05 &  4.65 $\pm$  1.75 \\
blue, brightened mocks & 21 > [3.6] > 20 &   85838 &  1.98 $\pm$  0.28 &  0.81 $\pm$  0.03 &  6.16 $\pm$  0.91 \\
blue, brightened mocks & 20 > [3.6] > 19 &   59967 &  4.60 $\pm$  0.38 &  0.93 $\pm$  0.03 &  8.42 $\pm$  0.74 \\
blue, brightened mocks & 19 > [3.6] > 18 &   20282 & 12.87 $\pm$  0.84 &  1.14 $\pm$  0.05 & 13.33 $\pm$  1.02 \\
\hline   
red, observed & 22 > [3.6] > 18 &  216848 &  1.32 $\pm$  0.15 &  0.75 $\pm$  0.03 &  7.83 $\pm$  0.95 \\
red, observed & 22 > [3.6] > 21 &  136986 &  0.87 $\pm$  0.17 &  0.72 $\pm$  0.03 &  7.10 $\pm$  1.38 \\
red, observed & 21 > [3.6] > 20 &   66538 &  3.32 $\pm$  0.40 &  0.84 $\pm$  0.04 &  9.88 $\pm$  1.25 \\
red, observed & 20 > [3.6] > 19 &   11970 &  6.76 $\pm$  1.59 &  0.88 $\pm$  0.08 & 13.55 $\pm$  3.45 \\
\hline   
red, brightened mocks & 22 > [3.6] > 18 &  220590 &  2.39 $\pm$  0.16 &  0.93 $\pm$  0.02 &  8.91 $\pm$  0.61 \\
red, brightened mocks & 22 > [3.6] > 21 &  146756 &  1.25 $\pm$  0.18 &  0.86 $\pm$  0.03 &  7.32 $\pm$  1.09 \\
red, brightened mocks & 21 > [3.6] > 20 &   60181 &  4.42 $\pm$  0.41 &  1.00 $\pm$  0.04 & 10.54 $\pm$  1.05 \\
red, brightened mocks & 20 > [3.6] > 19 &   12189 & 14.01 $\pm$  1.41 &  1.08 $\pm$  0.06 & 18.46 $\pm$  2.15 \\
\hline
\noalign{\vfilneg\vskip -0.2cm}
\hline
\end{tabular}
\caption{Clustering measures for subsamples in luminosity bins, observational as well as simulated
(see main text for details) for the two-parameter fits. This table also lists the actual sizes of these
subsamples, as well as the derived spatial clustering strength using Limber equation inversion technique.}
\label{tab:lumdep-results}
\end{table*}

\begin{figure*}
\centering
\begin{tabular}{cc}
\includegraphics[width=0.9\columnwidth]{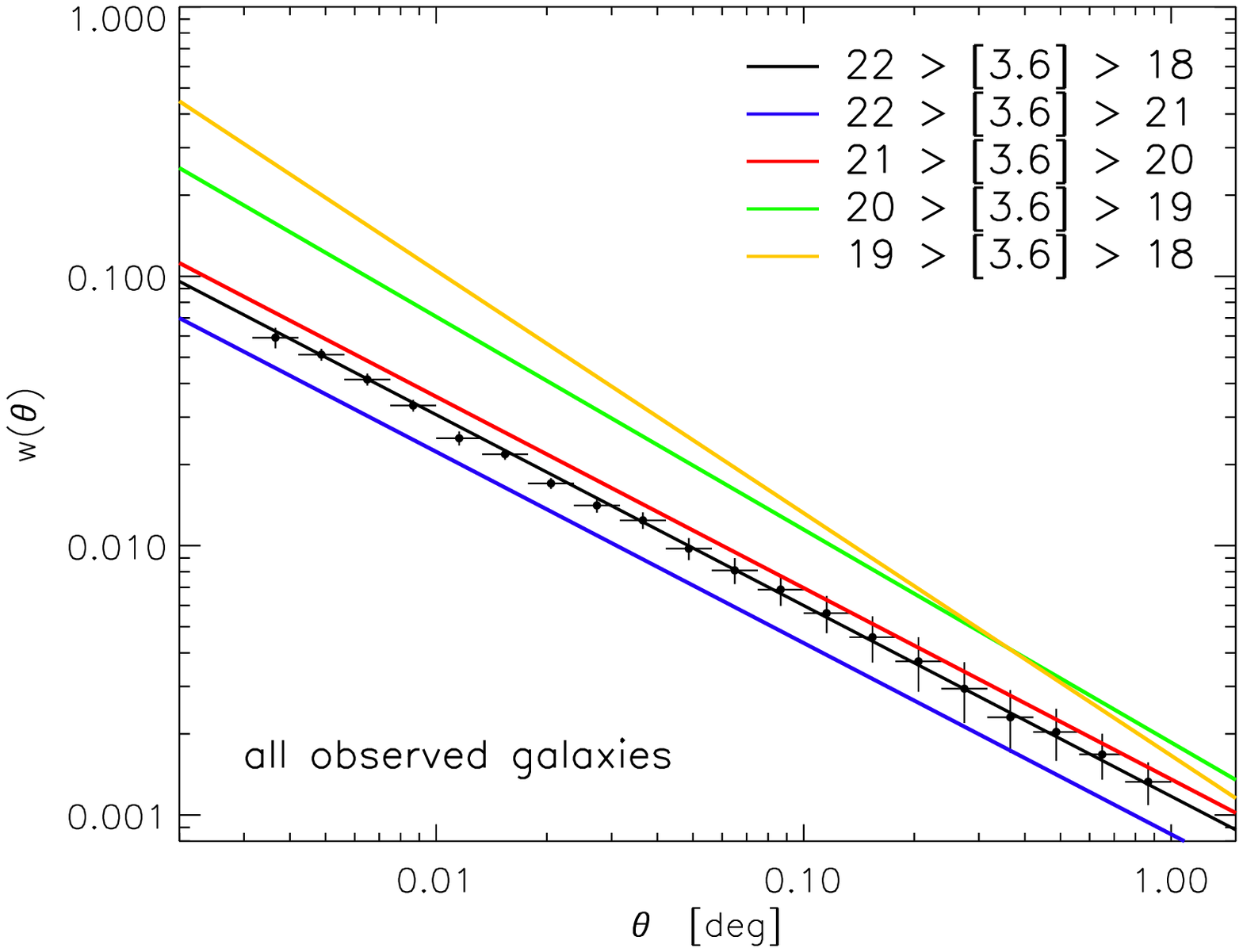} &
\includegraphics[width=0.9\columnwidth]{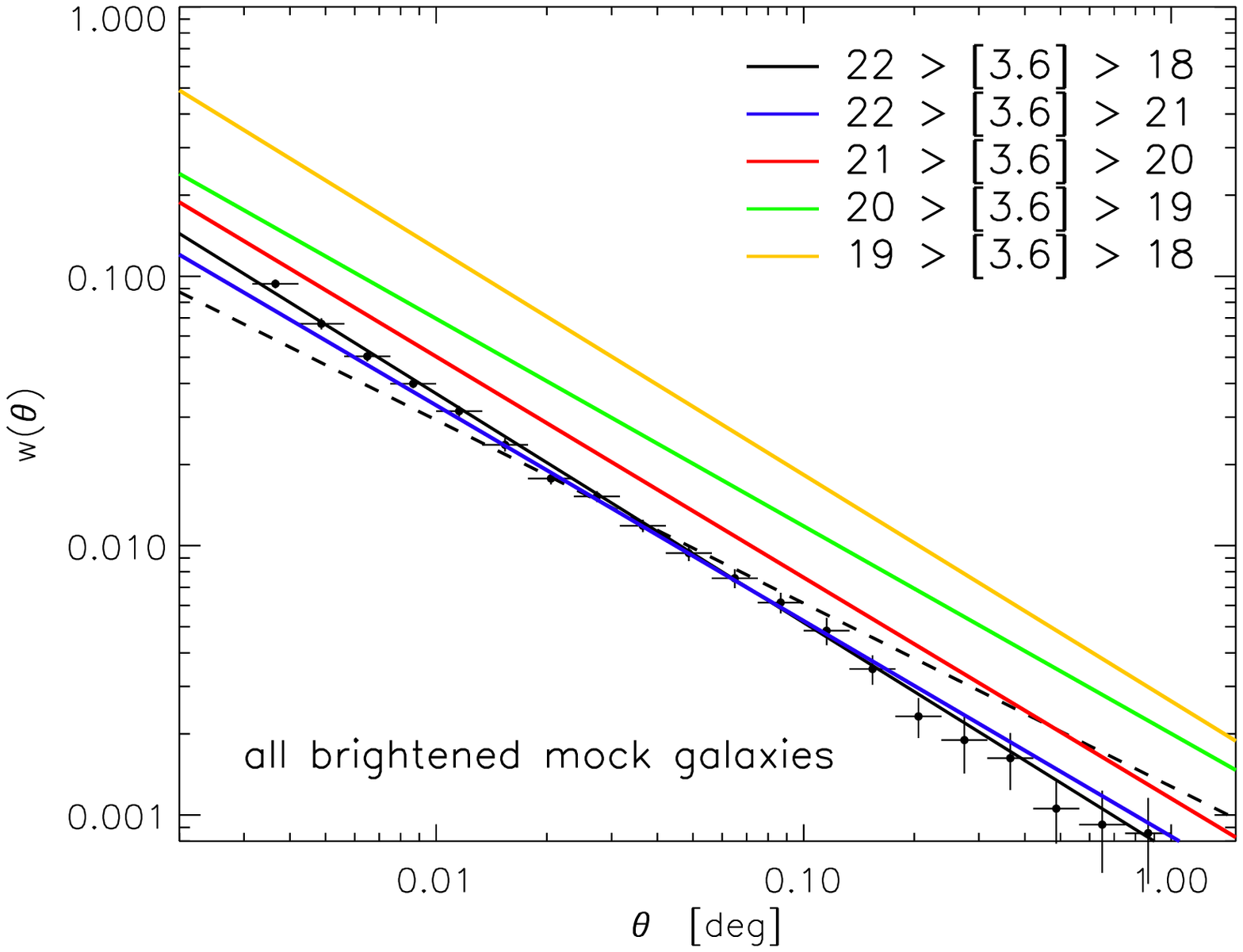} \\
\end{tabular}
\caption{Left-hand side: angular clustering in magnitude bins for the \servs+\deepdrill\ sample,
for the three fields combined, without colour selection.
Right-hand side: angular clustering in magnitude bins for the flux-corrected (brightened) SHARK models, for the
three mock fields combined, without colour selection. The dashed line denotes the power-law fit to the
observed angular correlation function for the same flux cut (the black solid line in the top panel),
to aid visual comparison.} 
\label{fig:angclus-lumbins}
\end{figure*}

\begin{figure*}
\centering
\begin{tabular}{cc}
\includegraphics[width=0.9\columnwidth]{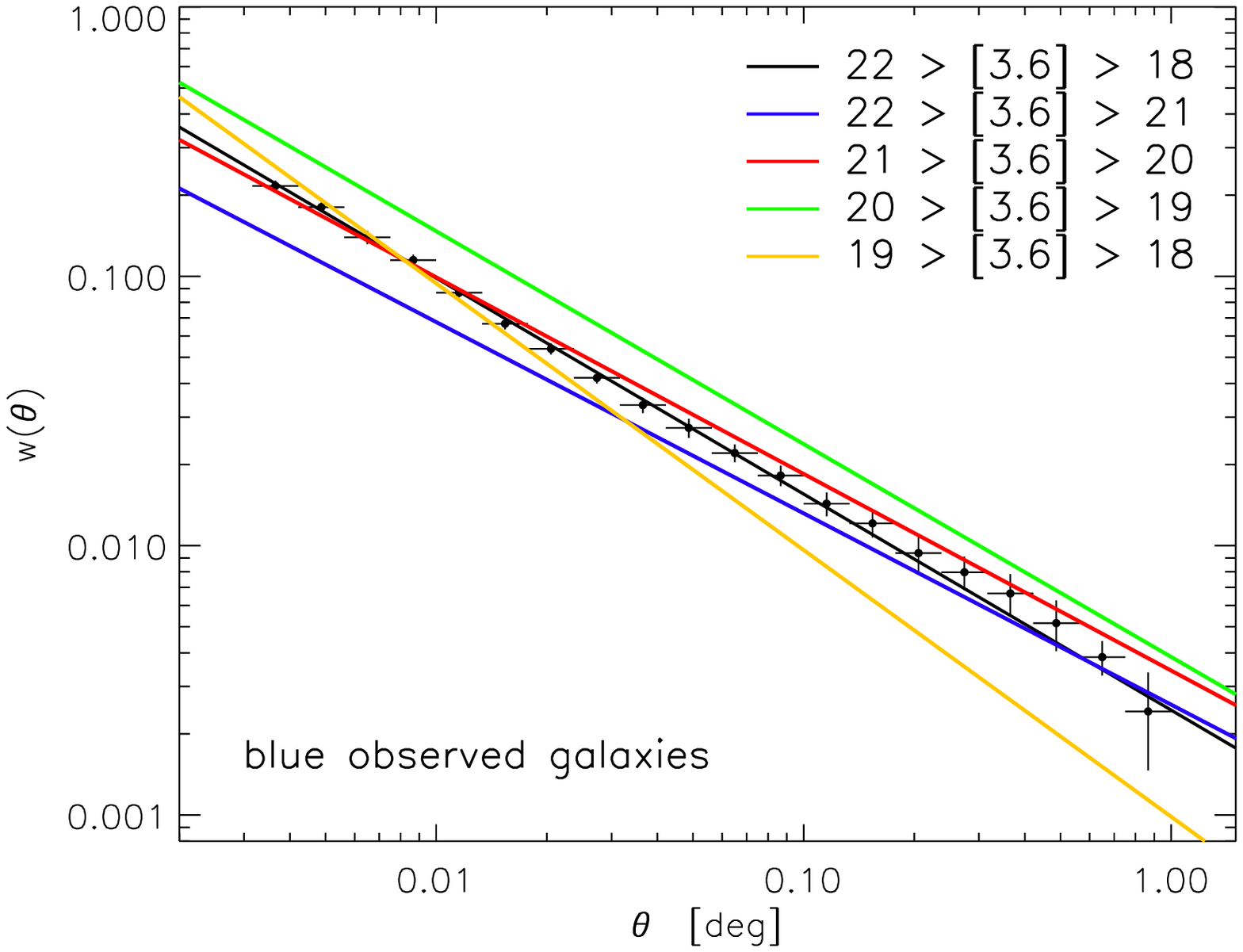} &
\includegraphics[width=0.9\columnwidth]{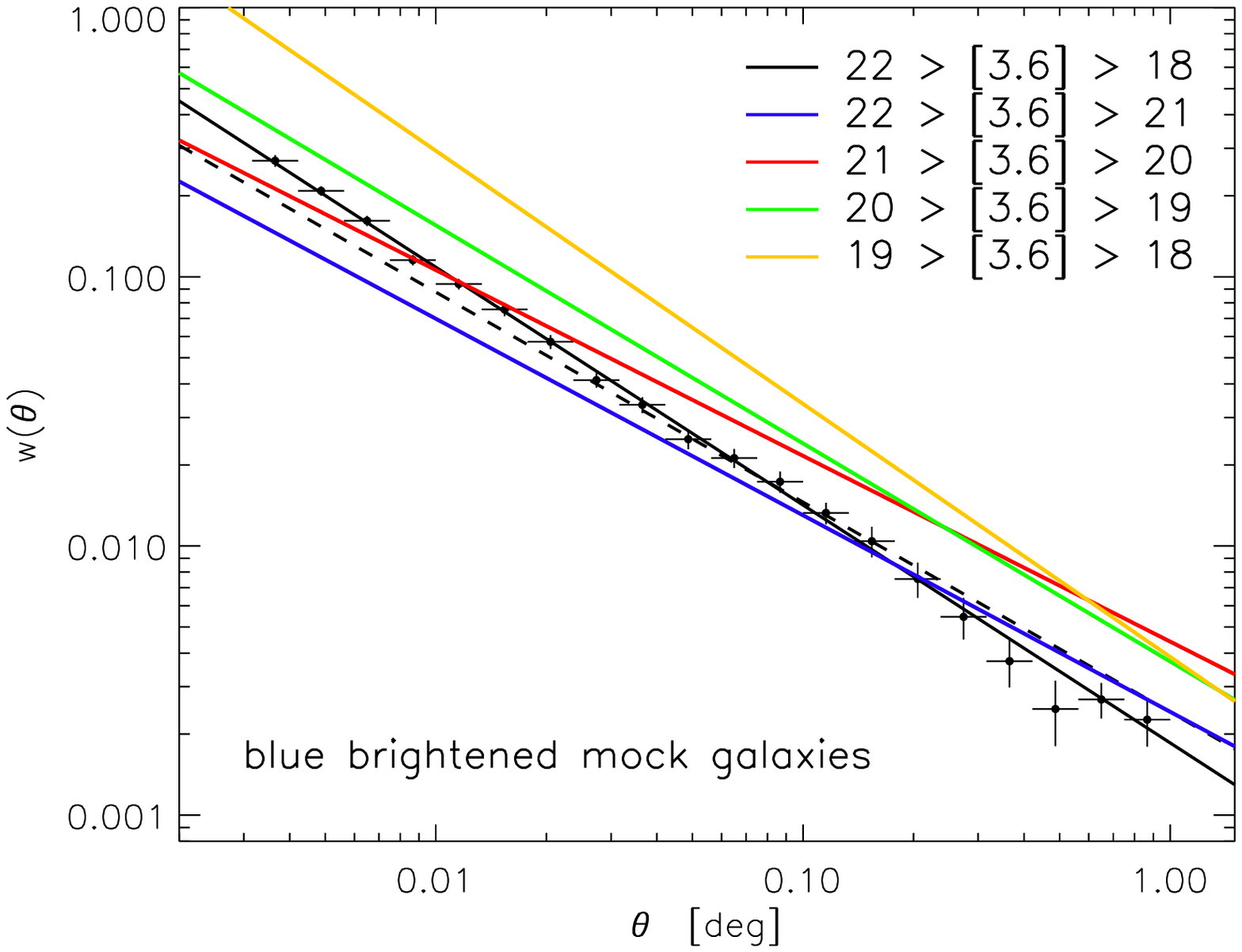} \\
\noalign{\vfilneg\vskip -0.4cm}
\includegraphics[width=0.9\columnwidth]{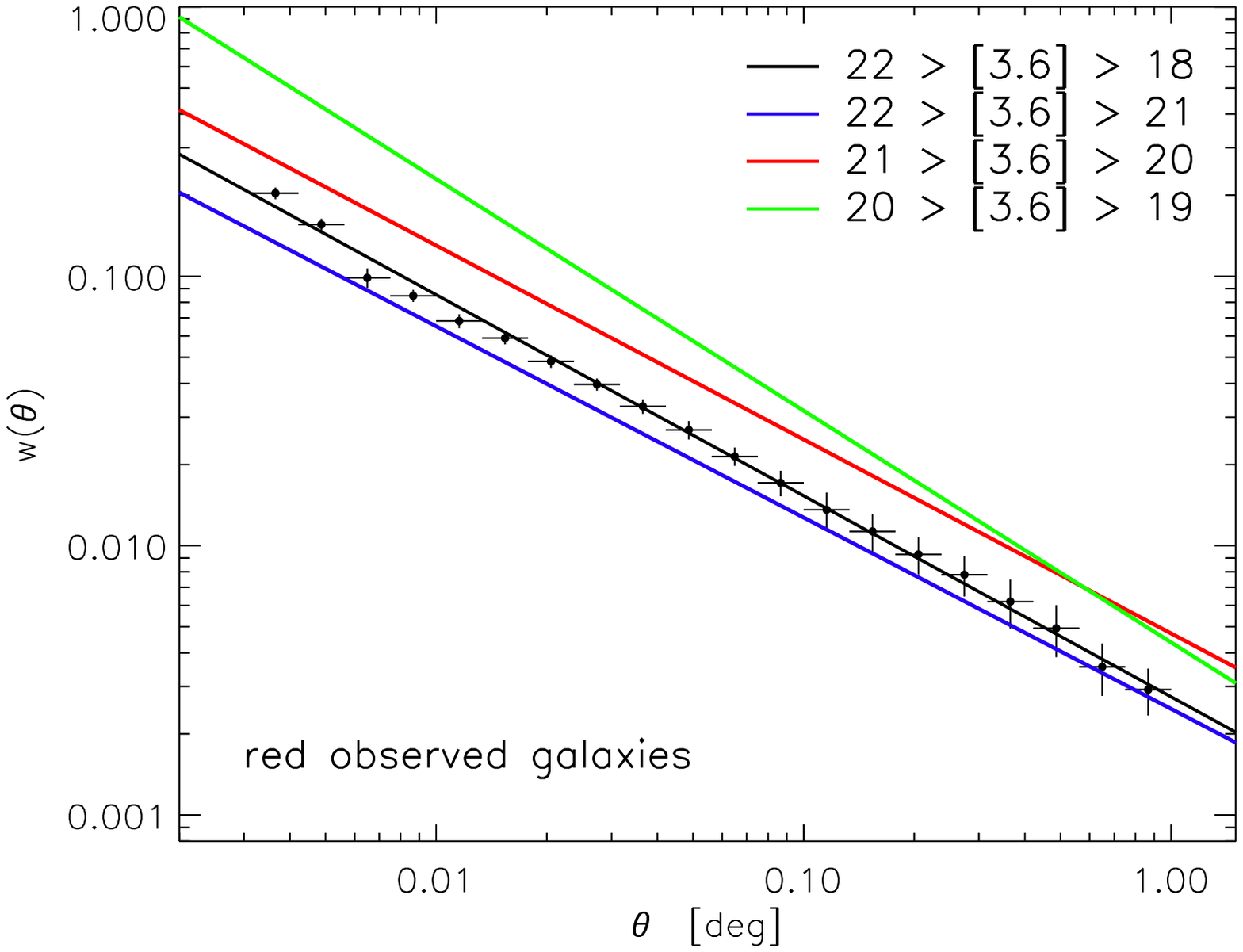} &
\includegraphics[width=0.9\columnwidth]{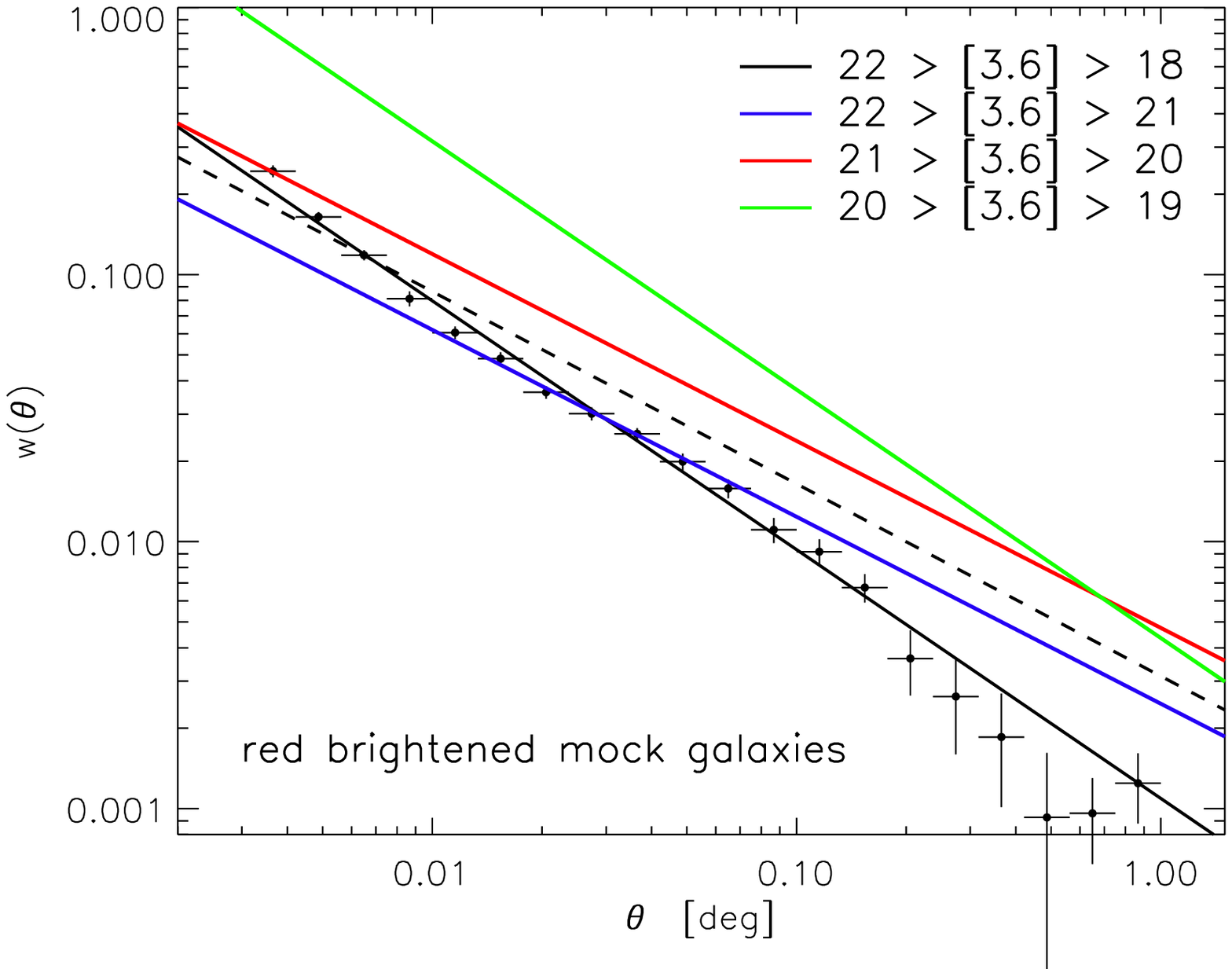} \\
\end{tabular}
\caption{Same as \autoref{fig:angclus-lumbins}, but for the blue and red subsamples. There are not
enough sources in the brightest bin for the red subsample to robustly estimate and fit an autocorrelation
function (see main text).} 
\label{fig:angclus-lumbins-cols}
\end{figure*}

\subsection{Estimates for magnitude selected subsamples}

In the following analysis we restrict ourselves to the second set of (brightened) mock samples, as
the first set does not match the observations well.
For the full sample as well as for the red and blue samples, for both \servs+\deepdrill\ and SHARK,
there are sufficient numbers of sources to also allow a selection in bins of IRAC [3.6] magnitude.
We use a bin size of one magnitude, starting at our faint flux limit of [3.6] = 22, and up to where the
number of sources becomes to small for a robust clustering estimate, which is the bin 19$>$[3.6]$>$18
for the full sample and the blue subsample, and the bin 20$>$[3.6]$>$19 for the red subsample.
The 19$>$[3.6]$>$18 bin for the red subsample only contains of order 1500 galaxies (of order 500 per
field), for which we would need to use a different estimate for clustering, such as the sky-averaged
autocorrelation function (eg. \citealt{vanKampen2005}), which makes comparison to the other subsamples
much harder. We do not select in the [4.5] micron band: just \changed{within the range 18$<$[4.5]$<$22,
as for the full sample.}

The results for the observed and simulated (sub)samples are presented in
\autoref{tab:lumdep-results}, and in more visual form in \autoref{fig:angclus-lumbins} and
\autoref{fig:angclus-lumbins-cols}, where the data points and the black solid line are the same
as for the corresponding panels at the bottom of \autoref{fig:angclus-all} and
\autoref{fig:angclus-rb1}, i.e. for the full luminosity range in both bands. The dashed line in
the right-hand side panel is just the fit to the observed data for all galaxies, in order to help visual
comparison. 

Looking at the fits to the clustering estimates for full sample divided in magnitude bins
(\autoref{fig:angclus-lumbins}), we see a similar trend for both observed and mock galaxies:
the estimate for the faintest bin (22$>$[3.6]$>$21, blue line) shows a clustering amplitude that is 
lower than the one for all galaxies (data points and black line). For each consecutive brighter magnitude
bin the angular clustering amplitude and the spatial clustering length get larger. For the brightest
magnitude bin (19$>$[3.6]$>$18, orange line) the observed subsample has a fitting function that is
steeper than the mock one, but that is the only clear difference between the data and the simulations.

For the red and blue subsamples we see something similar (\autoref{fig:angclus-lumbins-cols}):
in all four panel the estimate for the faintest bin (22$>$[3.6]$>$21, blue line) is below the
estimate for the red/blue subsample over the whole magnitude range (data points and black line).
Also, again for consecutive brighter magnitude bins the angular clustering amplitude and the spatial
clustering length get larger. For the red observed and red mock subsamples the steepening of the correlation
function already starts with the 20$>$[3.6]$>$19 bin (green line), but for the blue subsamples again
at 19$>$[3.6]$>$18 (orange line), this time for both the observed and mock data, with a clear difference
in amplitude.

Overall the differences in clustering estimates between observed and (brightened) mock samples are
surprisingly small, except for the smallest sample sizes for the brightest magnitude bins.
As the uncertainties are larger for these, shallower surveys over a larger area would be required
before drawing definite conclusions from such differences. 
However, it is worth stating that it is also in this magnitude range that the number counts of SHARK differ
most from the observations before magnitude corrections were made to the mocks, most likely
due to the model ignoring the AGN contribution to the galaxy SED, as shown in \autoref{fig:agns}.

\section{Discussion and conclusions}

We have measured the clustering strength for a large sample of near-infrared galaxies in all
fields of \servs+\deepdrill, with uncertainties, and for four colour-selected subsamples as well.
The full observational dataset is made up of three
round fields, which combine to a total survey area of just over 20 deg$^2$.
Angular clustering measures were obtained down to a flux limit as well as for
subsamples selected on the [3.6]-[4.5] colour, which effectively selects for redshift.
In order to obtain a spatial clustering estimate from the angular correlation function
we used the redshift distribution from S-COSMOS, with the same selection criteria as used
for our samples, and the Limber equation inversion technique.

We find that our angular correlation functions are robustly estimated for all (sub)samples,
which is not surprising given our sample size, but we also find that these are well fitted
by a power-law function over most of its range. 
We find a correlation length $r_0 = 4.96 \pm 0.99 h^{-1}$Mpc for the full sample down to an AB magnitude
limit of 22 in both bands ([3.6] and [4.5]), which is consistent with what has traditionally been
found in the optical as well for large samples with no redshift selection, and in various other wavebands,
such as 24 micron: \cite{Gilli2007} found a correlation length $r_0 = 4.0 \pm 0.4\ h^{-1}$Mpc
for all galaxies with $f_{24} > 20\ \mu$Jy (no redshift selection). In the following we also compare some
of our results to other measurements found in the literature, mostly those listed in
\autoref{tab:literature-results}.

\begin{table*}
\centering
  \begin{tabular}{lllr}
  \hline
  \noalign{\vfilneg\vskip -0.2cm}
  \hline
Reference & Sample              & Selection    &       $r_0$             \\
          &                     &              & [$h^{-1}$Mpc]           \\
\hline
\cite{Waddington2007} &  SWIRE   &  $S_{3.6} < 6.3 \mu$Jy  &  $3.18 \pm 0.94$ \\
\hline
\cite{Gilli2007}      &  GOODS    &  $f_{24 \mu m} > 20\ \mu$Jy  &  $4.0 \pm 0.4$  \\
\hline
\cite{delaTorre2007}  & VVDS-SWIRE  &  [3.6] < 21.5  &  $ 3.9 \pm 0.5 $   \\
                      &             &  [4.5] < 21    &  $ 4.4 \pm 0.5 $   \\
\hline
\noalign{\vfilneg\vskip -0.2cm}
\hline
\cite{Hatfield2016} & VIDEO, $K_s < 23.5$ & $0.75<z<0.9$ & $5.68${\raisebox{0.5ex}{\tiny$^{+0.13}_{-0.076}$}}  \\
                    &                     & $1.25<z<1.7$ & $6.47${\raisebox{0.5ex}{\tiny$^{+0.11}_{-0.14}$}}   \\
\hline
\cite{McCracken2015}  &    UltraVISTA-DR1  & $0.8<z<1.1$ & $3.5 \pm 0.4$  \\
\hline
\cite{Coil2008} &   DEEP2, optical, red subsample &   $0.7<z<0.925$  &   $5.25 \pm 0.26$  \\
\hline
\cite{Amvrosiadis2019} &  H-ATLAS, $S_{250 \mu m}> 33\ \mu$ Jy &     $1<z<2$      &   $8.1 \pm 0.5$   \\
\hline
\cite{Starikova2012}   &   SWIRE, $S_{24 \mu m}< 310\ \mu$Jy   &  $<z> = 0.7$  &   $ 4.98 \pm 0.28 $   \\ 
                       &                &  $<z> = 1.7$  &   $ 8.04 \pm 0.69 $   \\
\hline
\noalign{\vfilneg\vskip -0.2cm}
\hline
\end{tabular}
\caption{Compilation of clustering measurements from the literature, as discussed in the main text.
The top part of the table covers magnitude- or flux-selected samples, the bottom part lists clustering strengths
for samples also selected in redshift.}
\label{tab:literature-results}
\end{table*}

\cite{delaTorre2007} studied two samples based on data for the VVDS-SWIRE field: one for the full field
(0.82 deg$^2$) with photometric redshifts, and one for a sub-area (0.42 deg$^2$) for which spectroscopic
redshifts are available \cite{LeFevre2005}. Using these redshifts, they divided both
samples in redshift bins and estimated the spatial clustering strength for each bin.
Averaging over the full redshift range, they find a mean value of
$r_0 = 3.9 \pm 0.5\ h^{-1}$ Mpc for the galaxies selected down to [3.6] < 21.5
and a mean value of $r_0 = 4.4 \pm 0.5\ h^{-1}$ Mpc for the ones selected [4.5] < 21.
Even though they did not go quite as deep, these values are again consistent with what we found.

In order to study more specific populations or redshift ranges, we use subsamples
selected by colour, as we do not have sufficient numbers of photometric or spectroscopic redshift,
and those that we have \cite{Pforr2019} are not homogeneous over the whole survey. The colour
selection works best for blue colours, in the sense that this yields the most restricted redshift
distributions, with a clear peak at $z\approx 0.7$ for the most extreme cut ($[3.6]-[4.5] <$-0.45).
The most extreme red cut ($[3.6]-[4.5] >$0.3) shows a peak at $z\approx 1.5$, but with a tail
towards higher redshifts. Not surprisingly, the 'very blue' subsample (which has the narrowest redshift
distribution) shows the strongest 
clustering, with a clustering strength $r_0$ of $10.52 \pm 0.67 h^{-1}$Mpc. The other three cuts,
yield our blue, red, and 'very red' subsamples, with a broader redshift distribution.
For the 'very red' sample, corresponding to our highest redshift distribution (with a median redshift
of 1.89), we find $9.68 \pm 2.77 h^{-1}$Mpc. The less severe colour cuts (our red and blue samples)
contain more galaxies, thus providing more accurate estimates.
We found $r_0 = 7.83 \pm 0.95 h^{-1}$Mpc for the red ($[3.6]-[4.5]>$0.1) cut (median
redshift of 1.77), and $r_0 = 6.50 \pm 0.51 h^{-1}$Mpc for the blue ($[3.6]-[4.5]<$-0.3) cut
(median redshift of 0.72).

An interesting trend is that the clustering strength increases somewhat with redshift for
the blue, red, and 'very red' subsamples \changedtwo{(which have median redshifts of
0.72, 1.77, and 1.89, respectively)}, although this remains hard to interpret as the
colour selections are not selecting galaxy populations that are exactly comparable, 
\changedtwo{either differing in redshift (red vs. blue), or subsample size (red vs. 'very red')}.
\cite{Furusawa2011} looked at clustering as a function of stellar mass in the SXDS/UDS field,
which measures 0.83 deg$^2$ and includes galaxies down to $K_{AB} = 23.5$,
with photometric redshifts. As their subsample sizes are relatively small, they
fit a power-law with a constant slope of $\delta=0.8$ to their correlation functions.
They only present correlation lengths in bins of stellar mass, so we cannot directly compare
to what we find, but in their Figure 11 they plot the correlation length as a function of
redshift, for different stellar mass bins. They find an increase with redshift for fixed stellar
mass, for all stellar masses. \changedtwo{We find an increase with median redshift for our blue,
red and 'very red' subsamples, raising the question whether these three subsample might contain
galaxies of similar median stellar mass. Looking at \autoref{tab:properties}, this is not the case.
The median stellar mass depends mostly on subsample size, not redshift.
The subsample with the most severe colour cut (and highest median stellar masses) have the
largest clustering strengths. \cite{Furusawa2011} also find that
clustering strength increases with stellar mass.}

\begin{figure*}
\centering
\begin{tabular}{cc}
\includegraphics[width=0.99\columnwidth]{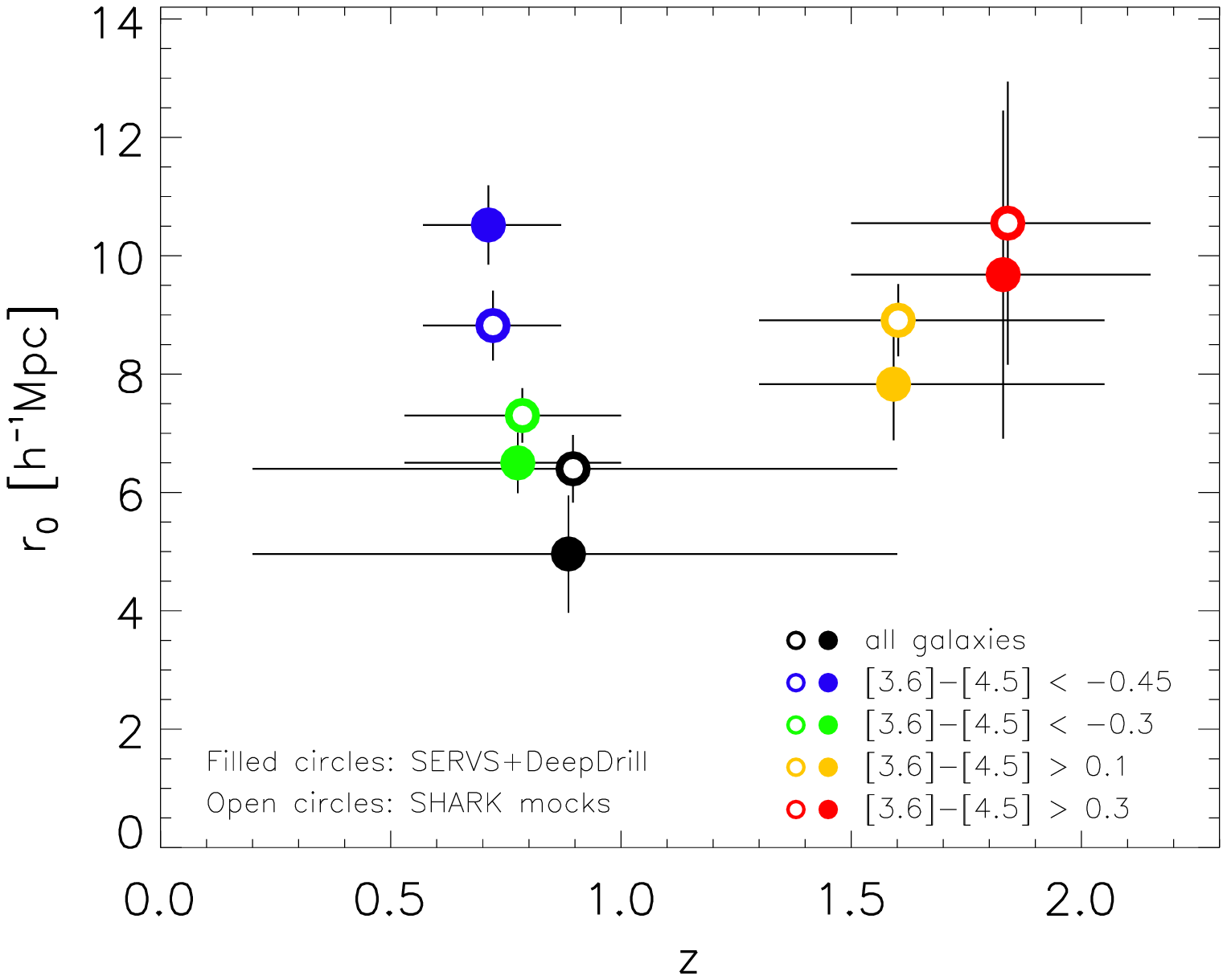} &
\includegraphics[width=0.99\columnwidth]{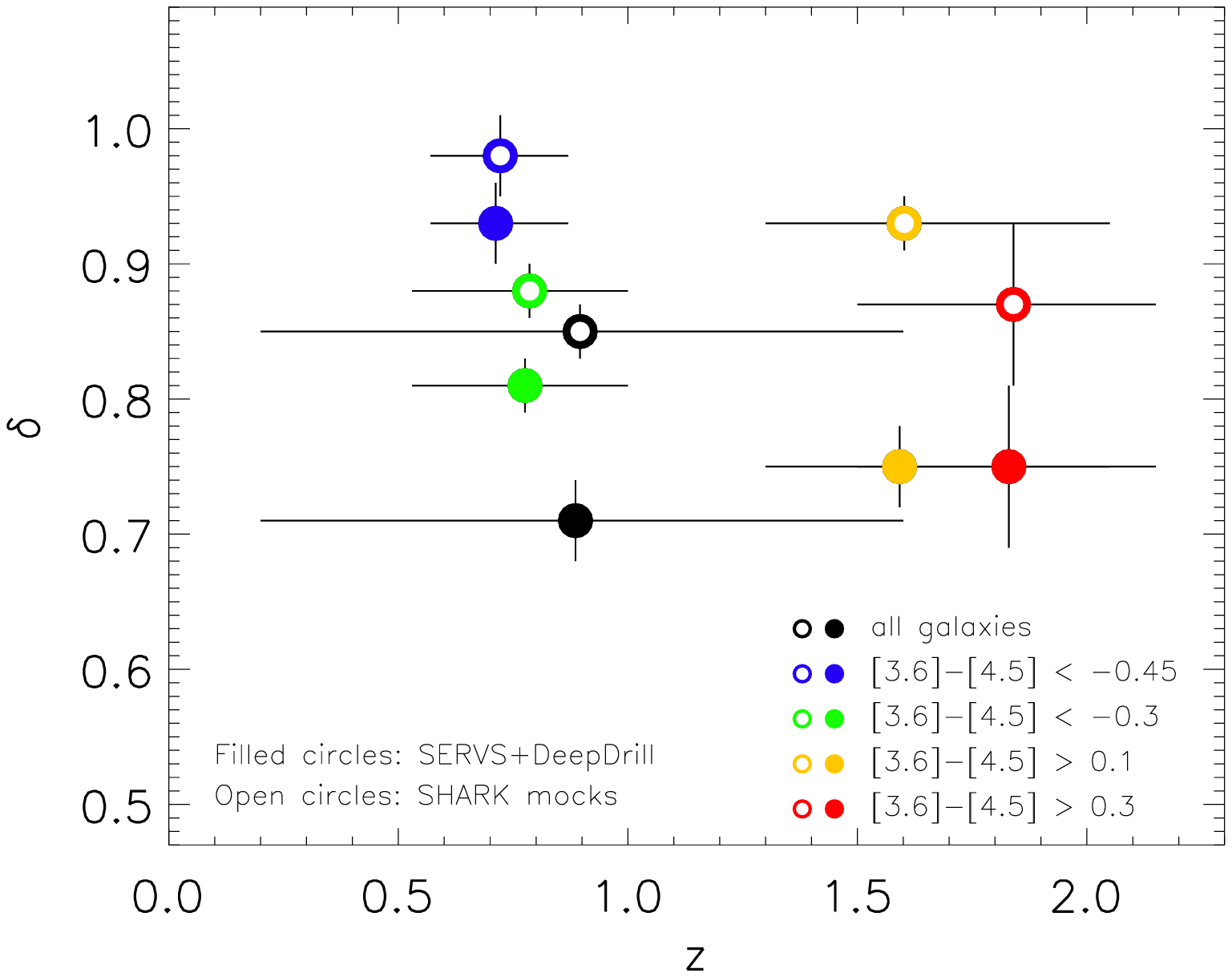} \\
\end{tabular}
\caption{Spatial clustering length and (fitted) clustering slope as a function of redshift
(through colour cuts), for the \servs+\deepdrill\ datasets (filled circles), and the brightened SHARK
mocks (open circles). The vertical bars indicate the uncertainty in the clustering estimate, but
the horizontal bar indicates the redshift range covered by the colour cut, as derived from the S-COSMOS
sample (see \autoref{fig:redshift-distribution} and \autoref{fig:colour-cuts-zdist}). } 
\label{fig:summary}
\end{figure*}

\cite{Hatfield2016} also see this trend, who
studied clustering as a function of stellar mass, amongst others, for the VIDEO survey.
The full galaxy sample used by \cite{Hatfield2016} for their clustering
analysis comprises 97052 sources, for $K_s < 23.5$, and various stellar mass cuts.
As we do not select for stellar mass, but just for redshift, we consider their estimates
for the lowest steller mass threshold they applied (they do not bin their data in stellar mass).
Two redshift ranges overlap with ours: $0.75<z<0.9$, where they have
9791 galaxies for their clustering analysis, and $1.25<z<1.7$, with 10800 galaxies remaining.
The corresponding correlation strength $r_0$ they estimated (see their Table 1) are 
$5.68${\raisebox{0.5ex}{\tiny$^{+0.13}_{-0.076}$} $h^{-1}$Mpc and
$6.47${\raisebox{0.5ex}{\tiny$^{+0.11}_{-0.14}$} $h^{-1}$Mpc, respectively, which compare
fairly well to what we find for our blue and red subsamples, respectively, given that the 
redshift distributions are different despite the overlap.

\changed{A comprehensive study of clustering of K-band (2.2 micron) selected galaxies
in the UltraVISTA-DR1 survey was
performed by \cite{McCracken2015}, who measured their clustering strength as a function of
stellar mass and star formation activity, amongst others. They found that at fixed redshift
and scale, clustering amplitude depends monotonically on the sample stellar mass threshold.
Looking at more specific results that allow for a comparison to our findings, their Figure 7
is most useful, which shows the correlation length as a function of sample median stellar mass.
One of their redshift ranges is $0.8<z<1.1$, which has a similar width as our 'blue' subsample.
We used the SHARK mocks (which have stellar masses listed) to obtain the median of $log M_*/M_\odot$,
as used by \cite{McCracken2015}, for the full 'blue' subsample. We find a value of 10.2, which
corresponds to a clustering length of $3.5 \pm 0.4 h^{-1}$Mpc according to Figure 7 of \cite{McCracken2015},
where the uncertainty is larger than the error bar in their figure to take into account the 
uncertainty in the value of 10.2 for the median of $log M_*/M_\odot$.
This is smaller (by almost a factor of two) than the clustering length of $6.5 \pm 0.51 h^{-1}$Mpc that we
found for our 'blue' subsample,
which is at somewhat lower redshifts, and has a more pronounced peak in its redshift distribution
as compared to the top-hat shape of their $0.8<z<1.1$ redshift selection. A larger spread over
redshift lowers the clustering strength, and their method used to estimate the clustering length
(using a halo model fit) is different from ours. However, it is unclear whether all this sufficiently
explains the difference, and merits further study. }
 
An interesting approach is that by \cite{Cochrane2018}, who use HiZELS to estimate clustering
at very specific redshifts using the H$\alpha$ line. This yields four subsamples, the largest being
over 2300 galaxies in 7.6 deg$^2$ at $z=0.81 \pm 0.011$. They split this in stellar mass bins,
where only galaxies with stellar masses in the range $4\times10^{10} M_\star$ to $10^{11} M\star$
produce a clustering strength close to our blue subsample (their Figure 4). What we can also derive
from their work is that our more extreme blue cut, with its higher clustering strength, again
seems to select higher stellar masses on average, as we found above. Also, their $z=1.47$
redshift sample (451 galaxies over an area of 2.3 deg$^2$), best corresponding to our red subsample,
also has a larger clustering strength for all stellar mass bins, as in \cite{Furusawa2011}.

In the optical,
\cite{Coil2008} measured clustering in the DEEP2 survey around $z\sim 1$, for
subsamples which all start at $z>0.7$, and go down to $z\approx 0.8-1.2$ depending on the selection
for $U-B$ colour in the optical. Their 'main' red and blue subsamples have a redshift range of
$0.7-0.925$ and $0.7-1.05$ respectively, which in both cases is 
somewhat beyond our blue selected galaxies ($[3.6]-[4.5]<$-0.3), which peak at $z\approx 0.7$,
and have a range that best matches their 'main' red (in the optical) subsample.
For this subsample they find a spatial clustering
length $r_0 = 5.25 \pm 0.26 h^{-1}$ Mpc, which is comparable to what we find for our blue
(in the NIR) subsample.

In the far-IR wavebands, \cite{Amvrosiadis2019} quote a correlation length $r_0 = 8.1 \pm 0.5$ $h^{-1}$Mpc
for a redshift range $1<z<2$. For our redshift ranges we find $r_0=7.83\pm0.95$ $h^{-1}$Mpc and
$r_0=9.68\pm2.77$ $h^{-1}$Mpc for our red and 'very red' subsamples, respectively.
As the redshift ranges are somewhat different (as are the observed bands),
\changed{these results are remarkably similar, keeping in mind that the NIR flux
is a proxy for the stellar mass while the FIR flux is a proxy for the star formation rate}.

\cite{Starikova2012} provides clustering estimates in two different redshift
ranges that are not too dissimilar to ours: they found $r_0 = 4.98 \pm 0.28 h^{-1}$ Mpc
for a population with a mean redshift $<z> = 0.7$, similar to our blue subsamples,
and $r_0 = 8.04 \pm 0.69 h^{-1}$ Mpc for and a population with $<z> = 1.7$,
which is similar to our red subsamples. Their first estimate quote above compares
well to our broader $[3.6]-[4.5]<$-0.3 subsample, the latter is consistent with our
'very red' $[3.6]-[4.5]>$0.3 subsample, and with the $1<z<2$ clustering length found
by \cite{Amvrosiadis2019}.

From these comparisons to existing estimates in the literature for a range of samples and
wavelength ranges, we find that our findings are consistent, although it remains difficult
to reach conclusions from such comparisons as the selected galaxy populations can be fairly
different, despite overlapping redshift ranges. There will be overlap in galaxy properties
as well, like stellar mass as we saw above, but there is likely to be significant spread,
making the overlap far from complete.

This is not an issue for simulated galaxy populations, like the ones derived from the SHARK
semi-analytical galaxy formation model \citep{Lagos2019}, as these are complete by definition
and the same selections as for the observations can easily be applied.
We had to offset the fluxes for the simulated galaxies somewhat to match overall
number densities found for our flux cuts, brightening each galaxy by -0.565 mag at 3.6 micron,
and -0.61 mag at 4.5 micron (thus also including a minor colour correction to get the number densities
right for the colour cuts). The reason this is necessary is likely to be the absence of AGN
modelling in the current SHARK semi-analytical models, as can be seen from the counts \citep{Lacy2021},
\changed{and from \autoref{fig:agns}, which is based on the work of \cite{Thorne2022}.}
There are other model components that can act in this direction as well, such as changes
in the star formation history, dust treatment, and adopted stellar population synthesis models.
We also applied a small completeness correction, as the models are complete down to our flux limit,
but the observed data set is 'only' $\sim90$ percent complete. We investigated mock samples
with and without brightening, and found that the mocks with brightening best match the observed
clustering estimates (and also match number counts and colours, by construction).
\changedtwo{This is encouraging, and we take this as a motivation to improve the models, through
the inclusion of AGN modelling, as well as different population synthesis and dust models.
The amount of brightening that we needed to apply will need to be matched by future mocks.}

The current mock datasets were treated in the same way as the observed ones, applying the
same flux and colour cuts, and the correlation functions obtained in a robust way. These were again
fitted to a single power-law (parameters are listed in \autoref{tab:corr-results}).
As different colour cuts select for different redshift ranges, we can compare the estimated
clustering parameters for these redshift ranges, which is is shown in \autoref{fig:summary} for the
spatial clustering length $r_0$ (left-hand side panel) and the fitted slope $\delta$
(right-hand side panel), for the \servs+\deepdrill\ sample (filled circles) as well as the
brightened SHARK mocks (open circles).
The vertical bars indicate the uncertainty in the fitting to the clustering estimates,
\changedtwo{but do not include systematic uncertainties like cosmic variance, although these
should be fairly small owing to the large angular size of our survey.}
The horizontal bar denotes the redshift range for each (sub)sample, as derived from the S-COSMOS
sample. For the full sample, this is the one shown in \autoref{fig:redshift-distribution}. For the
subsamples selected on colour, the redshift distributions from \autoref{fig:colour-cuts-zdist} are shown.
We find that for the full (brightened) mock sample (down to our flux limit) and the four colour
selected subsamples, the fitted slopes are all somewhat larger than for the
\servs+\deepdrill\ sample \changed{(see right-hand panel of \autoref{fig:summary})},
which translates to the spatial clustering strength $r_0$ being somewhat larger in most cases
(except for the 'very blue' subsample), \changed{as can be seen in the left-hand panel of 
\autoref{fig:summary}}. \changedtwo{The slopes for the full sample and the red sample show the
most significant mismatch between the data and the models, with a less significant mismatch
for the 'very red' subsample. However, below $z<1$ the slopes match well. This might indicate
that the brightening of the mocks should be redshift dependent, or there is another redshift
dependence in the treatment of the dust or the modelling of the stellar populations.}

In all cases the power-law fits for the mock and real data cross over each other at a given length scale,
which is on the order of a few arcminutes for the full sample and the blue subsample,
about an arcminute for both red samples, and below an arcminute for the 'very blue' subsample,
which is also the most clustered one. So there seem to be a systematic offset in clustering
strength, the origin of which is not readily apparent. This is not a large offset, and might be due
to the absence of an AGN contribution to the SHARK mock galaxy fluxes:
galaxy clustering depends fairly strongly on luminosity and colour (e.g. \citealt{Zehavi2011}),
so if the addition of AGN emission to the mock galaxy fluxes affects a specific environment
more than others, adding AGN emission to the models might well remove the small difference between
the brightened model and the data we currently find. Alternatively, other changes might need to be made
to the modelling, like aspects related to dust treatment and stellar population modelling.
\changedtwo{There is also an indication that these changes should be redshift dependent. All this will be
explored in a follow-up paper for newer versions of SHARK, but also for other models for galaxy
formation and evolution that are able to produce large mock catalogues of near-infrared galaxies.}

\section*{Acknowledgements} 

This work and S-COSMOS are based on observations made with the Spitzer Space Telescope,
which is operated by the Jet Propulsion Laboratory (JPL),
California Institute of Technology, under a contract with NASA.
Support for \servs\ was provided by NASA through an award issued by JPL/ Caltech.
\deepdrill\ was funded by a {\em Spitzer} grant associated with program ID 11086.
The National Radio Astronomy Observatory is a facility of the National Science 
Foundation operated under cooperative agreement by Associated Universities, Inc.
This publication also makes use of data products from the Two Micron All Sky Survey,
which is a joint project of the University of Massachusetts and the
Infrared Processing and Analysis Center/California Institute of Technology,
funded by the National Aeronautics and Space Administration and the
National Science Foundation. Basic research in Radio Astronomy at the
U.S. Naval Research Laboratory is supported by 6.1 Base Funding.
CL has received funding from the ARC Centre of Excellence for All Sky Astrophysics
in 3 Dimensions (ASTRO 3D), through project number CE170100013.
SHARK has been supported by resources provided by The Pawsey Supercomputing Centre with
funding from the Australian Government and the Government of Western Australia.
We are grateful to J.C. Mauduit and J. Pforr for their
work on the SERVS data processing and catalogues.

\section*{Data availability}

The data products from both \servs\ and \deepdrill\ used for this paper (comprising
the dual-band catalogues and bright star masks) are available for download from IRSA
({\tt irsa.ipac.caltech.edu/data/SPITZER/DeepDrill}). This resource makes available mosaic images,
coverage maps, uncertainty images and bright star masks. Each field has two single-band catalogues
cut at 5$\sigma$ and a dual-band catalogue requiring a detection at > 3$\sigma$ at both 3.6 and
4.5$\mu$m. The simulated light-cone catalogue from SHARK is also included in the release,
its columns are described in Table 11 of \cite{Lacy2021}.

\bibliographystyle{mnras}

\bsp	
\label{lastpage}
\end{document}